\documentclass[prd,aps,twocolumn,a4paper,floatfix,superscriptaddress]{revtex4}

\def\p{\partial}
\def\xb{\bar{x}}
\def\yb{\bar{y}}
\def\zb{\bar{z}}
\def\xh{\hat{x}}
\def\yh{\hat{y}}
\def\zh{\hat{z}}
\def\xti{\tilde{x}}
\def\yti{\tilde{y}}
\def\zti{\tilde{z}}
\def\iti{\tilde{i}}

\newcommand{\dV}{\mathrm{d}V\,}
\newcommand{\dA}{\mathrm{d}A\,}
\newcommand{\dx}{\mathrm{d}x\,}
\newcommand{\dy}{\mathrm{d}y\,}
\newcommand{\dz}{\mathrm{d}z\,}
\newcommand{\dVti}{\mathrm{d}\tilde{V}\,}

\newcommand{\dxti}{\mathrm{d}\tilde{x}\,}
\newcommand{\dyti}{\mathrm{d}\tilde{y}\,}
\newcommand{\dzti}{\mathrm{d}\tilde{z}\,}

\newcommand{\ud}{u^\dagger}

\newcommand{\del}{\partial}
\newcommand{\lie}{\mathcal{L}}
\newcommand{\bma}{\begin{pmatrix}}
\newcommand{\ema}{\end{pmatrix}}

\usepackage{graphicx,psfrag}
\usepackage{mathrsfs}
\usepackage{amsmath,amsfonts,amssymb,amsthm}
\usepackage{url}
\usepackage{comment}

\usepackage{color}
\usepackage{float}
\usepackage{ulem}

\definecolor{gray}{rgb}{0.8,0.8,0.8} 
\definecolor{cyan}{rgb}{0,0.9,0.9} 
\definecolor{orange}{rgb}{0.9,0.5,0} 
\definecolor{magenta}{rgb}{1,0,1}

\begin{document}
%%%%%%%%%%%%%%%%%%%%%%%%%%%%%%%%%%%%%%%%%%%%%%%%%%%%%%%%%%%%%

\title{A Pseudospectral Method for Gravitational Wave Collapse}

\author{David Hilditch}
\affiliation{Friedrich-Schiller-Universit\"at Jena, 07743 Jena, Germany}

\author{Andreas Weyhausen}
\affiliation{Friedrich-Schiller-Universit\"at Jena, 07743 Jena, Germany}

\author{Bernd Br\"ugmann}
\affiliation{Friedrich-Schiller-Universit\"at Jena, 07743 Jena, Germany}

\begin{abstract}
We present a new pseudospectral code, \texttt{bamps}, for numerical relativity 
written with the evolution of collapsing gravitational waves in mind. 
We employ the first order generalized harmonic gauge formulation. The 
relevant theory is reviewed and the numerical method is critically 
examined and specialized for the task at hand. In particular we investigate 
formulation parameters, gauge and constraint preserving boundary conditions 
well-suited to non-vanishing gauge source functions. Different types of 
axisymmetric twist-free moment of time symmetry gravitational wave initial data 
are discussed. A treatment of the axisymmetric apparent horizon condition 
is presented with careful attention to regularity on axis. Our apparent 
horizon finder is then evaluated in a number of test cases. Moving on to 
evolutions, we investigate modifications to the generalized harmonic 
gauge constraint damping scheme to improve conservation in the strong 
field regime. We demonstrate strong-scaling of our pseudospectral penalty 
code. We employ the Cartoon method to efficiently evolve axisymmetric data 
in our~$3+1$ dimensional code. We perform test evolutions of Schwarzschild 
perturbed by gravitational waves and by gauge pulses, both to demonstrate 
the use of our blackhole excision scheme and for comparison with earlier 
results. Finally numerical evolutions of supercritical Brill waves are 
presented to demonstrate durability of the excision scheme for the dynamical 
formation of a blackhole.
\end{abstract}

\pacs{
  95.30.Sf,   % relativity and gravitation
  04.25.D-   % numerical relativity
}

\maketitle

%\tableofcontents

%%%%%%%%%%%%%%%%%%%%%%%%%%%%%%%%%%%%%%%%%%%%%%%%%%%%%%%%%%%%%
\section{Introduction}\label{section:Introduction}
%%%%%%%%%%%%%%%%%%%%%%%%%%%%%%%%%%%%%%%%%%%%%%%%%%%%%%%%%%%%%

This is the first in a series of papers about the numerical 
treatment of collapsing gravitational waves using a new 
pseudospectral code developed for the problem. In the 
early~$1990$s critical phenomena were discovered in gravitational 
collapse~\cite{Cho91}, in spherical symmetry, with general relativity 
minimally coupled to a massless scalar field. One aspect of the finding 
was that, amazingly, the critical solution dividing the formation of 
a blackhole from dissipation of the field, was unique, in the sense 
that if one takes any one parameter family of initial data, with 
the parameter controlling somehow the strength of the data, and tunes 
this parameter to the threshold of blackhole formation, one finds 
that the same solution is always obtained, regardless of the 
family! Shortly thereafter similar phenomenology was reported in 
axisymmetric, vacuum general relativity~\cite{AbrEva92}, or in other 
words in the collapse of gravitational waves. Since then multiple 
studies have been performed to reproduce this finding, albeit with 
different initial data and numerical approaches, but without success. 
Perhaps most strikingly, in~\cite{Sor10}, numerical evidence of a 
different critical solution was presented. Even if one completely accepts 
the available evidence for criticality in vacuum collapse, this obviously 
begs the question whether or not the naive expectation of uniqueness 
of the critical solution in axisymmetric, rather than spherical, collapse 
holds.

Roughly speaking there are two types of code being used used in~3d 
numerical relativity. The first uses the moving puncture 
method~\cite{BakCenCho05,CamLouMar05}, 
which consists, in essence, of a clever choice of evolved variables and 
gauge conditions, normally treated numerically by finite differencing. 
Secondly is the pseudospectral method, most prevalently used with a 
first order generalized harmonic formulation of general relativity by the 
SpEC code~\cite{SpEC}. Recently, we presented a study of the collapse of 
gravitational waves using the moving puncture method~\cite{HilBauWey13}, in part 
to establish how close to the critical regime one can get with this standard 
approach. The conclusion being; not very. Major difficulties included the formation of 
coordinate singularities and a lack of accuracy. Therefore one would like to tackle 
the problem using the pseudospectral approach to establish what can be achieved 
in that setting. We have thus developed a new pseudospectral code along the lines 
of SpEC, specializing the continuum and numerical method as much as possible towards 
the problem of vacuum gravitational collapse. The present paper represents the 
first outcome of this maneuver. Herein we describe the formulation of GR employed, 
our boundary conditions, the code, calibration of the method, our initial data, 
our approach to axisymmetric apparent horizons, plus a suite of validation tests 
for gauge waves, gravitational waves, blackhole and collapse spacetimes. Throughout 
we compare our results carefully with those in the literature. We aim to give a 
body of evidence for the correctness of the method that the reader will find 
compelling. With this out of the way, in subsequent papers we turn to the problem 
of critical collapse. A summary follows before the main text. 

In section~\ref{section:GHG} we look at a slightly modified version of the first 
order generalized harmonic formulation of~\cite{LinSchKid05}. We consider constraint 
preserving, radiation controlling boundary conditions, paying special attention to 
the constraint preserving boundaries. By considering the reflection of outgoing waves
in the linear approximation we ultimately suggest modified conditions that should 
reduce spurious reflections caused by the use of constraint damping. We also suggest 
alternative gauge boundary conditions.

Next, in section~\ref{section:bamps}, we outline the \texttt{bamps} code, including our 
carefully constructed cubed-sphere grids, which avoid clustering of grid-points in 
unfortunate positions of the domain. For the discretization we employ a pure Chebyschev 
approach. We also discuss our `octant' symmetry implementation, the crucial patching-penalty 
approach for communicating data between neighboring coordinate patches, and finally the 
boundary implementation. In the follow-up section~\ref{section:EnergyPenalty} we complete 
the presentation of the penalty method by computing the penalty parameters appropriate for 
the semi-discrete system.

Given the difficulties in the literature in reproducing the results of~\cite{AbrEva92} it 
seems necessary to solve the problem in axisymmetry before moving to examine the collapse of 
fully~3d waves without symmetry. In our moving puncture gauge study~\cite{HilBauWey13} a major 
disadvantage in using the BAM code was that 3d grids were employed to evolve axisymmetric data. 
In section~\ref{section:Axi} we present our approach to evolving axisymmetric spacetimes with
the \texttt{bamps} code, for which we employ the Cartoon method~\cite{AlcBraBru99a} to reduce from the 
standard \texttt{bamps} 3d domains to a plane, by using the Killing vector to evaluate any angular 
derivatives. We discuss various flavors of axisymmetric 
moment-of-time-symmetry initial data and their numerical construction. These initial data sets 
are evolved in a forthcoming study. We also give a detailed description of our formulation of 
the apparent horizon conditions in axisymmetry. To the best of our knowledge this is the first 
time that the regularity conditions on the symmetry axis have been carefully taken care of. This 
is important in later work as the search for apparent horizons will be our key diagnostic tool.

The next three sections~(\ref{section:Experiments}-\ref{section:Supercritical}) contain a write-up
of our development and validation tests. The tests include evolutions with the proposed gauge 
boundary conditions, which we find are helpful when using large gauge source parameters, as 
desired. They also include runs comparing the fully 3d, octant symmetry and Cartoon evolutions, 
demonstrating that the various symmetry setups are well-behaved. In the evolution of single 
blackholes we test different gauges and boundary conditions, and following~\cite{LinSchKid05}, 
look at evolutions in which the blackhole is perturbed by a gravitational wave injected through 
the outer boundary. Our results are in good agreement with the earlier studies. We then examine the 
evolution of supercritical Brill waves, where, after the formation of an apparent horizon the 
run is continued after interpolation onto an excision grid, as used to evolve a single blackhole, 
which is needed to evolve data with a horizon for long-times. Finally we conclude in 
section~\ref{section:Conclusions}.

%%%%%%%%%%%%%%%%%%%%%%%%%%%%%%%%%%%%%%%%%%%%%%%%%%%%%%%%%%%%%
\section{The generalized harmonic formulation and boundary 
conditions}\label{section:GHG}
%%%%%%%%%%%%%%%%%%%%%%%%%%%%%%%%%%%%%%%%%%%%%%%%%%%%%%%%%%%%%

%%%%%%%%%%%%%%%%%%%%%%%%%%%%%%%%%%%%%%%%%%%%%%%%%%%%%%%%%%%%%
\subsection{GHG, constraints, boundary conditions}
\label{subsection:GHG}
%%%%%%%%%%%%%%%%%%%%%%%%%%%%%%%%%%%%%%%%%%%%%%%%%%%%%%%%%%%%%

%%%%%%%%%%%%%%%%%%%%%%%%%%%%%%%%%%%%%%%%%%%%%%%%%%%%%%%%%%%%%
\paragraph*{The evolution system:} We use the first order 
reduction of the generalized harmonic formulation with several 
free parameters. The full reduction from the second order 
Einstein equations is presented in detail elsewhere~\cite{LinSchKid05} 
so here we give only a brief overview to establish our notation. 
Throughout the paper in continuum equations we use the 
latin~$a,b,c\dots$ for four dimensional indices, but~$i,j,k\dots$ 
for spatial indices, with the exception of~$n$ and~$s$, whose meaning
when used as indices will be described shortly. Greek indices are 
used to refer to the position in a state-vector, grid indices, 
or where otherwise needed. 
We start from the vacuum generalized harmonic formulation in second order form,
\begin{align}
R_{ab}&=\nabla_{(a}C_{b)}+\gamma_4 \Gamma^c{}_{ab}C_c
-\tfrac{1}{2}\gamma_5 g_{ab}g^{cd}\Gamma^e{}_{cd}C_e\nonumber\\
&\quad-\gamma_0[n_{(a}C_{b)}-g_{ab}n^cC_c]\,,\label{eqn:GHG_22}
\end{align}
for the unknown spacetime metric~$g_{ab}$ with Christoffels~$\Gamma^c{}_{ab}$.
The constraints of the system are~$C_a=g^{bc}\Gamma_{abc}+H_a=0$, plus the standard 
Hamiltonian and momentum constraints of GR. The gauge source 
functions~$H_a$ are freely specifiable, provided that they do not include 
derivatives of the metric, which would affect the principal part of 
the PDE. The terms involving~$\gamma_0$ are included so as to damp 
away high-frequency constraint violations~\cite{GunGarCal05}. The 
parameters~$\gamma_4$ and~$\gamma_5$ control whether or not the constraint 
addition made in the construction of the formulation is done either with 
the covariant or the partial derivative, or some combination. The latter 
choice has the effect of simplifying the constraint subsystem. 
In the code we use a first order reduction by introducing the 
variables~$\Phi_{iab}$ and~$\Pi_{ab}$. The equations of motion are,
\begin{align}
\p_tg_{ab}&=\beta^i\p_ig_{ab}-\alpha\Pi_{ab}+\gamma_1\beta^iC_{iab}\,,
\nonumber\\
\p_t\Phi_{iab}&=\beta^j\p_j\Phi_{iab}-\alpha\p_i\Pi_{ab}
+\gamma_2 \alpha C_{iab}
+\tfrac{1}{2}\alpha n^cn^d\Phi_{icd}\Pi_{ab}\nonumber\\
&\quad+\alpha \gamma^{jk}n^c\Phi_{ijc}\Phi_{kab}\,,\nonumber\\
\p_t\Pi_{ab}&=\beta^i\p_i\Pi_{ab}-\alpha \gamma^{ij}\p_i\Phi_{jab}
+\gamma_1\gamma_2\beta^iC_{iab}\nonumber\\
&\quad+2\alpha g^{cd}\big(\gamma^{ij}
\Phi_{ica}\Phi_{jdb}-\Pi_{ca}\Pi_{db}-g^{ef}
\Gamma_{ace}\Gamma_{bdf}\big)\nonumber\\
&\quad-2\alpha\big(\nabla_{(a}H_{b)}+\gamma_4\,\Gamma^c{}_{ab}C_c
-\tfrac{1}{2}\gamma_5\,g_{ab}\Gamma^cC_c\big)\nonumber\\
&\quad-\tfrac{1}{2}\alpha n^cn^d\Pi_{cd}\Pi_{ab}
-\alpha n^c\gamma^{ij}\Pi_{ci}\Phi_{jab}\nonumber\\
&\quad+\alpha\gamma_0\big[2\delta^c{}_{(a}n_{b)}-g_{ab}n^c\big]C_c\,,
\label{eqn:GHG11evsys}
\end{align}
with shorthands to be defined momentarily.
The formulation here agrees with that of~\cite{LinSchKid05} except 
for the inclusion of the~$\gamma_4$ and~$\gamma_5$ parameters. We will 
either take the new parameters to vanish, or choose~$\gamma_4=\gamma_5=1/2$. 
The lapse and shift are denoted~$\alpha$ and~$\beta^i$ respectively. The 
unit normal to the spatial slices of constant coordinate time~$t$ is 
written~$n^a$. When the normal is contracted with a tensor we  
sometimes use the abbreviation~$S_{an}=S_{ab}n^b$, and likewise for the 
arbitrary unit spatial vector~$s^a$. The induced metric 
on the slice is written~$\gamma_{ij}$. In matrix notation this system
can be written as
\begin{align}
\p_tu^\mu&=A^{k\mu}{}_\nu\p_k u^\nu+S^\mu,
\end{align}
with~$u^\mu=(g_{ab},\Pi_{ab},\Phi_{iab})^T$, and principal matrix,
\begin{align}
{A^{p\mu}}_\nu =
\begin{pmatrix}
(1+\gamma_1) \beta^k        &  0                   &  0 \\
\gamma_1\gamma_2 \beta^k    & \beta^k              &  -\alpha \gamma^{ik} \\
\gamma_2 \alpha \delta^k_i  &  -\alpha \delta^k_i  &  \beta^k
\end{pmatrix}\,,
\end{align}
and~$S^\mu$ containing all source terms. We use the shorthand for 
the Christoffel symbols under the first order reduction,
\begin{align}
\Gamma_{abc}&=\gamma^i{}_{(b|}\Phi_{i|c)a}
-\tfrac{1}{2}\gamma^i{}_a\Phi_{ibc}+n_{(b}\Pi_{c)a}
-\tfrac{1}{2}n_a\Pi_{bc}\,,
\end{align}
and will frequently use the abbreviation~$\Gamma^a=g^{bc}\Gamma^a{}_{bc}$.
The system is symmetric hyperbolic, having the same principal part as a 
particular first order reduction of the wave equation. The characteristic 
variables are given by,
\begin{align}
   u^{\hat 0}_{ab}   &=   g_{ab}\,,\nonumber\\
   u^{\hat \pm}_{ab} &= \Pi_{ab} \mp s^i \Phi_{iab}- \gamma_2\,g_{ab}\,,\nonumber \\
   u^{\hat{\beta}}_{Aab}         &= q^i{}_A\Phi_{iab}\,,\label{eqn:GHG11_char_vars}
\end{align}
with the projection operator~$q^j{}_i=\delta^j{}_i-s^js_i$, and 
speeds,
\begin{align}
v^{\hat{0}}     = (1+\gamma_1)\beta^s\,,            &\quad\quad\quad
v^{\hat{\pm}}   = \beta^s\pm\alpha\,, &\quad
v^{\hat{\beta}} = \beta^s \,,\label{eqn:GHG11_char_speeds}
\end{align}
respectively. For future reference let us also note that a 
convenient way to transform to the characteristic variables is 
to write~$u^{\hat\alpha}=T^{-1\, \alpha}{}_\beta\,u^\beta$, where here 
the indices represent the position in the state-vector~$u^{\hat\alpha}$
and where the similarity matrix is,
\begin{align}
T^{-1\,\hat\alpha}{}_\mu&=
\left(\begin{array}{ccc}
 1 & 0 & 0\\
-\gamma_2 & 1 & -s^i\\
-\gamma_2 & 1 & s^i\\
 0 & 0 & q^i{}_{j}
\end{array}\right)\,,
\end{align}
which has left inverse~$T^{\mu}{}_{\hat\alpha}$. But note however 
that~$T^{-1\,\hat\alpha}{}_{\mu}\,T^{\mu}{}_{\hat{\beta}}\ne\delta^{\hat\alpha}{}_{\hat\beta}$.
The strength of this representation in practical terms is in 
avoiding special cases in the numerical implementation, like 
for example~$s^x=0$, in the characteristic decomposition.

%%%%%%%%%%%%%%%%%%%%%%%%%%%%%%%%%%%%%%%%%%%%%%%%%%%%%%%%%%%%%
\paragraph*{Gauge source functions:} For the gauge source 
functions~$H_a$ we choose,
\begin{align}
H_a&=\eta_L \log\left(\frac{\gamma^{p/2}}{\alpha}\right)n_a
-\frac{\eta_S}{\alpha^2}\gamma_{ai}\beta^i.\label{eqn:Ha}
\end{align}
Our convention differs from that of both~\cite{SziLinSch09} 
and~\cite{Sor10} in a trivial normalization of the spatial part 
with respect to the lapse function. Writing the resulting gauge
conditions in terms of the lapse and shift we get, 
\begin{align}
\p_t\alpha&=-\alpha^2 K+\eta_L\alpha^2\log\left(\tfrac{\gamma^{p/2}}
{\alpha}\right)
+\beta^i\p_i\alpha\,,\nonumber\\
\p_t\beta^i&=\alpha^2\,{}^{\textrm{\tiny{(3)}}}\Gamma^i-\alpha\,\p^i\alpha
-\eta_S\beta^i+\beta^j\p_j\beta^i\,,\label{eqn:3+1_ghg}
\end{align}
with~$K$ the trace of the extrinsic curvature 
and~${}^{\textrm{\tiny{(3)}}}\Gamma^i$ the contracted Christoffel symbol 
of the spatial metric. Before blackhole formation for the scalar 
functions~$\eta_L,\eta_S$ we choose,
\begin{align}
\eta_L&=\bar{\eta}_L\alpha^q\,,&\quad
\eta_S=\bar{\eta}_S\alpha^r\,,
\end{align}
with~$\bar{\eta}_L,\bar{\eta}_S,q,r$ some constants. By default we 
choose~$p=1$ and~$q=r=0$, which naturally maintains the shift damping 
term even if the lapse function is close to zero, in contrast to 
the standard condition employed in SpEC~\cite{SziLinSch09}, which 
takes~$r=1$. Since we wish to study near-singular gravitational effects 
in the computational domain and avoid run-away growth of the shift vector 
this seems reasonable. We will report in later work on adjustments to 
these choices when evolving near-critical data. When evolving blackholes 
by excision we follow~\cite{SziLinSch09} taking instead~$r=1$, although 
so far we have not found it necessary to use the~$\log^2$ form 
of~$\eta_L$. 

%%%%%%%%%%%%%%%%%%%%%%%%%%%%%%%%%%%%%%%%%%%%%%%%%%%%%%%%%%%%%
\paragraph*{The constraint subsystem:} The first order reduced 
harmonic constraints are,
\begin{align}
C_a &= H_a+\gamma^{ij}\Phi_{ija}
-\tfrac{1}{2}\gamma_a{}^ig^{cd}\Phi_{icd}
+n^b\Pi_{ab}-\tfrac{1}{2}n_ag^{bc}\Pi_{bc}\,.
\end{align}
The terms without~$H_a$ are simply~$\Gamma_a$. In these variables 
the vacuum ADM Hamiltonian and momentum constraints can be 
expressed as
\begin{align}
2G_{nn}&=\gamma^{ij}\gamma^{kl}
\big(\p_k\Phi_{ijl}-\p_k\Phi_{lij}
+\Gamma^a{}_{jk}\Gamma_{ail}\nonumber\\
&\quad\quad-\Gamma^a{}_{ij}\Gamma_{akl}\big)\,,\nonumber\\
-\gamma^a{}_i\,G_{na}&=\gamma^{jk}\big(\p_{[j}\Pi_{i]k}+\tfrac{1}{2}d_{j}\Phi_{kin}
-\tfrac{1}{2}d_{i}\Phi_{jkn}
\nonumber\\
&\quad\quad-\tfrac{1}{2}\Pi_{j[i}\Phi_{j]nn}+\gamma^{lm}\Phi_{mk[j}\Phi_{i]ln}\nonumber\\
&\quad\quad+2\Gamma_{an[i}\,\Gamma^a{}_{k]j}\big)\,.
\end{align}
As stated above we use a subscript~$n$ to denote contraction with the normal 
vector~$n^a$, but with the convention that~$d_i$ stands for the partial 
derivative, but {\it with any such contraction outside of the derivative.} We 
can put the Hamiltonian and momentum constraints together as a four-vector of 
constraints,
\begin{align}
M^a&=G^{an}\,.
\end{align}
Working with the first order system creates the reduction and closely related 
ordering constraints,
\begin{align}
C_{iab}&=\p_ig_{ab}-\Phi_{iab}=0\,,\nonumber\\
C_{ijab}&=\p_i\Phi_{jab}-\p_j\Phi_{iab}=2\p_{[j}C_{i]ab}=0\,.
\end{align}
The constraints~$C_a$ and~$C_{iab}$ evolve according to, 
\begin{align}
\p_tC_a&=(1+\gamma_1)\beta^i\p_iC_a-\gamma_1\beta^i\bar{\p}_iC_a+ \alpha\,G_a\nonumber\\
&\quad
 + (\gamma_4-\gamma_5)\alpha n_a \Gamma^bC_b 
 - \alpha(2\gamma_4-1)\Gamma^b{}_{an}C_b \nonumber\\
&\quad 
+ 2 \gamma_0 \alpha n^bn_{(a}C_{b)} + \alpha\gamma^{ij}\gamma^{kl}\Phi_{ikn}C_{ljn}n_a
\nonumber\\
&\quad-\alpha\,\gamma^i{}_aC_{ijn}\big[\tfrac{1}{2}g^{bc}\Phi^{j}{}_{bc}
+\Phi^{j}{}_{nn}\big]\nonumber\\
&\quad-\gamma_1\gamma_2\beta^i\big(\tfrac{1}{2}g^{cd}C_{icd}n_a-C_{ina}\big)\,,\nonumber\\
\p_tC_{iab}&=\beta^j\big(\p_jC_{iab}+\gamma_1\p_iC_{jab}\big)
+\alpha\,\big[(1+\gamma_1)d_ig_{jn}\,C^j{}_{ab}\nonumber\\
&\quad-\gamma_2C_{iab}+\Phi^j{}_{ab}\,C_{ijn}+\frac{1}{2}C_{inn}\Pi_{ab}\big]\,,
\end{align}
where we have introduced the constraint,
\begin{align}
G_a&=2M_a+ (n_a \gamma^{ib}-\gamma^i{}_a n^b)(\bar{\p}_iC_b-\Gamma^c{}_{ib}C_c)\nonumber\\
&\quad+\gamma_2(\delta^c{}_a\gamma^{ib}-\tfrac{1}{2}g^{bc}\gamma^i{}_a)C_{ibc}\,,\label{eqn:G_def}
\end{align}
and where the notation~$\bar{\p}_i$ means take the partial derivative, 
and afterwards replace all first derivatives of the metric with the 
reduction variable~$\Phi_{iab}$. Up to lower derivatives in the contraints 
we find,
\begin{align}
\p_tG_{a}&\approx\beta^i\p_iG_a+\alpha\gamma^{ij}\p_i\p_jC_a
-\alpha\gamma^{jk}\gamma^{li}\p_lC_{ijka}\nonumber\\
&\quad+\tfrac{1}{2}\alpha\gamma^j\!{}_a\gamma^{il}g^{cd}\p_lC_{ijcd}\,,
\end{align}
where~$\approx$ denotes equality up to non-principal 
terms, the remainder having been suppressed for brevity. The equation of 
motion for~$C_{ijab}$ is readily derived by taking derivatives of that 
of~$C_{iab}$. Notice that the parameter~$\gamma_2$ serves to damp the 
reduction constraint. In the description of~\cite{LinSchKid05} the equivalent 
reduction variable is called~$F_a$, with, including~$\gamma_4$ and~$\gamma_5$
in the natural way,
\begin{align}
F_a&= G_a-(1-\gamma_4)(n_a\,\Gamma^b-2\,\Gamma^b{}_{an})C_b
-\gamma_5\,n_a \Gamma^b\,C_b\,,\label{eqn:Fa_def}
\end{align}
in our variables. The difference is not substantial, being only that~$G_a$ 
appears slightly more naturally in the second order form of the equations. 
Note that in~\eqref{eqn:Fa_def}, the final term contains a piece which is 
simply the Harmonic constraint in the pure harmonic case, but will act as a 
non-zero coefficient otherwise.

%%%%%%%%%%%%%%%%%%%%%%%%%%%%%%%%%%%%%%%%%%%%%%%%%%%%%%%%%%%%%
\paragraph*{First order reduction of the constraint subsystem:} 
Following~\cite{LinSchKid05}, a first order reduction of the 
constraint subsystem is formally introduced by defining the 
new variable~$C_{ia}$ with,
\begin{align}
C_{ia}&= \gamma^{jk}\p_j\Phi_{ika}-\tfrac{1}{2}\gamma^j_a g^{cd}\p_j\Phi_{icd}
+d_i\Pi_{an}-\tfrac{1}{2}n_ag^{cd}\p_i\Pi_{cd}\nonumber\\
&+\p_iH_a+\tfrac{1}{2}\gamma^j_a\Phi_{j}{}^{cd}\Phi_{icd}
+\tfrac{1}{2}\gamma^{jk}\Phi_{jc}{}^c\Phi_{ikn}n_a\nonumber\\
&-\gamma^{jk}\gamma^{lm}\Phi_{jla}\Phi_{ikm}
+\tfrac{1}{2}\Phi_{icd}\Pi_{be}n_a(g^{cb}g^{de}+\tfrac{1}{2}g^{be}n^cn^d)
\nonumber\\
&-\Phi_{icn}\Pi_{ba}(g^{bc}+\tfrac{1}{2}n^bn^c)
+\tfrac{1}{2}\gamma_2(n_ag^{cd}-2\delta^c_a n^d)C_{icd}\,.
\end{align}
The principal part of this formal reduction is given by,
\begin{align}
\p_tC_a&\approx 0\,, \nonumber\\
\p_tG_{a}&\approx \beta^i\p_iG_a+\alpha\,\gamma^{ij}\p_iC_{ja}\,,\nonumber\\
\p_tC_{ia}&\approx\beta^j\p_jC_{ia}+\alpha\,\p_iG_a\,,\nonumber\\
\p_tC_{iab}&\approx (1+\gamma_1)\beta^j\p_jC_{iab}\,,\nonumber\\
\p_tC_{ijab}&\approx \beta^k\p_kC_{ijab}\,.
\end{align}
The characteristic variables of the constraint subsystem are then 
found to be 
\begin{align}
c^{\hat{\pm}}_a&=F_a\mp C_{sa}\,,\quad &
c^{\hat{0}}_a&=C_a\,,\nonumber\\
c^{\hat{\beta}}_{Aa}&=q^i{}_AC_{ia}\,,\quad &
c^{\hat{\gamma_1}}_{iab}&=C_{iab}\,,\nonumber\\
c^{\hat{\beta}}_{ijab}&=C_{ijab}\,,
\end{align}
with speeds~$\beta^s\mp\alpha,0,\beta^s,(1+\gamma_1)\beta^s$ and~$\beta^s$ 
respectively, where we use upper case latin indices to denote those projected
by~$q^a{}_b$. A suitable norm of the constraint violation is given by the 
constraint monitor which is defined as
\begin{align}
   C_\text{mon} &= \int \text{d}^3 x \sqrt{\gamma} \Big( 
      \delta^{ab} F_a F_b
   +  \delta^{ab} C_a C_b
   + \gamma^{ij} \delta^{ab} C_{ia} C_{jb} \nonumber\\
   &\qquad + \gamma^{ij} \delta^{ac} \delta^{bd} C_{iab} C_{jcd} 
   + \gamma^{ij} \gamma^{kl} \delta^{ac} \delta^{bd} C_{ikab} C_{jlcd}
    \Big).
\end{align}

%%%%%%%%%%%%%%%%%%%%%%%%%%%%%%%%%%%%%%%%%%%%%%%%%%%%%%%%%%%%%
\paragraph*{The gravitational wave degrees of freedom:} In vacuum 
the Weyl scalars~$\Psi_0,\Psi_4$ can be expressed as,
\begin{align}
\Psi_0&=m^Am^B[\perp^{(P)bd}\!\!\!{}_{AB}l^al^cR_{abcd}]\,,\nonumber\\
\Psi_4&=m^Am^B[\perp^{(P)bd}\!\!\!{}_{AB}k^ak^cR_{abcd}]\,,\label{eqn:Weyl_scalars}
\end{align}
respectively. Here we have introduced the null tetrad 
\begin{align}
l^a&=\tfrac{1}{\sqrt{2}}(n^a+s^a)\,,\quad & 
k^a&=\tfrac{1}{\sqrt{2}}(n^a-s^a)\,,\nonumber\\
m^a&=\tfrac{1}{\sqrt{2}}(v^a+iw^a)\,,\quad & 
\bar{m}^a&=\tfrac{1}{\sqrt{2}}(v^a-iw^a)\,,\label{eqn:NP_tetrad}
\end{align}
with~$s^a,v^a$ and~$w^a$ mutually orthogonal unit spatial vectors, 
and the projection operator,
\begin{align}
\perp^{(P)cd}\!\!_{ab}&=q^c{}_{(a}q^d{}_{b)}-\tfrac{1}{2}q^{cd}q_{ab}\nonumber\\
&=m_{(a}m_{b)}\bar{m}^{(c}\bar{m}^{d)}+\bar{m}_{(a}\bar{m}_{b)}m^{(c}m^{d)}\,.
\end{align}
In terms of the first order GHG variables we can express the 
principal part of the Riemann tensor as,
\begin{align}
R_{abcd}&\approx 
\gamma^j{}_{a}\p_i\Phi_{jb[c}\gamma_{d]}{}^i
-\gamma^j{}_{b}\p_i\Phi_{ja[c}\gamma_{d]}{}^i
+n_a\p_i\Pi_{b[c}\gamma_{d]}{}^i\nonumber\\
&\quad-n_b\p_i\Pi_{a[c}\gamma_{d]}{}^i
+\gamma^i{}_{a}\p_i\Pi_{b[c}n_{d]}
-\gamma^i{}_{b}\p_i\Pi_{a[c}n_{d]}\nonumber\\
&\quad- n_a \gamma^{ij}\p_i\phi_{jb[c}n_{d]} 
+ n_b \gamma^{ij}\p_i\phi_{ja[c}n_{d]}\nonumber\\
&\quad-\gamma_1\gamma_2 n_an^k\p_kg_{b[c}n_{d]}
+\gamma_1\gamma_2 n_bn^k\p_kg_{a[c}n_{d]}\nonumber\\
&\quad-\gamma_2\gamma^i{}_a\p_ig_{b[c}n_{d]}
+\gamma_2\gamma^i{}_b\p_ig_{a[c}n_{d]}\,.
\end{align}
Of course this expression is unique only up to constraint additions. 
Note that upon contraction with~$\perp^{(P)}$ and~$l$ to form the Weyl scalar~$\Psi_0$,
and after a single addition of~$C_{ijab}$, we naturally 
form a projection of the incoming characteristic 
variable~$d_su^{\hat{+}}_{ab}$. This is used in the construction of 
the boundary condition. The spatial vector~$s^i$ is taken to be 
the unit spatial normal to the boundary.

%%%%%%%%%%%%%%%%%%%%%%%%%%%%%%%%%%%%%%%%%%%%%%%%%%%%%%%%%%%%%
\paragraph*{Boundary conditions:} At the outer boundary we need 
to control incoming constraint violation, gauge perturbations 
and physical radiation. By default we initially impose, 
\begin{align}
F_a+C_{sa}+\tfrac{1}{r}C_a\,
\hat{=}\,0\,,\label{eqn:GHG_11_cpbc}
\end{align}
on the constraint subsystem assuming that the characteristic
variable~$c^{\hat+}_a$ is always incoming. These conditions are 
essentially those of~\cite{LinSchKid05}, with just the 
additional~$1/r$ term. Other conditions for this variable will 
be motivated and tested in what follows. The remaining constraint 
subsystem characteristic variables may or may not be incoming, and 
are dealt with on this basis as described in 
section~\ref{subsection:BC_imp}, but always according to the same 
prescription. For the gravitational wave degrees of freedom we 
choose,
\begin{align}
\Psi_0 &\,\hat{=}\,q_0\,,\label{eqn:GHG_11_Psi0_bc}
\end{align}
the lowest order member of a cascade of conditions on incoming 
radiation~\cite{BucSar06,BucSar07}, with given data~$q_0$. 
Examining~\eqref{eqn:Weyl_scalars} it is obvious that this is 
equivalent to setting,
\begin{align}
\perp^{(P)bd}\!\!\!{}_{AB}(l^al^cR_{abcd})
&=\,\perp^{(P)bd}\!\!\!{}_{AB}\,q^{(P)}_{bd}\,,
\end{align}
which is in practice how the conditions are implemented. For the 
remaining gauge degrees of freedom we choose either the improved 
gauge boundary conditions of~\cite{RinLinSch07},
\begin{align}
\perp^{(G)cd}_{ab}d_t\big[u^{\hat{+}}_{cd}
+(\gamma_2-r^{-1})g_{cd}\big]&\,\hat{=}\,0\,,\label{eqn:G_BCs}
\end{align}
or the alternative, 
\begin{align}
&\perp^{(G)cd}_{ab}\Big[\,
d_su_{cd}^{\hat{+}}-2\bar{d}_s[n_{(c}H_{d)}]+\gamma_2\Phi_{scd}\nonumber\\
&\qquad\qquad+r^{-1}(u_{cd}^{\hat{+}}-2n_{(c}H_{d)}+\gamma_2g_{cd})\,
\Big]\,\hat{=}\,0\,,\label{eqn:G_BCs_Freeze}
\end{align}
with given data~$q^{(G)}_{cd}$, which we will often take to vanish,
and where the overbar derivative notation has the same meaning as in 
equation~\eqref{eqn:G_def}. These conditions are similar to the `freezing'
gauge boundary conditions employed in~\cite{LinSchKid05}, but taking into 
consideration the discussion of gauge reflections given in~\cite{RinLinSch07}, 
and constructed so that the conditions are naturally applied to metric 
components (in ADM form) and their derivatives, but excluding contributions 
from the gauge sources. We will typically try to choose the given 
data to be fixed in time, and such that initially the time derivatives 
vanish for these quantities. Here we have introduced the gauge 
projection operator,
\begin{align}
\perp^{(G)cd}_{ab}&=l_{(a}k_{b)}l^{(c}k^{d)}+k_{a}k_{b}l^{c}l^{d}
-2k_{(a}q_{b)}{}^{(c}l^{d)}\,.
\label{eqn:Gauge_proj}
\end{align}
The above boundary conditions are implemented in \texttt{bamps} using the 
Bj\o rhus method~\cite{Bjo95} as in SpEC. Details of the method 
are explained in section~\ref{subsection:BC_imp}. For completeness
here the constraint projection operator~$\perp^{(C)}=I-\perp^{(P)}-\perp^{(G)}$ 
is,
\begin{align}
\perp^{(C)cd}_{ab}&=\tfrac{1}{2}\,q_{ab}\,q^{cd}-2\,l_{(a}q_{b)}{}^{(c}k^{d)}
+l_al_bk^ck^d\,,
\end{align}
and also plays an important role in the implementation of 
the boundary conditions, as they are again naturally written 
in the 
form~$\perp^{(C)cd}_{ab}d_su^{\hat{+}}_{cd}=\textrm{transverse derivatives}$. 

%%%%%%%%%%%%%%%%%%%%%%%%%%%%%%%%%%%%%%%%%%%%%%%%%%%%%%%%%%%%%
\subsection{Constraint preserving boundary conditions and 
damping}\label{subsection:CP}
%%%%%%%%%%%%%%%%%%%%%%%%%%%%%%%%%%%%%%%%%%%%%%%%%%%%%%%%%%%%%

%%%%%%%%%%%%%%%%%%%%%%%%%%%%%%%%%%%%%%%%%%%%%%%%%%%%%%%%%%%%%
\paragraph*{Generalized harmonic constraint subsystem:} We 
already saw the constraint subsystem of the first order 
reduction of the GHG system. But to get a better idea of the 
effect of the different constraint preserving boundary conditions 
let us consider now the subsystem without the reduction. We 
have,
\begin{align}
\nabla^bY_{ba}&=-R_{ab}C^b\,,\nonumber\\
Y_{ba}&=\nabla_bC_a+2\gamma_4\Gamma^c{}_{ab} C_c
-(\gamma_4-\gamma_5)g_{ab}\Gamma^c C_c \nonumber\\
&\quad- 2 \gamma_0 n_{(a}C_{b)}\,.\label{eqn:F2O_cons}
\end{align}
The shorthand~$Y_{ab}$ and the variable~$G_a$ that follows will be 
related to quantities present in the first order reduction of the 
GHG formulation shortly. We can equivalently express this as,
\begin{align}
n^b\p_bC_a&=G_a-(2\gamma_4-1)\Gamma^c{}_{ab}n^bC_c
-(\gamma_4-\gamma_5)n_a\Gamma^cC_c\nonumber\\
&+2\gamma_0n^bn_{(a}C_{b)}\,,\nonumber\\
n^b\p_bG_a&=\gamma^{bc}\nabla_{b}\big[\nabla_cC_a+2\gamma_4\Gamma^d{}_{ac} C_d
-(\gamma_4-\gamma_5)g_{ac}\Gamma^d C_d \nonumber\\
&- 2 \gamma_0 n_{(a}C_{c)}\big]+(n^b\nabla_bn^c)\big[\nabla_cC_a
+2\gamma_4\Gamma^d{}_{ac} C_d\nonumber\\
&-(\gamma_4-\gamma_5)g_{ac}\Gamma^d C_d - 2 \gamma_0 n_{(a}C_{c)}\big]
+\Gamma^c{}_{ab}n^bC_c\nonumber\\
&+R_{ab}C^b\,,\label{eqn:FT2S_cons}
\end{align}
where the variable,
\begin{align}
G_a&=n^bY_{ba}=2M_a+(n_a\gamma^{ib}-\gamma^i{}_an^b)\nabla_iC_b\,,
\end{align}
is used to allow for the most convenient form of these expressions,
and the final term of~\eqref{eqn:FT2S_cons} is in fact of second 
polynomial order in the constraints because of the vacuum field 
equations~\eqref{eqn:GHG_22}. Different choices of the constraint 
addition parameters~$\gamma_4,\gamma_5$ result in different behavior 
in terms of growth of the constraint fields. It is also obvious 
that different choices of these parameters can simplify the constraint 
subsystem, the natural choice apparently being~$\gamma_4=\gamma_5=1/2$. 

%%%%%%%%%%%%%%%%%%%%%%%%%%%%%%%%%%%%%%%%%%%%%%%%%%%%%%%%%%%%%
\paragraph*{Linearization:} Let us linearize and consider the 
behavior of a set of fields that satisfies these equations on a 
fixed constraint satisfying background. We start with 
equation~\eqref{eqn:F2O_cons} and use the tetrad consisting of 
the null vectors~$l^a,k^a,m^a,\bar{m}^a$ defined 
in~\eqref{eqn:NP_tetrad} to decompose the first index of~$Y_{ba}$. 
From this we obtain,
\begin{align}
&\nabla^b\bigg(k_b\,l^cY_{ca}+l_b\,k^cY_{ca}
-m_b\,\bar{m}^cY_{ca} - \bar{m}_b\,m^c Y_{ca}\bigg)=0\,,
\label{eqn:cpbc_cons_1}
\end{align}
for the linearization, where we are free to use the notation~$C_a$ 
for the linearized violation because the constraints are satisfied 
in the background. 

%%%%%%%%%%%%%%%%%%%%%%%%%%%%%%%%%%%%%%%%%%%%%%%%%%%%%%%%%%%%%
\paragraph*{Boundary conditions:}
Taking the standard setup at the outer boundary so that~$s^a$,
used in the construction of the tetrad, denotes the outward 
pointing spatial unit vector normal to the boundary.
Restricting our attention to boundary conditions that contain 
at most one derivative of the constraints, {\it geometrically} 
the most natural choice seems to be~$l^b\,Y_{ba}\,\hat{=}\,0\,.$
In the first order GHG language these conditions are,
\begin{align}
&G_a+\nabla_{s}C_a+2\gamma_4\Gamma^c{}_{as}C_c\nonumber\\
&+(\gamma_4-\gamma_5)s_a\Gamma^bC_b
-\gamma_0n_a C_s\,\hat{=}\,0\,.\label{eqn:geom_cpbc_1}
\end{align}
Whereas, discarding the first order reduction, those 
of~\eqref{eqn:GHG_11_cpbc} are instead,
\begin{align}
&G_a+\nabla_sC_a+\Gamma^c{}_{as}C_c-(2\gamma_4-1)\Gamma^c{}_{an}C_c
\nonumber\\
&-(\gamma_4-\gamma_5)n_a\Gamma^bC_b+\tfrac{1}{r}C_a\,\hat{=}\,0\,.
\label{eqn:GHG_11_cpbc_22form}
\end{align}
With either conditions one might guess that the choice~$\gamma_4
=\gamma_5=1/2$ reduces reflections from the boundary, 
especially when using a non-harmonic~$\Gamma_a=-H_a\ne0$ gauge.
Incidentally this choice also makes the two conditions almost 
coincident. Suppose all derivatives of~$C_a,G_a$ tangent to the 
boundary vanish, and that the background is flat. Then we can 
analyze the solutions in a plane wave approximation.

%%%%%%%%%%%%%%%%%%%%%%%%%%%%%%%%%%%%%%%%%%%%%%%%%%%%%%%%%%%%%
\paragraph*{Mode solutions on flat-space:} When linearized 
around flat-space this system takes the form,
\begin{align}
\Box C_a-2\,\gamma_0\,\p^b\big(n_{(a}C_{b)}\big)=0\,. 
\end{align}
The right-travelling mode solutions are,
\begin{align}
C_n&=\rho_1\,e^{s_1^+t+i\,\omega x}+\rho_2^s\,e^{s_2^+t+i\,\omega x}\,,\nonumber\\
C^i&=\rho_{2}^i\,e^{s_2^+t+i\,\omega x}\,,
\end{align}
with eigenfrequencies,
\begin{align}
s_1^+&=-\tfrac{1}{2}\gamma_0-\tfrac{i}{2}\,\sqrt{4\omega^2-\gamma_0^2}\,,
\nonumber\\
s_2^+&=-\gamma_0-i\,\sqrt{\omega^2-\gamma_0^2}\,.
\end{align}
A very desirable property for our boundary 
conditions would be that they absorb outward going waves perfectly, 
that is, without reflection. With this motivation high-order derivative 
boundary conditions on the gravitational wave degrees of freedom have been
studied~\cite{BucSar06,BucSar07}, and implemented in the SpEC 
code~\cite{RinBucSch08} in order to absorb higher spherical harmonics 
of the Weyl scalar~$\Psi_4$. In the current context absorption means 
that outgoing mode solutions, those associated with an~$s^+$, lie in 
the kernel of the boundary conditions. This is only the case if 
we switch off the damping~$\gamma_0=0$. Since the low order spherical 
harmonics might be expected to dominate in the gauge and constraint 
subsystems, optimizing against this phenomena may be more important than 
using high-order conditions for the gauge and constraint subsystems 
whilst neglecting the damping terms. 

%%%%%%%%%%%%%%%%%%%%%%%%%%%%%%%%%%%%%%%%%%%%%%%%%%%%%%%%%%%%%
\paragraph*{Remainder of mode solutions:} Substituting these mode 
solutions into the boundary conditions~\eqref{eqn:GHG_11_cpbc_22form}, 
or the natural geometric conditions~\eqref{eqn:geom_cpbc_1}, 
each after appropriate linearization, and expansion at large 
frequency~$\omega$ gives remainders of order~$O(\gamma_0\,C_a)$, 
indicating that neither is the optimal that can be obtained by 
adding source terms to the constraint boundary conditions. Taking 
instead,
\begin{align}
\big(\p_t+\p_s+\gamma_0\big)C_n+\tfrac{1}{2}\gamma_0C_x&\,\hat{=}\,0,\nonumber\\
\big(\p_t+\p_s+\tfrac{1}{2}\gamma_0\big)C_i&\,\hat{=}\,0\,,
\end{align}
the remainder is rather of order~$O(\gamma_0\,C_a\,\omega^{-1})$. 
There is some freedom in expressing these conditions in the first 
order GHG language, but we choose,
\begin{align}
&G_a+\bar{\nabla}_{s}C_a+2\gamma_4\Gamma^c{}_{as}C_c
+(\gamma_4-\gamma_5)s_a\Gamma^bC_b\nonumber\\
& + \frac{1}{2}\gamma_0\gamma_a{}^bC_b
-\gamma_0\,n_a\big(C_n+\tfrac{1}{2}C_s\big)
+\tfrac{1}{r}C_a\,\hat{=}\,0\,.\label{eqn:cpbc_ref_red}
\end{align}
The conditions~\eqref{eqn:geom_cpbc_1} can be similarly 
rewritten. A similar analysis can be performed using the pure 
gauge subsystem presented in~\cite{HilRic13}, but we currently find
that existing gauge boundary conditions are sufficient for our 
needs, so we do not present these calculations here. Tests with the 
various boundaries are presented in section~\ref{section:Experiments}.

%%%%%%%%%%%%%%%%%%%%%%%%%%%%%%%%%%%%%%%%%%%%%%%%%%%%%%%%%%%%%
\section{The BAMPS code}\label{section:bamps}
%%%%%%%%%%%%%%%%%%%%%%%%%%%%%%%%%%%%%%%%%%%%%%%%%%%%%%%%%%%%%

Having discussed the continuum system in the previous
section, we now discuss details of our numerical implementation
of the GHG system. For this we present the \texttt{bamps} code, 
which uses a pseudospectral method on cubed-sphere grids. The 
basic idea of the code is based on SpEC~\cite{SpEC}, but in 
many details, such as the actual grid implementation and the 
outer boundary treatment, \texttt{bamps} differences are present. 

%%%%%%%%%%%%%%%%%%%%%%%%%%%%%%%%%%%%%%%%%%%%%%%%%%%%%%%%%%%%%
\subsection{Grid setup}
\label{subsection:BAMPS_grid_setup}
%%%%%%%%%%%%%%%%%%%%%%%%%%%%%%%%%%%%%%%%%%%%%%%%%%%%%%%%%%%%%

%%%%%%%%%%%%%%%%%%%%%%%%%%%%%%%%%%%%%%%%%%%%%%%%%%%%%%%%%%%%%
\paragraph*{Grid types:} The numerical domain on which we solve 
the evolution equations in \texttt{bamps} is either a cubed-ball
or a cubed-shell grid. Each type is built up of multiple deformed 
cubes. Each patch is described by two fundamental overlapping charts. 
In~\textit{local coordinates}~$\xb,\yb$ and~$\zb$ it is a 
rectangular box $[\bar{x}_0,\bar{x}_1]\times[-1,1]\times[-1,1]$. 
In \textit{global Cartesian coordinates}~$x,y$ 
and~$z$ the cubes are transformed and rotated in such a way 
that when added together they build the desired domain. We give 
a detailed description in the following. The 
cubed-ball-grid includes the origin and has a spherical outer 
boundary. It consists of~$13$ coordinate patches:
\begin{description}
  \item[The central cube:] is centered around the origin
  and ranges from~$-r_{\text{cu}}$ to~$r_{\text{cu}}$ in the global
  Cartesian coordinate directions.
  \item[The transition shell:] transfers the grid
  from the inner cube grid to a spherical shell with
  radius $r_{\text{cs}}$. It contains six patches.
  \item[The outer shell:] consists of six patches
  which extends the grid with additional cubed-shells up to
  the outer grid boundary at~$r_{\text{ss}}$.
\end{description}
The cubed-shell-grid is an excision grid, meaning that it does 
not include the origin. It is a special case of the cubed-ball-grid, 
consisting only of the six outer shell coordinate patches. 

%%%%%%%%%%%%%%%%%%%%%%%%%%%%%%%%%%%%%%%%%%%%%%%%%%%%%%%%%%%%%
\paragraph*{Cubed-sphere coordinate transformation:} The coordinate 
transformation used in \texttt{bamps} to construct the
grids introduced above relies on the so called ``cubed sphere''
construction. It was introduced in~\cite{RonIacPao96} and
first applied in the context of numerical relativity in~\cite{Tho03a,Tho04}. 
Since then this idea was implemented in multi-patch 
approaches~\cite{PolReiSch11,ZinSchTig07,SchDieDor06,LehReuTig05}.
In contrast to many of the earlier examples, the numerical method of
\texttt{bamps} does not require overlapping grids, which simplifies the
discussion. In~\cite{RonIacPao96}, the coordinates are constructed by
considering great arcs parametrized by equidistant angles. Such angle
coordinates are used 
in~\cite{Tho03a,Tho04, PolReiSch11}, while~\cite{ZinSchTig07,LehReuTig05}
use an intermediate set of coordinates also given 
in~\cite{RonIacPao96} that does not have the equidistant angle property.
In \texttt{bamps} the latter type of coordinates are employed. The concrete
coordinate transformation is the following. First, the local
coordinates of each patch are transformed to temporary global
coordinates
\begin{align}
  x_t = \frac{\xb}{\bar s}, \quad
  y_t = \frac{\xb}{\bar s}  \yb, \quad
  z_t = \frac{\xb}{\bar s}  \zb.
\label{radproj:xyz}
\end{align}
This patch, which is orientated in positive $x$ direction, will
later be referred to as the master patch. From here, cyclic permutation
is used to rotate the patches to their location in the sphere.
The denominator~$\bar s$ depends on where the coordinate transformation
happens. For the patches of the outer shell it is
\begin{align}
  \bar s \equiv (1 + \yb^2 + \zb^2)^{1/2},
\label{radproj:s}
\end{align}
In the transition shell its definition includes a transition function $\lambda$
\begin{align}
  \bar s (\lambda)=
  \Big(\frac{1 + 2\lambda}{1 + \lambda (\yb^2 + \zb^2)}\Big)^{1/2}, \qquad
   \lambda = \frac{\xb^2-\xb_0^2}{\xb_1^2-\xb_0^2}\,.
\end{align}
This coordinate transformation is constructed to transition from the
inner cube to the outer shells. Note that this transformation is
uniform along the 3d diagonals, where the distance between inner and
outer shell boundary is smallest. This significantly improves the 
time-stepping restriction in the transition shell.

\begin{figure*}[t]
\centering
\includegraphics[]{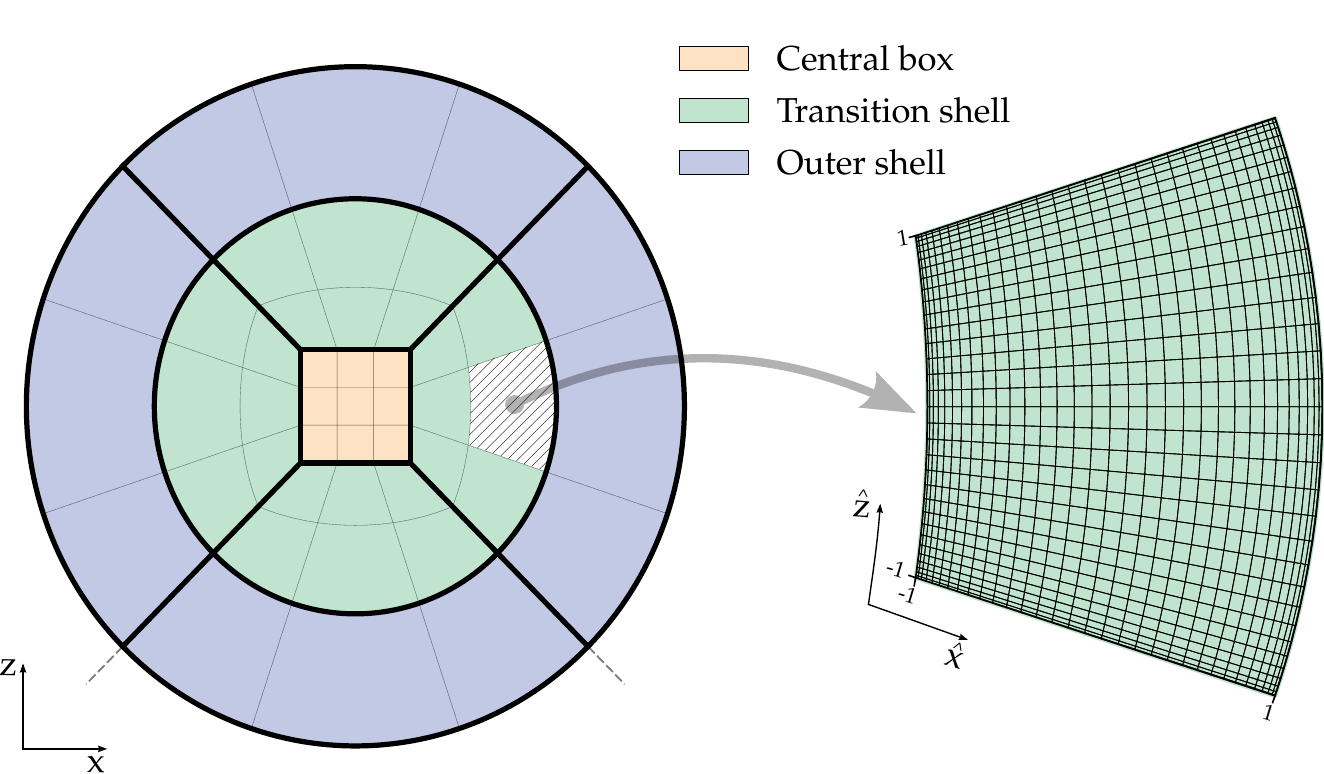}
\caption{The left part of the diagram gives a two 
dimensional sketch of the \texttt{bamps} cubedball grid layout. The ball 
is built up of several transformed cubes. These patches can 
further be divided in subpatches. In the example shown we 
have~$\mathcal{N}_\text{cu}=3$,$\mathcal{N}_\text{cs}=2$ 
and~$\mathcal{N}_\text{ss}=1$. On the right is shown that each 
subpatch is covered by Gauss-Lobatto grids ranging from~$-1$ 
to~$1$ in local coordinates.}
\label{fig:BAMPS_subpatch}
\end{figure*}

%%%%%%%%%%%%%%%%%%%%%%%%%%%%%%%%%%%%%%%%%%%%%%%%%%%%%%%%%%%%%
\paragraph*{Subpatches:} Each coordinate patch can be further 
divided into subpatches. Subpatches are helpful for increasing 
resolution, and form the backbone of the parallelism of \texttt{bamps}.
Each master patch can be split into~$\mathcal{N}_x \times 
\mathcal{N}_{y} \times \mathcal{N}_{z}$ subpatches with 
coordinates
\begin{align}
\xb^i \in [\xb^i_0 +  k^i \Delta\xb^i ,\xb_0^i + (k^i+1) \Delta\xb^i]\,,
\end{align}
with~$\Delta\xb^i = (\frac{\xb_1-\xb_0}{\mathcal{N}_x},\frac{2}{\mathcal{N}_y},
\frac{2}{\mathcal{N}_z})$ and~$k^i=0,\dots,\mathcal{N}_i-1$.
In practice we ensure subdivisions are made in such a way 
that subgrids of two neighboring patches match, and that 
neighboring patches and subpatches share grid-point positions 
on their respective boundaries. This is necessary because 
our current penalty-communication method does not deal with   
interpolating penalties. Concretely we split the inner cube 
into $\mathcal{N}_\text{cu}\times\mathcal{N}_\text{cu}\times
\mathcal{N}_\text{cu}$ subpatches. The transition and outer shell 
are divided in~$\mathcal{N}_\text{cs}$ or~$\mathcal{N}_\text{ss}$ 
subpatches in the radial direction. For the angular direction we 
choose the number of subpatches to be~$\mathcal{N}_\text{cu}\times
\mathcal{N}_\text{cu}$. In Fig.~\ref{fig:BAMPS_subpatch} we show 
a~2d sketch of the \texttt{bamps} cubed-ball grid subdivided into 
subpatches.

%%%%%%%%%%%%%%%%%%%%%%%%%%%%%%%%%%%%%%%%%%%%%%%%%%%%%%%%%%%%%
\paragraph*{Discussion:} It is straightforward to specify a mapping 
between a rectangular master patch and a cubed sphere, although 
some book keeping for the different patches and different types of 
shell transitions is involved.  It may be useful to examine 
different such mappings in terms of a numerical quality criterion, 
say the size of the Jacobian, and to minimize the distortions 
associated with the coordinate transformation.

%%%%%%%%%%%%%%%%%%%%%%%%%%%%%%%%%%%%%%%%%%%%%%%%%%%%%%%%%%%%%
%%%%%%%%%%%%%%%%%%%%%%%%%%%%%%%%%%%%%%%%%%%%%%%%%%%%%%%%%%%%%

%%%%%%%%%%%%%%%%%%%%%%%%%%%%%%%%%%%%%%%%%%%%%%%%%%%%%%%%%%%%%
\subsection{Numerical method}
%%%%%%%%%%%%%%%%%%%%%%%%%%%%%%%%%%%%%%%%%%%%%%%%%%%%%%%%%%%%%

%%%%%%%%%%%%%%%%%%%%%%%%%%%%%%%%%%%%%%%%%%%%%%%%%%%%%%%%%%%%%
\paragraph*{Spatial discretization:} \texttt{bamps} uses the method of 
lines with a standard ODE integrator to integrate in time.
The right-hand-sides are approximated using a pseudospectral 
method. We use a linear transformation to map the local 
coordinates of each subpatch $\xb^i$ into a unit cube~$\xti^i 
= (\xti,\yti,\zti)^T \in  [-1,1]^3$. We discretize the subpatch 
by choosing Gauss-Lobatto collocation points in each dimension,
for example,
\begin{align}
\xti_{\alpha} &= - \cos\big(\frac{\pi}{N_x-1} \alpha\big)\,,
\label{eqn:GaussLobattoPoints}
\end{align}
with~$\alpha=0\,,\cdots,N_x-1$, and similarly in the other directions. 
The number of of grid points $N$ depends on the patch location in the 
grid. The central cube is discretized with~$N_\text{cu}\times N_\text{cu} \times
N_\text{cu}$ points. The radial directions of the transition
and outer shell are filled with~$N_\text{cs}$ and~$N_\text{ss}$ points 
respectively. The number of angular points we 
chose to be the same as in the central cube to assure that we have 
matching grids. In Fig.~\ref{fig:BAMPS_subpatch} we show on the 
right the Gauss-Lobatto discretization of a subpatch.

%%%%%%%%%%%%%%%%%%%%%%%%%%%%%%%%%%%%%%%%%%%%%%%%%%%%%%%%%%%%%
\paragraph*{Basis expansion:} On the collocation points we 
expand all evolution fields~$u$ in each dimension in a spectral 
basis using Chebyshev polynomials~$T_n(x)$
\begin{align}
u_{\alpha\beta\delta} = u(\xti_\alpha,\yti_\beta,\zti_\delta)
        = \sum_{n=0}^{N_x-1} c^x_n(\yti_\beta,\zti_\delta) T_n(\xti_\alpha)\,
        ,
\end{align}
and analogously in the remaining two directions.
We use the pseudospectral approach and store not the expansion coefficients
$c^x, c^y, c^z$ but the function values $u_{\alpha\beta\delta}$ at the collocation points
$\xti^i_{\alpha\beta\delta}$.

%%%%%%%%%%%%%%%%%%%%%%%%%%%%%%%%%%%%%%%%%%%%%%%%%%%%%%%%%%%%%
\paragraph*{Derivatives:} The spatial derivatives of the evolution 
fields are computed by a matrix multiplication. For example in 
the~$\xti$-direction we have
\begin{align}
(\p_{\xh} u)_{\alpha\beta\delta} = \sum_{n=0}^{N_x-1} D_{\alpha n} u_{n \beta\delta}
\label{eqn:spec_deriv}
\end{align}
with the Gauss-Lobatto derivative matrix,
\begin{align}
D_{\alpha\beta} = \begin{cases}
-\frac{2(N_x-1)^2+1}{6} & \alpha=\beta=0\\
\frac{q_\alpha}{q_\beta}\frac{(-1)^{\alpha+\beta}}{\xti_\alpha-\xti_\beta} & \alpha\neq 
\beta\\
\frac{-\xti_\beta}{2(1-\xti_\beta^2)} & \alpha=\beta=1,\cdots,N_x-2\\
\frac{2(N_x-1)^2+1}{6} & \alpha=\beta = N_x-1
\end{cases}
\end{align}
where~$q_\alpha=2$ at boundary points and~$q_\alpha=1$ elsewhere. In practice we do 
not compute diagonal terms of the derivative matrix by the analytic formulas 
stated above but use the identity
\begin{align}
D_{\alpha\alpha} = - \sum_{n=0,n\neq \alpha}^{N_x-1} D_{\alpha n}.
\end{align}
This \textit{negative-sum-trick} maps a constant function explicitly
to zero and is known to give the derivative matrix better stability
as regards rounding errors~\cite{BalTru03}. In preliminary 
experiments we found that this gives slightly more accurate 
derivatives, but have not studied the influence on the accuracy 
of the simulations presented later in the paper.

%%%%%%%%%%%%%%%%%%%%%%%%%%%%%%%%%%%%%%%%%%%%%%%%%%%%%%%%%%%%%
\paragraph*{Filtering:} We find that a crucial ingredient for 
numerical stability is the use of a filter against high-frequency 
growth. For this we follow~\cite{SziLinSch09} exactly. After 
every full time-step we apply the filter in each dimension.
The filter is easily implemented as a matrix multiplication.
For example, in the $\xti$ direction, we filter  the function values
by,
\begin{align}
(\mathcal{F}u)_{\alpha\beta\gamma}=\sum_n^{N_x-1}\mathcal{F}_{n\alpha} u_{n\beta\gamma}\, ,
\end{align}
with the filter matrix
\begin{align}
\label{eqn:filter}
\mathcal{F}_{\alpha\beta} = \sum_n S_{\alpha n} e^{-36(n/n_{\textrm{max}})^{64}} A_{n\beta}\,,
\end{align}
where~$n_\text{max}=N_x-1$ and~$S_{\alpha\beta}$ and $A_{\alpha\beta}$ 
are the Chebyshev synthesis and analysis matrices respectively.

%%%%%%%%%%%%%%%%%%%%%%%%%%%%%%%%%%%%%%%%%%%%%%%%%%%%%%%%%%%%%
\paragraph*{Time integration:} We integrate the fields forward 
in time using a 4th order Runge-Kutta scheme. Unless otherwise stated
we fix the time-step,~$\Delta t = \frac{1}{4} \Delta x_\text{min}$,
with $\Delta x_\text{min}$ being the minimal Cartesian spatial grid
spacing of the whole domain. Empirically we find that this choice
for the time step always leads to stable numerical evolutions,
in the sense that increasing resolution results in smaller 
errors. We have not not carried out a stability analysis of the 
fully discrete system.

%%%%%%%%%%%%%%%%%%%%%%%%%%%%%%%%%%%%%%%%%%%%%%%%%%%%%%%%%%%%%
\paragraph*{BAMPS Octant grid: }
When evolving octant symmetric data in \texttt{bamps}, it is possible
to only evolve one eighth of the cubed ball grid. This saves
computational and memory costs. In the \texttt{bamps} octant mode we 
choose an odd number of subpatches~$\mathcal{N}_\text{cu}$ and 
a odd number of grid points~$N_\text{cu}$ and reduce the numerical 
domain to~$x\ge0$,~ $y\ge0$ and~$z\ge0$. This means that all 
subpatches containing one of the Cartesian axes are cut in half 
along them. In these patches we use the symmetry conditions to 
construct special matrices which compute the derivatives and 
filters.

%%%%%%%%%%%%%%%%%%%%%%%%%%%%%%%%%%%%%%%%%%%%%%%%%%%%%%%%%%%%%
\subsection{Patching boundary conditions}
\label{subsection:Penalties}
%%%%%%%%%%%%%%%%%%%%%%%%%%%%%%%%%%%%%%%%%%%%%%%%%%%%%%%%%%%%%

To glue all subpatches together, we have to impose
appropriate conditions at the connecting boundaries of the 
subpatches.  For this we apply the penalty method as is
described in~\cite{Hes00,HesGotGot07,TayKidTeu10}.
The main idea of this method is to add penalty terms for each
incoming characteristic variable at the boundary to the right
hand side of the evolution equations.
We use the characteristic variables of the evolution system
to formulate boundary conditions. 
On the boundary surface we define the outward pointing spatial
normal vector $s^i$. The characteristic variables of the evolution 
system are given in equation (\ref{eqn:GHG11_char_vars}) with speeds 
(\ref{eqn:GHG11_char_speeds}).
In vector notation we write
\begin{align}
u^\mu &= 
\begin{pmatrix}
  g_{ab}\\
  \Pi_{ab}\\
  \Phi_{iab}
\end{pmatrix}\, ,
\quad
u^{\hat \alpha}=
\begin{pmatrix}
  u^{\hat 0}_{ab}\\
  u^{\hat \pm}_{ab}\\
  u^{\hat{\beta}}_{Aab}
\end{pmatrix}.
\end{align}
Incoming characteristic variables to the subpatch boundary have positive speeds.
On these we want to impose the condition that they are equal to the outgoing
characteristic variables of the neighboring patch. Table~\ref{tab:in_out_char} 
summarizes all incoming and outgoing characteristic depending on the lapse 
function~$\alpha$ and the shift in $s^i$ direction, $\beta^s$.
%%%%%%%%%%%%%%%%%%%%%%%%%%%%%%%%%%%%%%%%%%%%%%%%%%%%%%%%%%%%%
% table in text:
%%%%%%%%%%%%%%%%%%%%%%%%%%%%%%%%%%%%%%%%%%%%%%%%%%%%%%%%%%%%%
\begin{table*}[t]
\centering
\begin{tabular}{|c|c||c|c|c|c|c|}
\hline
 & 
 &
 $\beta^s > \alpha  > 0$ &
 $\alpha  > \beta^s > 0$ &
 $\beta^s = 0          $ &
 $-\alpha < \beta^s < 0$ &
 $\beta^s<-\alpha  < 0 $ \\
\hline\hline 
$u^{\hat 0}_{ab}$&
$0$ &
zero&
zero&
zero&
zero& 
zero\\
\hline
$u^{\hat -}_{ab}$&
$\beta^s-\alpha$&
incoming&
outgoing&
outgoing&
outgoing& 
outgoing\\
\hline
$u^{\hat +}_{ab}$&
$\beta^s+\alpha$&
incoming&
incoming&
incoming&
incoming& 
outgoing\\
\hline
$u^{\hat{\beta}}_{Aab}$&
$\beta^s$&
incoming&
incoming&
zero&
outgoing&
outgoing \\
\hline
\end{tabular}
\caption{Incoming and outgoing characteristic variables to a subpatch boundary
with spatial normal vector $s^i$ depending on the gauge variables.}
\label{tab:in_out_char}
\end{table*}
%%%%%%%%%%%%%%%%%%%%%%%%%%%%%%%%%%%%%%%%%%%%%%%%%%%%%%%%%%%%%
As an example, let us now consider the boundary between two patches, 
patch~$L$ and patch $R$, and the case $-\alpha<\beta^s<0$. With respect 
to the spatial normal vector $s^i$ at the boundary pointing outwards 
of subpatch~$L$ and inwards in subpatch $R$, the incoming characteristic 
variables of~$L$ are the outgoing ones of $R$. In the chosen 
case,~$u^{\hat+}_{ab}$ are incoming to L and outgoing of $R$. We want 
to impose the condition, 
\begin{align}
u^{\hat+\,L}_{ab}\,\hat{=}\,u^{\hat+\,R}_{ab}\,.
\end{align}
Multiplying the first order GHG evolution equations from the left 
with the matrix of eigenvectors $T^{-1\,\hat\alpha}{}_\beta$, we obtain 
evolution equations for the characteristic variables.
\begin{align}
    d_t u^{\hat \alpha\,L} = 
    T^{-1\,\hat\alpha}{}_\mu {A^{k\mu}}_{\nu} \p_k u^{\nu\,L} + 
    T^{-1\,\hat\alpha}{}_\mu S^{\mu}
\end{align} 
Here the~$d$ again denotes that the similarity 
matrix~$T^{-1\,\hat\alpha}{}_\mu$ stands outside the partial time derivative.
At the boundary we now add a penalty to the right hand side of the
evolution equation of the incoming characteristic. This is often called
{\it weakly imposing} the boundary condition,
\begin{align}
d_t u^{\hat+\,L}_{ab} 
&\,\hat{=}\,T^{-1\,\hat+}{}_\mu {A^{k\mu}}_{\nu} \p_k u^{\nu\,L}_{ab} 
+  T^{-1\,\hat+}{}_\mu S^\mu \nonumber\\
&\quad+  p\left(u^{\hat+\,R}_{ab}-u^{\hat+\,L}_{ab} \right)\,.
\end{align}
Afterwards we use the inverse transformation to get back to the 
evolution equations enhanced with the necessary penalty terms
at the boundary. These are also the equations we implement in the code.
We treat all six boundaries of the subpatches independently from each other.
This means that on the edges we have to consider penalty contributions from
two and on the corner from three directions. The size of the penalty 
parameter~$p$ can be derived from an energy estimate of the semi-discrete 
evolution system. This we present in a section~\ref{section:EnergyPenalty}.

%%%%%%%%%%%%%%%%%%%%%%%%%%%%%%%%%%%%%%%%%%%%%%%%%%%%%%%%%%%%%
\subsection{Outer boundary implementation}
\label{subsection:BC_imp}
%%%%%%%%%%%%%%%%%%%%%%%%%%%%%%%%%%%%%%%%%%%%%%%%%%%%%%%%%%%%%

At the spherical outer boundary of the domain we use the
Bj{\o}rhus method~\cite{Bjo95,LinSchKid05} to impose the constraint, 
physical and gauge conditions given in section~\ref{section:GHG}. 
As for the patching boundaries, we impose conditions on the 
incoming characteristic to the boundary surface. However, this time
instead of adding penalty terms we modify the right hand side of 
the evolution equations at the boundary in such a way that the 
boundary conditions are satisfied. We define the outward pointing 
spatial normal unit vector~$s^i$ and use the projection 
operator~$q^j{}_i=\delta^j{}_i-s^js_i$ to split the principal part 
of the evolution equation in a part normal and tangential to the 
boundary surface
\begin{align}
\p_t u^\mu 
&\approx  A^{k\mu}{}_\nu (s_k s^j + q_k^j) \p_j u^\nu\nonumber\\ 
& = A^{s\mu}{}_\nu \p_s u^\nu +  A^{A\mu}{}_{\nu} q_A^B \p_B u^\nu\,.
\end{align}
Expressed in characteristic variables the normal part is
\begin{align}
 d_t u^{\hat{\alpha}} \sim T^{-1\,\hat\alpha}{}_\mu A^{s\mu}{}_\nu 
T^{\nu}{}_{\hat\beta} T^{-1\hat\beta}{}_\xi \p_s u^\xi 
= \Lambda^{s\hat\alpha}{}_{\hat\beta} d_s u^{\hat \beta}
\end{align}
The matrix~$\Lambda^{s\hat\alpha}{}_{\hat\beta}$ is a diagonal matrix containing 
the characteristic speeds. At the outer boundary we assume that the absolute 
value of the shift $\beta^s$ is always smaller than the size of the lapse~$\alpha$. 
This leads to two cases to be considered.
\paragraph*{Case $-\alpha<\beta^s<0$:}
In this case the incoming characteristic at the outer boundary condition is
$u^{\hat +}$. According to section~\ref{section:GHG} we impose the following boundary 
conditions, which we give here only schematically:
\begin{enumerate}
\item One of the constraint preserving boundary 
conditions~\eqref{eqn:GHG_11_cpbc},~\eqref{eqn:geom_cpbc_1} 
or~\eqref{eqn:cpbc_ref_red},
 \begin{align}
 \perp^{(C)} d_s u^{\hat +} + P^{(C)} + NP^{(C)} \,  \hat{=}\, 0\,.
 \end{align}
\item One of the gauge boundary conditions~\eqref{eqn:G_BCs} 
or~\eqref{eqn:G_BCs_Freeze}, which become either,
 \begin{align}
 \perp^{(G)} d_t u^{\hat +}+P^{(G)}+NP^{(G)}&\,\hat{=}\,0\,,\nonumber\\
 \perp^{(G)} d_s u^{\hat +}+P^{(G)}+NP^{(G)}&\,\hat{=}\,q^{(G)}\,.
 \end{align}
\item The physical boundary condition~\eqref{eqn:GHG_11_Psi0_bc},
 \begin{align}
 \perp^{(P)} d_s u^{\hat +}_{ab} + P^{(P)} + NP^{(P)}\,\hat{=}\,q^{(P)}\,.
 \end{align}
 \end{enumerate}
Here we labeled principal terms with derivatives tangent to the boundary~$P^{(x)}$ 
and non-principal terms with $NP^{(x)}$. At the boundary surface we project the 
evolution equation of the incoming characteristic $u^{\hat +}$ into the constraint, 
the physical and gauge part.
\begin{align}
d_t u^{\hat +}_{ab} \approx v^{\hat +} (\perp^{(C)}_{ab}{}^{cd}
+ \perp^{(P)}_{ab}{}^{cd} + \perp^{(G)}_{ab}{}^{cd}) d_s u^{\hat +}_{cd}\,.
\end{align}
All three parts have to be replaced using the boundary conditions. We do this
by subtracting the conditions from the bulk right hand side $D_t$,
\begin{align}
  d_t u^{\hat +}_{ab}\,\hat{=}\,D_t u^{\hat +}_{ab} - v^{\hat +} (\text{Conditions})_{ab}\,,
\end{align}
with the special case~\eqref{eqn:G_BCs} treated in the obvious way. Transforming 
back this modified right hand side leads to modified evolution equations 
at the boundary.

%%%%%%%%%%%%%%%%%%%%%%%%%%%%%%%%%%%%%%%%%%%%%%%%%%%%%%%%%%%%%%
\paragraph*{Case $0<\beta^s<\alpha$:}
In this case also the characteristic~$u^{\hat{\beta}}_{Aab}$ is incoming.
As described in~\cite{LinSchKid05}, we impose the additional constraint
preserving boundary condition
\begin{align}
 d_s u^{\hat{\beta}}_{Aab} - {q^B}_A d_B \Phi_{sbc} \,\hat{=}\, 0\,,
\end{align}
by subtracting it from the evolution equation of $u^{\hat{\beta}}_{Aab}$
\begin{align}
d_t u^{\hat{\beta}}_{Aab}  = D_t u^{\hat{\beta}}_{Aab} - v^{\hat{\beta}} ( \text{Condition}_{Aab} )\,.
\end{align}
After we have modified the right hand sides at the boundary
we transform back to the evolution equations for the primitive
fields.

%%%%%%%%%%%%%%%%%%%%%%%%%%%%%%%%%%%%%%%%%%%%%%%%%%%%%%%%%%%%%
\subsection{Code implementation details}
\label{subsection:BAMPS_implementation}
%%%%%%%%%%%%%%%%%%%%%%%%%%%%%%%%%%%%%%%%%%%%%%%%%%%%%%%%%%%%%

%%%%%%%%%%%%%%%%%%%%%%%%%%%%%%%%%%%%%%%%%%%%%%%%%%%%%%%%%%%%%
\paragraph*{Code structure:}
The \texttt{bamps} code is written in the C programming language in
a modular fashion. The code is designed in such a way that the 
technical layer is separated from {\it projects} for solving 
physics problems. Inside physics projects we use a Mathematica script, 
MathToC, which translates equations written in tensor notation into C 
code. As a standalone program we have developed an axisymmmetric apparent 
horizon finder, AHloc, which is typically used to search apparent 
horizons in \texttt{bamps} generated data at the post-processing step. It 
is also possible to run the finder in a daemon-like mode in which it 
searches horizons in data of a running instance of \texttt{bamps}. We 
describe the apparent horizon in subsection~\ref{subsection:AH}.

%%%%%%%%%%%%%%%%%%%%%%%%%%%%%%%%%%%%%%%%%%%%%%%%%%%%%%%%%%%%%
\paragraph*{Parallelization:} 
\texttt{bamps} is programmed to run in parallel on several computing nodes using 
the message passing interface (MPI). The $N_{\text{sub}}$ subpatches of 
a \texttt{bamps} grid are distributed on~$M$ MPI processes as evenly as 
possible. This means that each process has to handle at 
least~$n = \lfloor\frac{N_{\text{sub}}}{M}\rfloor$ subpatches. As in 
general the total number of grids is not divisible by the number of 
MPI processes without remainder, $N_{\text{sub}}\mod M$ processes have 
to take care of one additional grid. In practice we choose the number 
of MPI processes in such a way that the number of processes which 
have to compute one grid less is minimized.

%%%%%%%%%%%%%%%%%%%%%%%%%%%%%%%%%%%%%%%%%%%%%%%%%%%%%%%%%%%%%
\section{Energy estimate for penalty factor}
\label{section:EnergyPenalty}
%%%%%%%%%%%%%%%%%%%%%%%%%%%%%%%%%%%%%%%%%%%%%%%%%%%%%%%%%%%%%

In this section we derive an estimate for the right choice
of penalty factor at the patching boundaries of the \texttt{bamps}
domains. The actual technical implementation of the patching
condition was already described in subsection~\ref{subsection:Penalties}. 
The following calculation is based on the one presented 
in~\cite{TayKidTeu10}. However we present it for a general 
hyperbolic system in curvilinear  coordinates, albeit under 
rather restrictive assumptions.

%%%%%%%%%%%%%%%%%%%%%%%%%%%%%%%%%%%%%%%%%%%%%%%%%%%%%%%%%%%%%
\subsection{The continuum case}
%%%%%%%%%%%%%%%%%%%%%%%%%%%%%%%%%%%%%%%%%%%%%%%%%%%%%%%%%%%%%

We view the GHG system as a general symmetric hyperbolic
system of partial differential equations, but suppress all 
non-principal terms, and work in the linear, constant coefficient 
approximation, so we have, 
\begin{align}\label{eqn:evSys}
  \p_t u^\mu = {A^{p\mu}}_{\nu} \p_p u^{\nu},\hspace{0.75cm} p\in{x,y,z}\,,
\end{align}
where, in matrix notation,
\begin{align}
u^\mu =
\begin{pmatrix}
g_{ab} \\
\Pi_{ab} \\
\Phi_{iab}
\end{pmatrix}\,,
\quad
{A^{p\mu}}_\nu =
\begin{pmatrix}
(1+\gamma_1) \beta^k        &  0                   &  0 \\
\gamma_1\gamma_2 \beta^k    & \beta^k              &  -\alpha \gamma^{ik} \\
\gamma_2 \alpha \delta^k_i  &  -\alpha \delta^k_i  &  \beta^k
\end{pmatrix}\,.
\end{align}
For clarity we suppress the state vector indices~$\mu,\nu$.
For this system there is a symmetrizer~$H$ such that~$H A^p s_p$
is Hermitian for every unit spatial vector~$s^p$. The energy of the 
system is,
\begin{align}
  E^2 = \int_V \dV (u^\dagger H u) \,.
\end{align}
with the volume form $\dV = \dx\dy\dz \sqrt{\gamma}$.
As discussed in section~\ref{subsection:BAMPS_grid_setup}, 
each subpatch of \texttt{bamps} has a set of global Cartesian 
coordinates~$x^i=(x,y,z)$ and a set of
local coordinates $\xti^i=(\xti, \yti, \zti)$. The Jacobian
$J^i_{\tilde{i}} = \tfrac{\p x^i}{\p \xti^i}$ transforms
between the two charts. To formulate boundary conditions at 
the patching boundaries which control the energy in the patch, we study 
the time derivative of the energy, using the evolution equations we 
replace the time derivatives by spatial derivatives,
\begin{align}
\p_t E^2 &= \int\dV \p_p \left[ \ud H A^p u \right ]\,.
\end{align}
In the constant coefficient approximation we can commute the
determinant of the three metric in the volume form with the
partial derivative and end up a divergence in flat Cartesian
coordinates
\begin{align}
\p_t E^2 &= \int\dx\dy\dz \p_p \left[ \ud H A^p u \sqrt{\gamma}\right]\,.
\end{align}
In the next step we change to the patch local coordinates 
$\xti, \yti $ and $\zti$,
\begin{align}
\p_t E^2 &= \int\dVti \frac{1}{\det{J^i_{\iti}}}
            \p_{\tilde{p}}
              \left[ \ud H A^{\tilde{p}} u \sqrt{\gamma}\det J^i_{\iti}
              \right] \nonumber\\
 &= \int\dxti\dyti\dzti \p_{\tilde{p}} \Phi^{\tilde{p}}\,.
\end{align}
Here we have defined $\sqrt{\tilde{\gamma}} := \sqrt{\gamma}\det{J^i_{\iti}}$ and 
the flux~$\Phi^{\tilde{p}} = \ud H A^{\tilde{p}} u \sqrt{\tilde{\gamma}}$. Now we integrate 
over all boundary surfaces of the patch,
\begin{align}
\p_t E^2 &= \int\limits_{-1}^{1}\int\limits_{-1}^{1}\dyti\dzti
          \Phi^{\xti}|_{\xti=-1}^{1} 
          + \int\limits_{-1}^{1}\int\limits_{-1}^{1}\dxti\dzti
          \Phi^{\yti}|_{\yti=-1}^{1} \nonumber\\
          &+ \int\limits_{-1}^{1}\int\limits_{-1}^{1}\dxti\dyti
          \Phi^{\zti}|_{\zti=-1}^{1}.
\end{align}
At a boundary surface, for example~$\xh = \text{const}$, we can write
the unit normal vector as,
\begin{align}
\tilde{s}^{i} = \underbrace{(\tilde{\gamma}^{jk}
          \tilde{\p}_{j}\xti\tilde{\p}_{k}\xti)^{-\tfrac{1}{2}}}_{\equiv l}
          \tilde{\gamma}^{il}\tilde{\p}_{l} \xti = l\,\tilde{\p}^{i} \xti\,,
\end{align}
and~$2+1$ split the spatial metric~$\tilde\gamma_{ij}$,
\begin{align}
\tilde{\gamma}_{ij} = \begin{pmatrix}
l^2 + \gamma_{\xti\tilde{A}}\tilde{\gamma}_{\xti}^{\tilde{A}} &
    \tilde\gamma_{\xti\tilde{A}} \\
\tilde{\gamma}_{\xti\tilde{B}} & \tilde{q}_{\tilde{A}\tilde{B}}
\end{pmatrix}\,.
\end{align}
The relationship between the determinant of~$\tilde{\gamma}_{ij}$ and the 
metric in the boundary surface~$\tilde{q}_{\tilde{A}\tilde{B}}$ 
is,~$\sqrt{\tilde{\gamma}}=l\sqrt{\tilde{q}}\,.$
We rewrite,
\begin{align}
\Phi^{\xh} = \Phi^{\tilde{p}}\p_{\tilde{p}}\xti
           = \ud H A^{\tilde{p}} u\,l \sqrt{\tilde q}\,\p_{\tilde{p}} \xti 
           = \sqrt{\tilde q}\underbrace{\ud H A^{s} u}_{\tilde{\Phi}^s}\,,
\end{align}
and express the time derivative of the energy as the sum of boundary
surfaces integrals over the fluxes~$\tilde{\Phi}^s$,
\begin{align}
  \p_t E^2 &=
  \int\limits_{-1}^{1}\int\limits_{-1}^{1}\dA_{\yti\zti}\,
  \tilde{\Phi}^s \big|_{\xti=-1}^{1} 
  + \int\limits_{-1}^{1}\int\limits_{-1}^{1}\dA_{\xti\zti}\,
  \tilde{\Phi}^s \big|_{\yti=-1}^{1} \nonumber\\
  &+ \int\limits_{-1}^{1}\int\limits_{-1}^{1}\dA_{\xti\yti}\,
  \tilde{\Phi}^s \big|_{\zti=-1}^{1}\,.
  \label{eqn:PenInt}
\end{align}
The area element is~$\dA_{\yti\zti}=\sqrt{\tilde q}\,\dyti\dzti$.
The fluxes can be rewritten in terms of incoming and outgoing
characteristic variables at the boundary surface. The system is symmetric 
hyperbolic. Therefore the principal symbol has a full set of Eigenvectors which 
we write as columns of the similarity matrix $T_s$. With the inverse of this 
matrix,~$T_s^{-1}$, we transform the vector of evolution variables to the 
characteristic variables of the system~$v = T_s^{-1} u$.
The flux expressed in the language of characteristic variables is,
\begin{align}
\tilde{\Phi}^s = \underbrace{\ud (T_s^{-1})^\dagger}_{v^\dagger}\;
          \underbrace{T_s^\dagger H T_s}_{\tilde{H}}\;
          \underbrace{(T_s)^{-1} A^s T_s}_{\Lambda_s}\;
          \underbrace{T_s^{-1} u}_{v}
        = v^\dagger \tilde{H} \Lambda_s v\,.
\end{align}
The diagonal matrix~$\Lambda_s$ contains all the speeds of the characteristic
variables
\begin{align}
\Lambda_s = \begin{pmatrix}
\Lambda_I & 0 \\
0 & -\Lambda_{II}
\end{pmatrix}\,.
\end{align}
Where we have ordered the characteristic variables in such a way that we group
all incoming with positive speeds $\Lambda_I$ and outgoing with negative 
speeds $-\Lambda_{II}$. In this partition it follows that
\begin{align}
v = \begin{pmatrix}
v_I \\
v_{II}
\end{pmatrix}\,,
\qquad
\tilde{H} = \begin{pmatrix}
\tilde{H}_I & 0 \\
0         & \tilde{H}_{II}
\end{pmatrix}\,,
\end{align}
and with this 
\begin{align}
\tilde{\Phi}^s &= v_I^\dagger \tilde{H}_I \Lambda_I v_I
 - v_{II}^\dagger\tilde{H}_{II}\Lambda_{II} v_{II}\,.
\end{align}
If all integrands in (\ref{eqn:PenInt}) are negative 
semi-definite, the energy of the system does not grow over time.
For the boundary conditions we use the ansatz~$v_I = \kappa v_{II} + g$,
which means that at the boundary surface we set the incoming
characteristic variables equal to a linear combination of the outgoing
characteristic variables plus some given data $g$. Choosing the 
matrix~$\kappa^\dagger\kappa$ small, we obtain,
\begin{align}
\tilde{\Phi}^s &= (g^\dagger + v_{II}^\dagger \kappa^\dagger)
                  \tilde{H}_I \Lambda_I (\kappa v_{II} + g)
                  -v_{II}^\dagger
                  \tilde{H}_{II}\Lambda_{II}v_{II}
                  \nonumber\\
               &\lesssim g^\dagger H_I \Lambda_I g
               + v_{II}^\dagger\left[ \kappa^\dagger \tilde{H}_I \Lambda_I \kappa
               - \tilde{H}_{II} \Lambda_{II}\right] v_{II}\,.
\end{align}
The first term only depends on the given data. As we are free to
choose it we have full control over this term. The second we can
make negative again by choosing~$\kappa^\dagger\kappa$ sufficiently 
small.

%%%%%%%%%%%%%%%%%%%%%%%%%%%%%%%%%%%%%%%%%%%%%%%%%%%%%%%%%%%%%
\subsection{The semi-discrete case}
%%%%%%%%%%%%%%%%%%%%%%%%%%%%%%%%%%%%%%%%%%%%%%%%%%%%%%%%%%%%%

In this subsection we carry out the energy estimate for a
semi-discrete system. In our case this means that we discretize
the evolution variables in space using Gauss-Lobatto collocation 
points according to equation (\ref{eqn:GaussLobattoPoints}). 
The semi-discrete evolution equations are 
 \begin{align}
   \p_t u_{\alpha\beta\delta} = A^p [\p_p u]_{\alpha\beta\delta}
                = A^p [J_p^{\tilde{p}}]_{\alpha\beta\delta} [\p_{\tilde p} u]_{\alpha\beta\delta}\,.
 \end{align}
The energy of this system is defined using Gauss-Lobatto quadrature
with the appropriate integration weights $\omega_\alpha,\omega_\beta,\omega_\delta,$
\begin{align}
   E^2 = \sum_{\alpha\beta\delta} \omega_\alpha \omega_\beta \omega_\delta 
         \sqrt{\tilde{\gamma}}_{\alpha\beta\delta} \ud_{\alpha\beta\delta} H u_{\alpha\beta\delta}\,.
\end{align}
Again we compute the time energy of the system,
with~$\tilde{\omega}_{\alpha\beta\delta}=\omega_\alpha \omega_\beta 
\omega_\delta [\sqrt{\tilde{\gamma}}]_{\alpha\beta\delta}$, using the 
inverse product rule to write,
\begin{align}
\p_t E^2  &= \sum_{\alpha\beta\delta} \tilde{\omega}_{\alpha\beta\delta} \p_{p} \left[
\ud_{\alpha\beta\delta} H_{\alpha\beta\delta} A^p  u_{\alpha\beta\delta} \right]\,,
\end{align}
and transform to local coordinates.
For this we assume that~$\p_{\tilde{p}}\sqrt{\tilde{\gamma}_{\alpha\beta\delta}}=0$ 
and obtain,
\begin{align}
\p_t E^2  &= \sum_{\alpha\beta\delta} \omega_{\alpha\beta\delta}
\p_{\tilde{p}} \left[
\ud_{\alpha\beta\delta} H_{\alpha\beta\delta} A^p  u_{\alpha\beta\delta} \sqrt{\tilde{\gamma}}_{\alpha\beta\delta}\right]\,.
\end{align}
Using an expansion in Legendre polynomials we can use the summation by parts 
property to write,
\begin{align}
\p_t E^2 =& \sum_{\beta\delta} \omega_{\beta\delta} \left. \ud_{\alpha\beta\delta} H_{\alpha\beta\delta}
 A^{\xh} u_{\alpha\beta\delta} \sqrt{\tilde{\gamma}}_{\alpha\beta\delta}\right|_{\alpha=0}^{N_{x}-1}\nonumber\\
+& \sum_{\alpha\delta} \omega_{\alpha\delta} \left. \ud_{\alpha\beta\delta} H_{\alpha\beta\delta} A^{\yh} u_{\alpha\beta\delta} \sqrt{\tilde{\gamma}}_{\alpha\beta\delta}\right|_{\beta=0}^{N_{y}-1}\nonumber\\
+& \sum_{\alpha\beta} \omega_{\alpha\beta} \left.\ud_{\alpha\beta\delta} H_{\alpha\beta\delta} A^{\zh} u_{\alpha\beta\delta} \sqrt{\tilde{\gamma}}_{\alpha\beta\delta}\right|_{\delta=0}^{N_{z}-1}\,.
\end{align}
As in the continuum case we introduce the normal outward pointing~$s^i$ vector 
at the boundary and write,
\begin{align}
\p_t E^2 =& \sum_{\beta\delta} \tilde{\omega}_{\beta\delta} \left. \ud_{\alpha\beta\delta} H_{\alpha\beta\delta} A^{\tilde{p}} [s^{\xh}_{\tilde{p}}]_{\alpha\beta\delta} u_{\alpha\beta\delta} \right|_{\alpha=0}^{N_{x}-1}\nonumber\\
+&\sum_{\alpha\delta} \tilde{\omega}_{\alpha\delta} \left. \ud_{\alpha\beta\delta} H_{\alpha\beta\delta} A^{\tilde{p}} [s^{\yh}_{\tilde{p}}]_{\alpha\beta\delta} u_{\alpha\beta\delta} \right|_{\beta=0}^{N_{y}-1}\nonumber\\
+&\sum_{\alpha\beta} \tilde{\omega}_{\alpha\beta} \left. \ud_{\alpha\beta\delta} H_{\alpha\beta\delta} A^{\tilde{p}} [s^{\zh}_{\tilde{p}}]_{\alpha\beta\delta} u_{\alpha\beta\delta} \right|_{\delta=0}^{N_{z}-1}
\end{align}
with $\tilde{\omega}_{\beta\delta} 
\equiv \sqrt{\tilde q} \omega_{\beta\delta}$.
We define the flux 
\begin{align}
\tilde{\Phi}_{\alpha\beta\delta} = \ud_{\alpha\beta\delta} H_{\alpha\beta\delta} A^p [s_p]_{\alpha\beta\delta} u_{\alpha\beta\delta}\,,
\end{align}
and transform it to characteristic variables in the obvious way. This gives us for 
the semi-discrete case the analogue expression for the time derivative 
of the energy at the boundary~\eqref{eqn:PenInt}. In case of patching the boundaries between 
two subpatches we apply the penalty method to impose boundary conditions. For simplicity 
we restrict ourselves to the~$\alpha=0$ boundary. For each incoming characteristic variable we add 
a penalty term to the right hand side of the evolution equations,
\begin{align}
 \p_t u_{\alpha\beta\delta} = A^P [J_p^{\tilde p}]_{\alpha\beta\delta} [\p_{\tilde p} u]_{\alpha\beta\delta} 
+ \delta_{\alpha,0} [T_s]_{\beta\delta} P_{\beta\delta} \delta v_{\alpha\beta\delta}\,,
\end{align}
with the penalty matrix
\begin{align}
P_{\beta\delta} = \begin{pmatrix}
p_{\beta\delta} & 0\\
0          & 0
\end{pmatrix}\,,
\end{align}
and $\delta v_{\alpha\beta\delta} = [v^{BC}]_{\alpha\beta\delta} - [v_{I}^R]_{\alpha\beta\delta}$, with~$v^{BC}$
the desired boundary data. The time derivative of the energy splits into two parts
\begin{align}
 \p_t E^2 = \p_t E^2_{\text{bulk}} + \p_t E^2_{\text{pen}}\,.
\end{align}
The first part is the contribution from the bulk,
\begin{align}
\p_t E_{\text{bulk}}^2 =& \sum_{\beta\delta} \tilde{\omega}_{\beta\delta}
 [v_I^\dagger]_{0\beta\delta} [\tilde{H}_I]_{0\beta\delta} [\Lambda^s_I]_{0\beta\delta} 
[v_I]_{0\beta\delta} \nonumber\\
&-\sum_{\beta\delta} \tilde{\omega}_{\beta\delta}
[v_{II}^\dagger]_{0\beta\delta} [\tilde{H}_{II}]_{0\beta\delta} 
[\Lambda^s_{II}]_{0\beta\delta} [v_{II}]_{0\beta\delta}\,.
\end{align}
The second part changes the time derivative of the energy because of the
additional penalty terms in the evolution equation at the boundary,
\begin{align}
\p_t E^2_{\text{pen}} = \sum_{\beta\delta}\tilde{\omega}_{0\beta\delta}
&([u_{0\beta\delta}]^\dagger H_{0\beta\delta} T_{0\beta\delta} P_{\beta\delta} 
\delta v_{0\beta\delta}\nonumber\\
&+ [\delta v_{0\beta\delta}]^\dagger p^\dagger_{\beta\delta} T^\dagger_{0\beta\delta} 
H_{0\beta\delta} u_{0\beta\delta})\,.
\end{align}
By inserting the identity $TT^{-1}=I$ into the appropriate
places we transform the state vector~$u$ to the vector of
characteristic variables. Then multiplying out the penalty matrix and 
rearranging leads to,
\begin{align}
\p_t E^2_{\text{pen}}
&=\sum_{\beta\delta} p_{\beta\delta} \tilde{\omega}_{0\beta\delta}
( [v^{BC}]^\dagger_{0\beta\delta} [\tilde H_I]_{0\beta\delta} [v^{BC}]_{0\beta\delta}\nonumber\\
&- [v_I]^\dagger_{0\beta\delta} [\tilde{H}_I]_{0\beta\delta} [v_I]_{0\beta\delta}
- [\delta v]^\dagger_{0\beta\delta} [\tilde{H}_I]_{0\beta\delta} [\delta v]_{0\beta\delta})\,.
\end{align}
In total the change of energy at the boundary surface is,
\begin{align}
\p_t E^2 &= \sum_{\beta\delta} [v_I]^\dagger_{0\beta\delta}  (\tilde{\omega}_{\beta\delta} \Lambda^s_I 
- p_{\beta\delta} \tilde{\omega}_{0\beta\delta}) [\tilde H_I]_{0\beta\delta} [v_I]_{0\beta\delta} \nonumber\\
&-\sum_{\beta\delta} \tilde{\omega}_{\beta\delta}
 [v_{II}^\dagger]_{0\beta\delta} [\tilde{H}_{II}]_{0\beta\delta} [\Lambda^s_{II}]_{0\beta\delta} [v_{II}]_{0\beta\delta} \nonumber \\
&+\sum_{\beta\delta} p_{\beta\delta} \tilde{\omega}_{0\beta\delta} [v^{BC}]^\dagger_{0\beta\delta} [\tilde{H}_I]_{0\beta\delta} [v^{BC}]_{0\beta\delta} \nonumber\\
&-\sum_{\beta\delta} p_{\beta\delta} \tilde{\omega}_{0\beta\delta} [\delta v]^\dagger_{0\beta\delta} [\tilde{H}_I]_{0\beta\delta} [\delta v]_{0\beta\delta}\,. 
\label{eqn:energy_subpatch}
\end{align}
We now consider two neighboring subpatches which we label
L (for left) and R (for right). Let us assume they have a common 
boundary at~$\alpha=N-1$ for the left patch and~$\alpha=0$ for the right patch.
For each subpatch we can write down the change of energy as in
equation~\eqref{eqn:energy_subpatch}. As boundary conditions we
set the incoming characteristic variables of one patch to be the
outgoing one of the neighboring grid,
\begin{align}
v_{BC}^R = v_{II}^L\,, \qquad
v_{BC}^L = v_{II}^R\,,
\end{align}
and demand that the change of energy of the sub patches in time due to the patching 
boundary is not growing.
Sufficient conditions for this are given by,
\begin{align}
p^R_{\beta\delta} = \frac{\tilde{\omega}_{\beta\delta} \Lambda^s_I}{\tilde{\omega}_{0\beta\delta}}\,, \qquad\qquad
p^L_{\beta\delta} = \frac{\tilde{\omega}_{\beta\delta} \Lambda^s_I}{\tilde{\omega}_{(N-1)\beta\delta}}\,.
\end{align}
In \texttt{bamps} we use these penalty parameters, but our discretization 
is made with Chebyschev rather than Legendre polynomials, the equations 
we solve are not linear with constant coefficients and nor are the 
Jacobians mapping from the master coordinates to our global 
Cartesian coordinates constant. Therefore it is to be determined 
empirically that the implemented method is in an appropriate sense 
stable. These facts may contribute to the necessity of employing 
the filter~\eqref{eqn:filter}.

%%%%%%%%%%%%%%%%%%%%%%%%%%%%%%%%%%%%%%%%%%%%%%%%%%%%%%%%%%%%%
\section{Axisymmetric considerations}\label{section:Axi}
%%%%%%%%%%%%%%%%%%%%%%%%%%%%%%%%%%%%%%%%%%%%%%%%%%%%%%%%%%%%%

Although \texttt{bamps} is a fully 3d code we are often interested in 
evolving axially symmetric data, which requires special 
attention for efficient treatment. In this section we collect 
together the relevant developments undertaken for axisymmetric 
initial data, apparent horizons and time evolution with the 
\texttt{bamps} code. 

%%%%%%%%%%%%%%%%%%%%%%%%%%%%%%%%%%%%%%%%%%%%%%%%%%%%%%%%%%%%%
\subsection{Brill wave initial data}\label{subsection:Brill}
%%%%%%%%%%%%%%%%%%%%%%%%%%%%%%%%%%%%%%%%%%%%%%%%%%%%%%%%%%%%%

Brill wave initial data is described in detail in many other 
sources. For completeness we give a bare-bones summary to 
highlight the particular choices that we make.

%%%%%%%%%%%%%%%%%%%%%%%%%%%%%%%%%%%%%%%%%%%%%%%%%%%%%%%%%%%%%
\paragraph*{Metric ansatz:} Following~\cite{Bri59,Epp77}, we 
start from a spatial metric of the form,
\begin{align}
\textrm{d}l^2&=\gamma_{ij}\textrm{d}x^i\textrm{d}x^j
=\Psi^4\big[e^{2q}(\textrm{d}\rho^2+\textrm{d}z^2)
+\rho^2\textrm{d}\phi^2\big]\,,\label{eqn:Brill_ansatz}
\end{align}
in cylindrical polar coordinates, and take the extrinsic curvature 
to vanish. Note that the assumption of conformal flatness in 
the~$\rho$-$z$ sector of the metric can be made in axisymmetry without
loss of generality. Under these assumptions the momentum constraints 
are trivially satisfied and the remaining Hamiltonian constraint takes 
the form
\begin{align}
D^2\Psi&=-\frac{\Psi}{4}\left(\frac{\p^2q}{\p \rho^2}+\frac{\p^2q}
{\p z^2}\right)\,.
\end{align}
We then make the parametrized ansatz,
\begin{align}
q(\rho,z)&=A\rho^2e^{-[(\rho-\rho_0)^2+(z-z_0)^2]}\,.\label{eqn:brill_seed}
\end{align}
for the seed function~$q(\rho,z)$ and solve for~$\Psi$ with boundary 
conditions~$\Psi\,\hat{=}\,0$ for asymptotic flatness at spatial 
infinity. This ansatz is the same as that studied in a number of 
other studies~\cite{HilBauWey13,Sor10,Rin06,HolMilWak93}. We call data 
with~$A>0$ geometrically prolate, and that with~$A<0$ geometrically 
oblate. In the initial data an apparent horizon can first be found 
at~$A=11.82$ with horizon mass~$M_H=4.8$. For geometrically oblate data 
an apparent horizon can first be found at~$A=-5.30$ with mass~$M_H=4.4$.
The pseudospectral method we use to solve the constraints is discussed 
a little more in section~\ref{subsection:Spec_id}. Our apparent horizon
search is explained in~\ref{subsection:AH}.

%%%%%%%%%%%%%%%%%%%%%%%%%%%%%%%%%%%%%%%%%%%%%%%%%%%%%%%%%%%%%
\begin{figure}[t]
\centering
\includegraphics[]{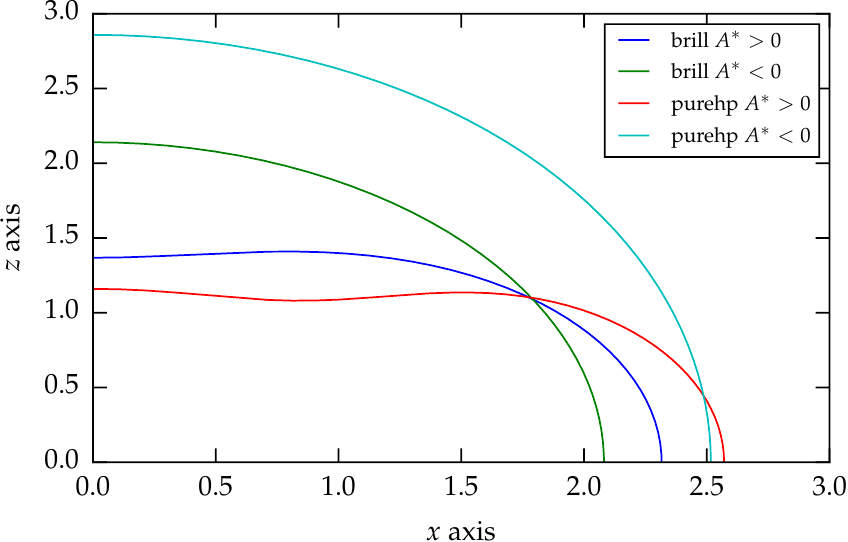}
\caption{The apparent horizons for centered Brill data 
with~$A=11.82$ and~$A=-5.3$ and for pure plus polarization data 
with~$A=2.381$ and~$A=-2.28$.
\label{fig:AH_first}}
\end{figure}
%%%%%%%%%%%%%%%%%%%%%%%%%%%%%%%%%%%%%%%%%%%%%%%%%%%%%%%%%%%%%

%%%%%%%%%%%%%%%%%%%%%%%%%%%%%%%%%%%%%%%%%%%%%%%%%%%%%%%%%%%%%
\subsection{Pure plus polarization wave data}
\label{subsection:H_plus_data}
%%%%%%%%%%%%%%%%%%%%%%%%%%%%%%%%%%%%%%%%%%%%%%%%%%%%%%%%%%%%%

%%%%%%%%%%%%%%%%%%%%%%%%%%%%%%%%%%%%%%%%%%%%%%%%%%%%%%%%%%%%%
\paragraph*{Metric ansatz:} Observers distant from a compact 
object see gravitational waves in the form,
\begin{align}
\textrm{d}l^2&=\textrm{d}r^2+r^2(1+h_+)\textrm{d}\theta^2
+ r^2\sin^2\theta(1-h_+)\textrm{d}\phi^2\nonumber\\
&\quad+2r^2\sin\theta\, h_\times \textrm{d}\theta \textrm{d}\phi
\,,\label{eqn:hplushcross}
\end{align}
with the wave polarizations~$h_+$ and~$h_\times$ small 
perturbations of the Minkowski metric. This suggests modifying 
the ansatz~\eqref{eqn:Brill_ansatz} to
\begin{align}
\textrm{d}\tilde{l}^2&=\textrm{d}r^2+r^2(e^{2q}\textrm{d}\theta^2
+ e^{-2q}\sin^2\theta\textrm{d}\phi^2)\,,\label{eqn:PurehplusAnsatz}
\end{align}
so that if we choose the seed function small and centered far
from the origin we will have initial data that represents a 
pure plus polarization gravitational wave. One could similarly 
make an ansatz for pure cross polarization waves, or indeed 
make other choices completely like~\cite{Shi97a} which we have 
also implemented and tested. 

%%%%%%%%%%%%%%%%%%%%%%%%%%%%%%%%%%%%%%%%%%%%%%%%%%%%%%%%%%%%%
\paragraph*{The constraints:} Again we start with moment of time
symmetry initial data, so the remaining constraint takes the 
form,
\begin{align}
\tilde{\Delta}\Psi&=\frac{1}{8}\Psi \tilde{R}\,.
\label{eqn:HamCMomTimeSym}
\end{align}
The conformal Ricci scalar is, 
\begin{align}
\tilde{R}&= \frac{2}{r^2}\big[e^{-2q}-1-(r\,\p_rq)^2\big]
-\frac{1}{r^2\sin^3\theta}\p_\theta(\sin^3\theta\p_\theta e^{-2q})\,,
\end{align}
and the Laplacian of the conformal metric is,
\begin{align}
\tilde{\Delta}\Psi&=\frac{1}{r^2}
\p_r\big(r^2\p_r\Psi\big)
+\frac{e^{2q}}{\sin^2\theta}
\p_\theta\big(e^{-2q}\sin^2\theta\,\p_\theta\Psi\big)\,.
\end{align}
Once more we impose the obvious boundary conditions for asymptotic
flatness at spatial infinity, and choose the seed function,
\begin{align}
q(r,\theta)&=A\,r^4\sin^2\theta\,e^{-[r^2-2r\rho_0 \sin\theta+\rho_0^2]}\,,
\label{eq:purehp}
\end{align}
which makes the conformal metric regular on axis. 

%%%%%%%%%%%%%%%%%%%%%%%%%%%%%%%%%%%%%%%%%%%%%%%%%%%%%%%%%%%%%
\paragraph*{Apparent horizons:} Taking centered data with~$A<0$
we first find an apparent horizon at around~$A=-2.28$, with 
mass~$M_H=5.47$. Looking for apparent horizons in centered data 
when~$A>0$, we find the curious result that there is a 
region~$[2.381,2.568]$ in which apparent horizons are first found.
Curiously, in the range~$[2.569,3.006]$, the data again seemed
to be horizonless. Continue at~$A=3.007$ we find horizons
again up to~$A=3.750$ where we stopped our search. We searched
for horizons using the resolution~$\Delta A=0.001$.
A closer look at the data at the boundaries of the `horizonless' region 
shows that the shape of the horizon is very nearly not a ray-body, 
and we expect that our method simply can not find the horizons 
in this range of amplitudes (see section~\ref{subsection:AH}). We 
expect that this could be remedied by implementing an offset in~$\rho$ 
in the parametrization of the surface similar to that in~$z$ which 
we already have, but we leave this improvement for the future. 
The first apparent horizon for this data, found at~$A=2.381$, is 
plotted in Fig.~\ref{fig:AH_first}. It has a mass of~$M_H=4.8$. 

%%%%%%%%%%%%%%%%%%%%%%%%%%%%%%%%%%%%%%%%%%%%%%%%%%%%%%%%%%%%%
\subsection{Teukolsky wave initial data}
\label{subsection:Teukolsky_data}
%%%%%%%%%%%%%%%%%%%%%%%%%%%%%%%%%%%%%%%%%%%%%%%%%%%%%%%%%%%%%

%%%%%%%%%%%%%%%%%%%%%%%%%%%%%%%%%%%%%%%%%%%%%%%%%%%%%%%%%%%%%
\paragraph*{Initial data for numerical relativity:} Teukolsky 
waves~\cite{Teu82,Rin08} are an exact solution to GR 
linearized around flat-space, and were used as a seed function 
in~\cite{AbrEva92}, the first numerical study of the critical 
collapse of gravitational waves, in the construction of full 
solutions to the constraints. In particular the waves 
were taken to be centered at some~$r_0\ne0$, with a radial width 
much less than $r_0$, with an~$l=2$, $m=0$ spherical harmonic 
dependence, and mostly incoming. Since we are restricting to 
moment of time symmetry data, we can not satisfy the last of 
these conditions, but we expect that if the waves are placed 
at some sufficiently large~$r_0$ then they will initially be 
weakly self-interacting, and roughly one half of the wave will 
simply propagate outwards. One could use the ansatz made in the 
Teukolsky wave initial data to construct incoming boundary data, 
but we leave this for future work. The construction of these data 
are well-described in~\cite{BauSha10} and were employed 
in~\cite{HilBauWey13}. See also~\cite{PfeKidSch04}. Therefore 
here we want only to describe a subtlety that was overlooked in 
both of these references.

%%%%%%%%%%%%%%%%%%%%%%%%%%%%%%%%%%%%%%%%%%%%%%%%%%%%%%%%%%%%%
\paragraph*{Regularity of the conformal metric:} Let us consider
the `polar' Teukolsky data. A similar discussion holds for axial
data. The conformal metric for the solution of the Hamiltonian 
constraint is, in spherical polar coordinates, 
\begin{align}
\tilde{\gamma}_{rr}&=1+a f_{rr}\,,\qquad\quad
\tilde{\gamma}_{r\theta}= b f_{r\theta}\,r\,,\nonumber\\
\tilde{\gamma}_{\theta\theta}&=(1+c\,f_{\theta\theta}-a)r^2\,,\nonumber\\
\tilde{\gamma}_{\phi\phi}&= (1-c\,f_{\theta\theta}+a\,f_{\phi\phi})
r^2\sin^2\theta\,,
\end{align}
with the remaining components vanishing. Here we have already 
restricted the ansatz by removing terms that vanish for~$l=2$ 
and~$m=0$ spherical harmonics. The angular 
functions~$f_{rr},f_{\theta\theta},f_{r\theta}$ and~$f_{\phi\phi}$ are,
\begin{align}
f_{rr}&= 2-3\sin^2\theta\,,&\quad f_{r\theta}=-3\sin\theta\cos\theta\,,\nonumber\\
f_{\theta\theta}&= 2-3\sin^2\theta\,,&\quad f_{\phi\phi}=3\sin^2\theta-1\,,
\end{align}
whilst the remaining radial functions~$a,b,c$ are constructed 
according to the recipe of~\cite{BauSha10}, so that,
\begin{align}
a&=3\left[\frac{F^{(2)}}{r^3}+\frac{3F^{(1)}}{r^4}
+\frac{3F}{r^5}\right]\,,\nonumber\\
b&=-\left[\frac{F^{(3)}}{r^2}+\frac{3F^{(2)}}{r^3}+\frac{6F^{(1)}}{r^4}+\frac{6F}{r^5}
\right]\,,\\
c&=\frac{1}{4}\left[\frac{F^{(4)}}{r}+\frac{2F^{(3)}}{r^2}+\frac{9F^{(2)}}{r^3}
+\frac{21F^{(1)}}{r^4}+\frac{21F}{r^5}\right]\,.\nonumber
\end{align}
In this expression we have used the shorthand,
\begin{align}
F^{(n)}=\left[\frac{\textrm{d}^nF(x)}{\textrm{d}x^n}\right]_{x=-r}
-(-1)^n\left[\frac{\textrm{d}^nF(x)}{\textrm{d}x^n}\right]_{x=r}\,,
\end{align}
and finally the seed function is~$F(x)$. In~\cite{HilBauWey13} the seed 
function was taken to be,
\begin{align}
F(x)&=\frac{A}{2}
\frac{x^p}{\sigma}\Big(e^{-[(x+r_0)/\sigma]^2}+e^{-[(x-r_0)/\sigma]^2}\Big)\,,
\label{eq:Teuk_irreg_seed}
\end{align}
with~$p=1$. For local flatness however it is necessary~\cite{Rin06} 
that the combinations,
\begin{align}
&\cos^2\theta\,\tilde{\gamma}_{rr}+r^{-2}\sin^2\theta\,
\tilde{\gamma}_{\theta\theta}
-r^{-1}\sin2\theta\,\tilde{\gamma}_{r\theta}\,,\nonumber\\
&r^{-1}\cos\theta\,\tilde{\gamma}_{rr}-r^{-3}\cos\theta\,
\tilde{\gamma}_{\theta\theta} 
+r^{-2}\sin^{-1}\theta\cos2\theta\,
\tilde{\gamma}_{r\theta}\,,\nonumber\\
&\sin^2\theta\,\tilde{\gamma}_{rr} + r^{-2}\cos^2\theta\,
\tilde{\gamma}_{\theta\theta} 
+r^{-2}\sin^2\theta\,\tilde{\gamma}_{\phi\phi}
+r^{-1}\sin\theta\,\tilde{\gamma}_{\theta\theta}\,,\nonumber\\
&r^{-2}\,\tilde{\gamma}_{rr}+r^{-4}\tan^{-2}\theta\,\tilde{\gamma}_{\theta\theta}
-r^{-4}\tilde{\gamma}_{\phi\phi}+2r^{-3}\tan^{-1}\theta\,
\tilde{\gamma}_{r\theta}\,,\nonumber
\end{align}
of the metric components are regular functions of~$z=r\cos\theta$ 
and~$\rho^2=r^2\sin^2\theta$. Therefore one may worry about 
the high powers of~$r^{-1}$ present in the recipe. This worry is 
justified, because for the particular seed function~\eqref{eq:Teuk_irreg_seed} 
the resulting conformal metric has a conical singularity at the 
origin, since for example the latter combination diverges like~$r^{-5}$
as~$r\to0$. Therefore the seed function used in~\cite{HilBauWey13} 
is not suitable to construct regular gravitational wave initial data.
Regularity is obtained if one chooses instead takes~$p=9$. The fact 
that such a high power of~$x$ is required in the seed function 
illustrates the depth of the singularity that was present 
beforehand.

%%%%%%%%%%%%%%%%%%%%%%%%%%%%%%%%%%%%%%%%%%%%%%%%%%%%%%%%%%%%%
\paragraph*{Comments on the numerical results of~\cite{HilBauWey13}
with Teukolsky initial data:}
Since the seed function~\eqref{eq:Teuk_irreg_seed} gives rise 
to an irregular conformal metric, the corresponding initial data 
evolved in~\cite{HilBauWey13} were in principle wrong, as more
careful convergence testing may have revealed. Should we then 
discard those results? Probably not; since the parameters 
taken were~$r_0=2$ and~$\sigma=1/2$, both the seed function and its 
derivatives were highly suppressed at the origin, and therefore 
in practical terms it is unlikely that the leading error in the 
simulations was caused by this problem. We will not attempt
however to evolve the older data with \texttt{bamps}, not least because
constructing irregular data with our spectral elliptic solver
is troublesome. These issues do not affect the seed function 
used in~\cite{AbrEva92} which was of compact support.

%%%%%%%%%%%%%%%%%%%%%%%%%%%%%%%%%%%%%%%%%%%%%%%%%%%%%%%%%%%%%
\subsection{Solving the constraints}
\label{subsection:Spec_id}
%%%%%%%%%%%%%%%%%%%%%%%%%%%%%%%%%%%%%%%%%%%%%%%%%%%%%%%%%%%%%

%%%%%%%%%%%%%%%%%%%%%%%%%%%%%%%%%%%%%%%%%%%%%%%%%%%%%%%%%%%%%
\paragraph*{Compactified coordinates:} To solve for moment of 
time symmetry initial data, we write the spatial metric in 
spherical polar coordinates~$(r,\theta,\phi)$, and compactify 
the radial coordinate, leaving us with 
coordinates~$(A,\theta,\phi)$. The compactification is defined 
either by,
\begin{align}
r&=\frac{m\,A}{2\,(1-A)}\,,\label{eq:abt_comp}
\end{align}
as suggested in~\cite{AnsBruTic04}, and used in~\cite{HilBauWey13} 
in the same elliptic solver employed presently, or
\begin{align}
r&=\frac{m\,A}{2\,(1-A^2)}\,,\label{eq:reg_comp}
\end{align}
similar to that employed for example in~\cite{CalGunHil05}. The 
parameter~$m$ partially controls the rate of compactification,
but in either case spatial infinity corresponds to~$A=1$. 

%%%%%%%%%%%%%%%%%%%%%%%%%%%%%%%%%%%%%%%%%%%%%%%%%%%%%%%%%%%%%
\paragraph*{Numerical solution:} To discretize we employ a Chebyschev 
discretization in the radial~$A$ direction, and a Fourier grid 
in the angular directions. Since the Hamiltonian constraint in 
this context is linear, solving the constraints amounts to a 
matrix inversion. With our particular method we find that the 
choice~\eqref{eq:abt_comp} leads to slightly worse constraint 
violations at a fixed resolution. One possible cause of this is 
that the coordinates~\eqref{eq:abt_comp} are irregular at the 
origin. Perhaps it is possible to use the alternative 
compactification in the construction of trumpet or puncture 
blackhole initial data, but we leave this for future consideration.

%%%%%%%%%%%%%%%%%%%%%%%%%%%%%%%%%%%%%%%%%%%%%%%%%%%%%%%%%%%%%
\subsection{Axisymmetric apparent horizons}\label{subsection:AH}
%%%%%%%%%%%%%%%%%%%%%%%%%%%%%%%%%%%%%%%%%%%%%%%%%%%%%%%%%%%%%

%%%%%%%%%%%%%%%%%%%%%%%%%%%%%%%%%%%%%%%%%%%%%%%%%%%%%%%%%%%%%
\paragraph*{Formulation of the AH conditions:} An apparent horizon 
is a closed two surface in the spatial slice, with unit outward 
pointing normal~$s^i$, with expansion,
\begin{align}
H=D_is^i-K+s^is^jK_{ij}=0\,,\label{eqn:AH}
\end{align}
where~$s^i$ is the unit normal to the surface. Our approach 
to the apparent horizon search is based on that of~\cite{AlcBraBru98} 
as also presented in~\cite{Tho06,Alc08}. First given the spatial 
metric and extrinsic curvature~$\gamma_{ij},K_{ij}$ in Cartesian 
coordinates, we transform to work in spherical polar coordinates 
defined by
\begin{align}
r^2&=x^2+y^2+(z-z_0)^2\,,\quad
\theta=\arccos\Big(\frac{z-z_0}{r}\Big)\,.
\end{align}
with~$\theta\in[0,\pi]$ and where we take the z-axis to be the 
symmetry axis. In axisymmetry without twist, the spatial metric and 
extrinsic curvature then take the form
\begin{align}
S_{ij}&=\left(\begin{array}{ccc}
S_{rr}&r\sin\theta S_{rT}&0\\
r\sin\theta S_{rT}&r^2S_{\theta T}&0\\
0&0&r^2\sin^2\theta S_{\phi T}
\end{array}\right)\,,
\end{align}
in the~$\phi=0$ plane. Local flatness on axis implies that the 
components~$S_{rr},S_{rT},S_{\theta T}$ and~$S_{\phi T}$ are even functions 
of~$\theta$ around the symmetry axis, with~$S_{\theta T}-S_{\phi T}\sim\theta^2$ 
around~$\theta=0$, and similar dependence around~$\theta=\pi$. Working 
in the~$\rho$-$z$ plane we may parametrize an apparent horizon by the 
level set~$s=0$ of 
\begin{align}
s=r-F(\theta)\,,
\end{align}
in terms of which the apparent horizon condition~\eqref{eqn:AH} 
can be rewritten as a first order ODE system,
\begin{align}
F'&=G \,,\nonumber\\
G'&= (\sin^2\theta\,\gamma_{rT}^2-\gamma_{rr}\gamma_{\theta T})F^2L^2
q^{ij}(\Gamma^k{}_{ij}D_ks+LK_{ij})
\,.\label{eqn:AH_axi}
\end{align}
for~$F(\theta)$ and~$G(\theta)$, where the unit spatial vector~$s^i$
and magnitude~$L$ are given by,
\begin{align}
s^i&=\gamma^{ij}LD_js\,,&\quad L^{-2}=\gamma^{ij}(D_is)(D_js)\,,
\end{align}
and~$q_{ij}=\gamma_{ij}-s_is_j$ is the induced metric in the level set. 
These expressions are evaluated in spherical polar coordinates. As noted 
elsewhere this parametrization is not completely general, only being 
sufficient if the apparent horizon is a ray-body containing the 
point~$z_0$. Regularity of an apparent horizon means 
that~$G(0)=G(\pi)=0$ 

%%%%%%%%%%%%%%%%%%%%%%%%%%%%%%%%%%%%%%%%%%%%%%%%%%%%%%%%%%%%%
\paragraph*{Search strategy:} Given the metric and extrinsic 
curvature we go about searching for an apparent horizon in the 
following way. First we choose~$z_0,r_0$ and integrate the 
ODE~\eqref{eqn:AH_axi} from~$\theta=0$ to~$\theta=\pi/2$, with 
initial conditions~$F(0)=r_0$ and~$G(0)=0$. We simultaneously
integrate backwards from~$\theta=\pi$ to~$\theta=\pi/2$ taking 
as initial conditions~$F(\pi)=r_0$ and~$G(\pi)=0$. If we have an 
apparent horizon the forwards~$(F^+, G^+)$ and backwards~$(F^-, G^-)$ 
solutions will satisfy,
\begin{align}
\Delta F &= F^+(\pi/2)-F^-(\pi/2)=0\,,\nonumber\\
\Delta G &= G^+(\pi/2)-G^-(\pi/2)=0\,.
\end{align}
This gives a non-linear root finding task on the 
function~$S:\mathbb{R}^2\to\mathbb{R}^2$ defined by 
\begin{align}
S(z_0,r_0)=(\Delta F,\Delta G).
\end{align}
One complication is that the ODE system~\eqref{eqn:AH_axi} needs to be 
regularized on the axis to impose our initial conditions. This is 
straightforwardly done by using the regularity conditions above, 
resulting in,
\begin{align}
G'&=\left(\frac{\gamma_{\theta T}}{2\gamma_{rr}}
-\frac{\gamma_{r T}}{2\gamma_{rr}}\right)F
+\left(\frac{\p_r\gamma_{\theta T}}{4\gamma_{rr}}
-\frac{K_{\theta T}}{2\sqrt{\gamma_{rr}}}\right)F^2\,.
\end{align}
at~$\theta=0$ and similarly at~$\theta=\pi$. To arrive at this 
expression we have explicitly used the regularity 
condition~$S_{\theta T}-S_{\phi T}\sim\theta^2$. In our numerical 
implementation we transform from Cartesian components, so this 
condition is automatically satisfied and we can instead use 
the condition in a slightly more complicated form 
involving~$\gamma_{\phi T}$ and~$K_{\phi T}$. To the best of our 
knowledge this regularization of the coefficients has not 
been used before. The second step of our search is to iterate 
on~$(z_0,r_0)$ until we find a solution, or until the method 
fails. As an alternative strategy, it is normally proposed to
integrate the ODE from~$\theta=0$ to~$\theta=\pi$ then 
perform a bisection search on~$G(\pi)$. We were unable to obtain 
satisfactory results this way because every surface except the 
apparent horizon itself diverges near~$\theta=\pi$, making the 
bisection hopeless. Reasonable first guesses for~$z_0$ would seem 
to be the position of the maximum of the Kretschmann scalar, or, 
if an apparent horizon was already found in a previous time-slice, 
the coordinate center of the previous horizon.

%%%%%%%%%%%%%%%%%%%%%%%%%%%%%%%%%%%%%%%%%%%%%%%%%%%%%%%%%%%%%
\paragraph*{Horizon mass:} In twist-free axisymmetry the apparent 
horizon mass~$M_H$ is related to the area of the apparent 
horizon~$A_H$ as,
\begin{align}
M_H^2&=\frac{A_H}{16\pi}\,.\label{eqn:M_H_defn}
\end{align}
We can compute the area of the apparent horizon as a simple 
integral, 
\begin{align}
A_H&=2\pi\int_{0}^{\pi}L^{-1}\sqrt{\gamma}\,
r^2\sin\theta\,\textrm{d}\theta\,.
\end{align}
where we have used the fact that apparent horizon is a surface 
of revolution. Here~$\gamma$ is the determinant of the 
spatial metric in Cartesian coordinates.

%%%%%%%%%%%%%%%%%%%%%%%%%%%%%%%%%%%%%%%%%%%%%%%%%%%%%%%%%%%%%
\paragraph*{Simplifying assumptions:} We are often interested 
in finding apparent horizons centered at the origin in spacetimes 
that are additionally reflection symmetric about the~$z=0$ plane. 
In this case we can trade our root-finding search above for a 
bisection search by simply fixing~$z_0=0$ and integrating the 
ODE~\eqref{eqn:AH_axi} from~$\theta=0$ to~$\theta=\pi/2$. Here 
we start the integration from different initial radii~$r_0$ 
until we find points about which which~$G(\pi/2)$ changes sign. 
We then bisect in~$r_0$ to find the apparent horizon, 
where~$G(\pi/2)=0$. We typically choose the 
criterion~$G(\pi/2)<10^{-8}$ to end the search. As in the more 
general case, if we find many such surfaces we take the 
outermost as the apparent horizon. 

%%%%%%%%%%%%%%%%%%%%%%%%%%%%%%%%%%%%%%%%%%%%%%%%%%%%%%%%%%%%%
\begin{figure*}[t]
\centering
\includegraphics[]{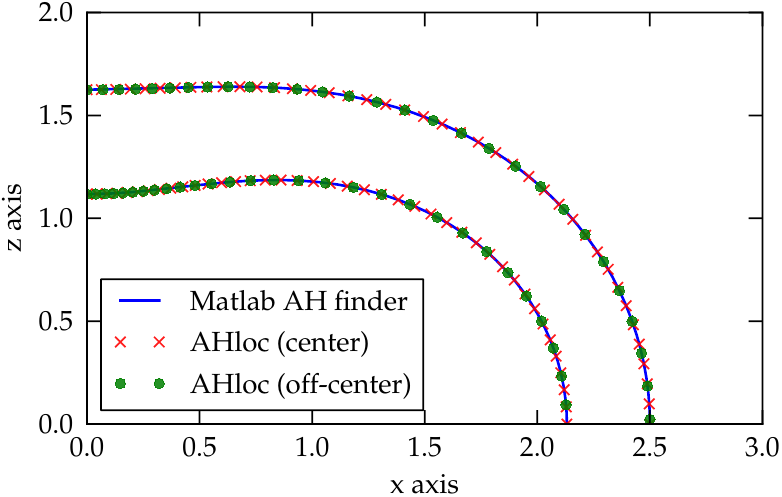}
\includegraphics[]{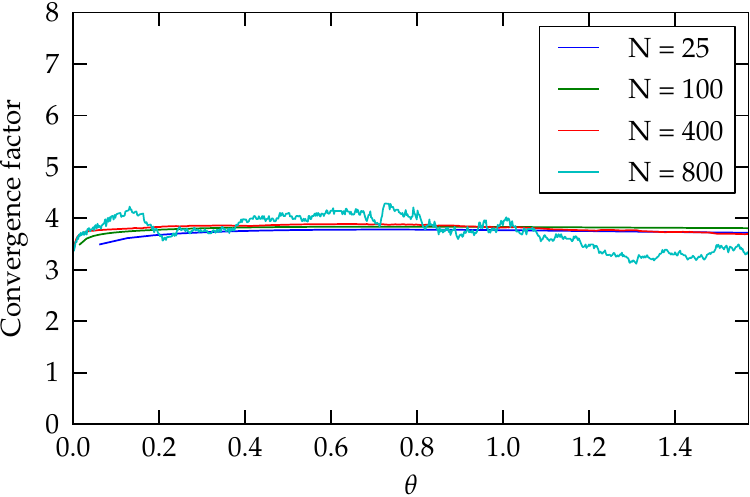}
\caption{In the left hand panel the apparent horizon for a 
centered~$A=12$ Brill wave, as found by our apparent horizon 
finder and a bespoke Brill-wave apparent horizon finder, are 
plotted. This data has been used as a standard test case 
elsewhere in the literature~\cite{AlcBraBru98,OliRod12}. 
We compute the ADM mass as~$M_{\textrm{ADM}}=4.67$, which compares 
perfectly with~$M_{\textrm{ADM}}=4.67\pm0.01$ given in~\cite{AlcBraBru98}.
The horizon mass is~$M_H=4.66$, again in agreement with the 
literature. In the right panel we show pointwise self-convergence 
labelled by~$N=25,100,400$ and~$800$, with~$N+1$ the lowest number 
of points in the series, and where we evolved with~$2N+1$ and~$4N+1$ 
to make the plot. Note that very few points are needed to show 
clean convergence because the surface varies slowly in~$\theta$. 
This also means that one can not reliably convergence test at 
high resolutions because the difference between the computed 
surfaces are essentially at the level of round-off.}
\label{fig:AH_test}
\end{figure*}
%%%%%%%%%%%%%%%%%%%%%%%%%%%%%%%%%%%%%%%%%%%%%%%%%%%%%%%%%%%%%

%%%%%%%%%%%%%%%%%%%%%%%%%%%%%%%%%%%%%%%%%%%%%%%%%%%%%%%%%%%%%
\paragraph*{Numerical implementation:} In practice we search 
for an apparent horizon as follows. During a \texttt{bamps} evolution 
we output the necessary components of the spatial metric and 
extrinsic curvature in the~$y=0$ plane at different coordinate
times. For the integration of the ODE, we use various ODE 
integrators in the GSL~\cite{gsl_web}. To determine the apparent 
horizon accurately as fast as possible we use the {\it explicit 
embedded Runge-Kutta Prince-Dormand (8, 9) method}, a high-order 
adaptive step integrator. When convergence testing we use a 
simple fourth order Runge-Kutta integrator. To evaluate the 
metric and extrinsic curvature at each point~$(r=F,\theta)$ 
along the level set we use barycentric Lagrange interpolation 
inside each \texttt{bamps} sub-grid. For the root-finding we again
use the GSL, now choosing one of the `hybrid' algorithms that do
not need the Jacobian of the system of equations being solved. In 
Fig.~\ref{fig:AH_test} we present the apparent horizon found 
using our method for a centered~$\rho=0$, amplitude~$A=12$, Brill 
wave initial data set, comparing it with that which we find using 
a stand-alone apparent horizon finder implemented in the MATLAB 
initial data code employed in~\cite{DieBru13}. 

%%%%%%%%%%%%%%%%%%%%%%%%%%%%%%%%%%%%%%%%%%%%%%%%%%%%%%%%%%%%%
\subsection{The analytic Cartoon method}
\label{section:Cartoon}
%%%%%%%%%%%%%%%%%%%%%%%%%%%%%%%%%%%%%%%%%%%%%%%%%%%%%%%%%%%%%

Here we discuss the implementation of the so-called Cartoon method
\cite{AlcBraBru99a} for axisymmetry in a pseudospectral method for the
Einstein equations.  We assume that we are given the 3d system in a
Cartesian coordinate system $x^i$ in which all variables are smooth,
$T\in C^\infty$.  The basic idea of the Cartoon method is to apply
wherever possible the same coordinates and discretization that lead to
stable evolutions in 3d.  Hence we compute the axisymmetrically
reduced system in Cartesian coordinates and with Cartesian tensor
components, without adapting coordinates and thereby avoiding the
coordinate singularity at the axis.

Concretely, the computational domain is chosen to be the $x$-$z$-plane
defined by $y=0$. Partial derivatives $\del_x$ and $\del_z$ are
computed as for the 3d system. What is missing are the points and the
numerical data in the $y$-direction for the computation of
$\del_y$. However, we can obtain the $y$-derivative by invoking
axisymmetry, since the fields in the $y=0$, $x$-$z$-plane determine
the fields for $y\ne0$ by the rotation symmetry. Similarly, it
suffices to consider only the half-plane $x\ge0$ and $y=0$ while still
using the same stencils for $\del_x$ and $\del_z$ as in 3d. 

The Cartoon method was first introduced for a Cartesian 
BSSNOK~\cite{BauSha98,ShiNak95,NakOohKoj87} code
using finite differencing \cite{AlcBraBru99a}. The $\partial_y$
derivative was computed by adding ghost points in the $y$ direction,
so that identical 3d stencils could be used for 3d and axisymmetric 2d
calculations. For a spectral collocation method, we could
do the same and populate a 3d spectral element by rotation. There
would still be significant gains in efficiency since only a 2d subset 
of a 3d spectral grid consisting of many patches needs to be
populated. However, it is also possible to derive analytical formulas
for $\partial_y$ in terms of quantities in the $y=0$ plane only, so this is
clearly the preferred way to proceed. To our knowledge this was first
implemented in~\cite{Pre04}, in that case for finite differences and 
the second order GHG system.
For an arbitrary smooth tensor~$T$, axisymmetry is given by the
vanishing of its Lie derivative along the rotational vector,~$\lie_\phi T = 0\,$.

%%%%%%%%%%%%%%%%%%%%%%%%%%%%%%%%%%%%%%%%%%%%%%%%%%%%%%%%%%%%%
\paragraph*{Off-axis, $x\neq0$.}
Let us consider various tensor types of interest, suppressing their~$t$ 
and~$z$ dependence. For a scalar,
\begin{align}
  \del_y u(x,0) = 0\,.
\end{align}
The second derivative does not vanish in general.
For vectors and covectors~($x\neq0$),
\begin{align}
  \del_y v^x(x,0) = -\frac{1}{x} v^y(x,0)\,, \quad
  \del_y v^y(x,0) = \frac{1}{x} v^x(x,0)\,,
\nonumber\\
  \del_y w_x(x,0) = -\frac{1}{x} w_y(x,0)\,, \quad
  \del_y w_y(x,0) = \frac{1}{x} w_x(x,0)\,.
\label{delyvw}
\end{align}
the derivative is equal to the components of the vector divided by
radius, with~$x$ and~$y$ components interchanged. For a symmetric~$(0,2)$ 
tensor (say, the four-metric $g_{ab}$), at~$y=0$, $x\neq0$,
\begin{align}
& 
\del_yg_{tt}=0, \quad 
\del_yg_{tz}=0, \quad
\del_yg_{zz}= 0,
\nonumber\\
&
\del_yg_{tx}=-\frac{1}{x}g_{ty}, \quad
\del_yg_{ty}= \frac{1}{x}g_{tx},
\nonumber\\
&
\del_yg_{xz}= -\frac{1}{x}g_{yz}, \quad
\del_yg_{yz}=  \frac{1}{x}g_{xz},
\nonumber\\
&
\del_yg_{xx}= -\frac{2}{x}g_{xy}, \quad
\del_yg_{yy}=  \frac{2}{x}g_{xy}, \quad
\nonumber\\
&
\del_yg_{xy}=  \frac{1}{x} (g_{xx} - g_{yy}). \quad
\label{delyg1}
\end{align}
Some components behave like scalars, some like covectors, and some
show the two terms occurring in the Lie derivative, which may result
in a factor two due to symmetry.

%%%%%%%%%%%%%%%%%%%%%%%%%%%%%%%%%%%%%%%%%%%%%%%%%%%%%%%%%%%%%
\paragraph*{On-axis, $x=0$.} Axisymmetry by itself does not imply 
differentiability on the axis. Consider, for example,~$u(x,y)=\rho$. 
We combine axisymmetry with the condition that in Cartesian 
coordinates~$T\in C^\infty$ in two ways. First, consider parity 
under~$(x,y)\rightarrow(-x,-y)$, which corresponds to a rotation 
by~$\pi$ around the $z$-axis. Because of axisymmetry, we 
have~$T(\rho,0) = \pm T(-\rho,0)$ and~$\del_yT(\rho,0)=\mp 
\del_yT(-\rho,0)$.  Since~$\del_yT$ is continuous, the 
limit~$\rho\rightarrow0$ exists. Hence for tensors that are even 
under this type of parity, the derivative vanishes,~$\del_yT_{even}(0,0)=0$. 
For tensors that are odd, the tensor vanishes,~$T_{odd}(0,0)=0$, 
and~$\del_yT_{odd}(0,0)$ is a regular, finite value. We therefore impose 
that~$\del_y$ vanishes on the axis for even quantities and 
ask how we can compute the value for the odd quantities.

From vanishing of the Lie derivative, we obtain relations for the 
tensor components themselves, not for their derivative. For a scalar, 
there is no extra condition. Examples for relations obtained 
from~\eqref{delyvw}--\eqref{delyg1} are,
\begin{align}
   & v^i(0,0) = 0, \quad w_i(0,0) = 0 
\nonumber\\
   & g_{tx}(0,0) = g_{ty}(0,0) = g_{xz}(0,0) = g_{yz}(0,0) = 0, 
\nonumber\\
   & g_{xy}(0,0) = 0, \quad g_{xx}(0,0) = g_{yy}(0,0).
\label{gxxgyy}
\end{align}
Although we obtain some of the same information that we already
discussed for~$(x,y)\rightarrow(-x,-y)$ parity,
for even parity quantities with two or more indices there are
additional relations. For the metric components these are related to 
covariance under rotation by~$\pi/2$, or~$(x,y)\rightarrow(-y,x)$. 

To find the derivative $\partial_y$ at (0,0), we invoke l'Hopital's rule. 
Basically, in (\ref{delyvw})--(\ref{delyg1}) the~$\frac{1}{x}$ factors
become a partial derivative in~$x$ because the other terms vanish. 
For example,
\begin{align}
  \del_y v^x(0,0) = -\del_x v^y(0,0), \quad
  \del_y v^y(0,0) = \del_x v^x(0,0).
\label{delyv00}
\end{align}
Notice that starting with two-index components this is not entirely 
trivial since there is more than just one term on the right-hand side.

%%%%%%%%%%%%%%%%%%%%%%%%%%%%%%%%%%%%%%%%%%%%%%%%%%%%%%%%%%%%%
\paragraph*{Axisymmetry for partial derivatives of tensors.}
There also are objects like~$\Phi_{iab}=\partial_ig_{ab}$, which are 
not tensors, but partial derivatives of tensors. The Lie derivative
$\lie_\phi\partial_ig_{ab}$ is in general not defined for non-tensors,
and a priori it is not clear whether~$\lie_\phi\partial_ig_{ab}=0$ 
implies axisymmetry. However, we can obtain the required formulas 
by computing
\begin{align}
  \del_i\lie_\phi g_{ab}
&=
  \hat\lie_\phi \del_ig_{ab} 
+ g_{cb} \del_a \del_i \phi^c
+ g_{ac} \del_b \del_i \phi^c,
\end{align}
where~$\hat\lie_\phi$ is introduced to collect those terms that correspond
to the Lie derivative of a tensor, and the remaining terms are the deviation 
from the tensor formula. Note how the last term in 
$\del_i (\phi^c\del_c g_{ab}) = \phi^c\del_c\del_i g_{ab} + \del_c g_{ab} \del_i\phi^c$ 
provides precisely the term that would otherwise
be missing in the sum over index locations in $\hat\lie_\phi \del_ig_{ab}$.

The key observation is that in the case of a {\it rigid} rotation in
adapted coordinates generated by~$\phi^a = (0,-y,x,0)^T$, the second
derivatives of $\phi^a$ vanish,
\begin{align}
  \del_a \del_b \phi^c = 0\,.
\label{deldelphi}
\end{align}
Therefore, in this special case we obtain the correct result using the
tensor formula, 
\begin{align}
  \del_i\lie_\phi g_{ab} = \hat\lie_\phi \del_ig_{ab},
\label{liehatlie}
\end{align}
as was also noted in~\cite{AlcBraBru99a}. This generalizes immediately to 
partial derivatives of arbitrary tensors, and also includes the case of the 
Christoffel symbol required for the BSSNOK or Z4c system, 
compare~\cite{AlcBraBru99a}. Eqn.~\eqref{deldelphi} furthermore simplifies 
the computation of second derivatives.

%%%%%%%%%%%%%%%%%%%%%%%%%%%%%%%%%%%%%%%%%%%%%%%%%%%%%%%%%%%%%
\section{Code validation}\label{section:Experiments}
%%%%%%%%%%%%%%%%%%%%%%%%%%%%%%%%%%%%%%%%%%%%%%%%%%%%%%%%%%%%%

In this section we present a set of numerical experiments 
performed to try and obtain an optimal setup for the first 
order generalized harmonic system for our gravitational 
wave collapse evolutions that follow in later work. 

%%%%%%%%%%%%%%%%%%%%%%%%%%%%%%%%%%%%%%%%%%%%%%%%%%%%%%%%%%%%%
\subsection{Gauge boundary}
\label{section:Exp_G_BCs}
%%%%%%%%%%%%%%%%%%%%%%%%%%%%%%%%%%%%%%%%%%%%%%%%%%%%%%%%%%%%%

%%%%%%%%%%%%%%%%%%%%%%%%%%%%%%%%%%%%%%%%%%%%%%%%%%%%%%%%%%%%%
\begin{figure*}[t]
\centering
\includegraphics[]{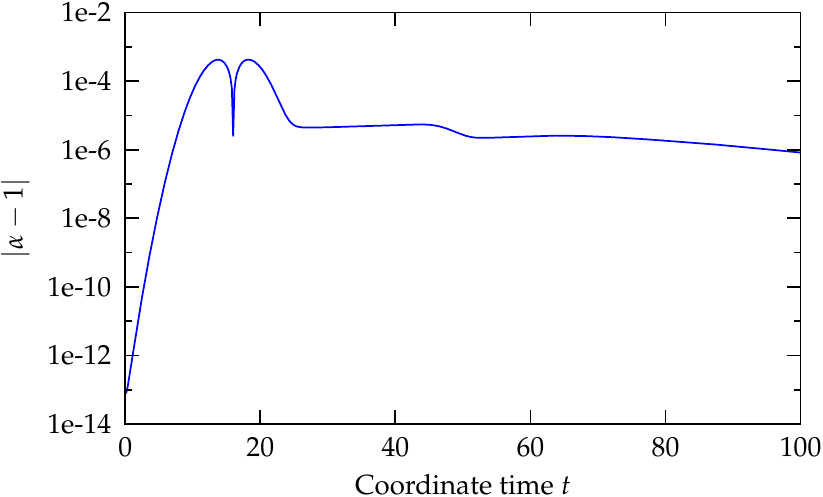}
\includegraphics[]{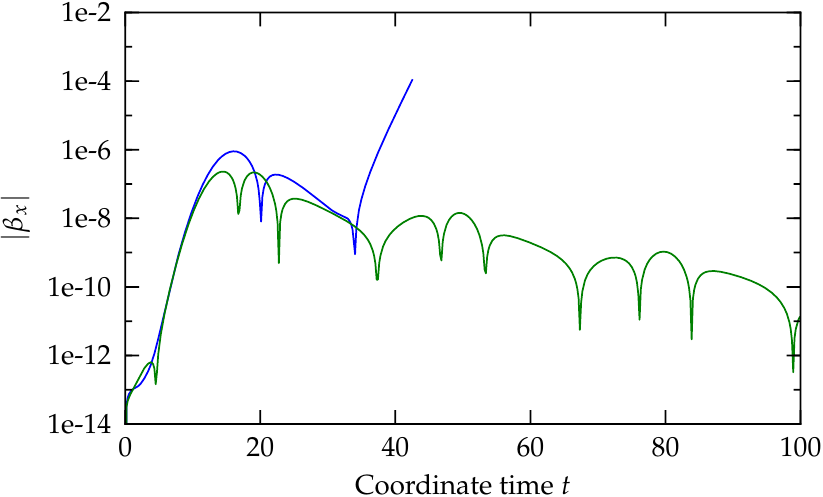}
\caption{In the left panel we plot the$~|\alpha-1|$ in the 
outer boundary as a function of time, obtained in the 
evolution of a gauge pulse on flat space, initially centered 
at the origin. The coordinates eventually seem to settle on,
or very close to Minkowski slices. On the right we plot the 
magnitude of the shift in the outer boundary using the 
harmonic damped wave gauge to evolve the same gauge pulse with 
either the gauge boundary condition~\eqref{eqn:G_BCs} 
or~\eqref{eqn:G_BCs_Freeze}. In the former case the shift rapidly
grows, causing the code to crash.
\label{fig:pgcp_1_2_16}}
\end{figure*}
%%%%%%%%%%%%%%%%%%%%%%%%%%%%%%%%%%%%%%%%%%%%%%%%%%%%%%%%%%%%%

%%%%%%%%%%%%%%%%%%%%%%%%%%%%%%%%%%%%%%%%%%%%%%%%%%%%%%%%%%%%%
\paragraph*{Gauge wave initial data:} We evolve the Minkowski 
line-element with a perturbation initially placed in the lapse, 
so that,
\begin{align}
\alpha(t=0)&=1+A\,e^{-[x^2+y^2+z^2]/\sigma}\,. 
\label{eqn:gauge_wave}
\end{align}
In the following set of experiments we always take~$A=0.01$ 
and~$\sigma=10$, and fix the grid setup. We take the standard 
formulation used in the SpEC code, namely~$\gamma_0=-\gamma_1=\gamma_2=1$,
and $\gamma_4=\gamma_5=0$. We impose outer boundary conditions at
a coordinate radius of~$r=16$ and evolve in~3d with octant 
symmetry imposed. 

%%%%%%%%%%%%%%%%%%%%%%%%%%%%%%%%%%%%%%%%%%%%%%%%%%%%%%%%%%%%%
\paragraph*{Harmonic gauge:} Starting with the pure harmonic 
gauge~$H_a=0$, we find that the outgoing gauge wave is harmlessly 
absorbed using either the gauge boundary condition~\eqref{eqn:G_BCs} 
or~\eqref{eqn:G_BCs_Freeze}. At the particular resolution and 
grid-setup that we chose for this test the harmonic constraint 
violation at the end of the evolution,~$t=100$, is around~$10^{-14}$ 
and shows no sign of increasing with either choice of gauge boundary 
condition. The difference between the results with the two gauge 
boundary conditions is rather small, the maximum difference in the
shift being around~$10^{-7}$ at the end of the run. But here the 
initial pulse is very weak, and this is of no concern. In the 
left panel of Fig.~\ref{fig:pgcp_1_2_16} we plot~$|\alpha-1|$ in the 
outer boundary, to demonstrate how the coordinates settle down.

%%%%%%%%%%%%%%%%%%%%%%%%%%%%%%%%%%%%%%%%%%%%%%%%%%%%%%%%%%%%%
\paragraph*{Generalized harmonic gauge:} Switching now to use the 
generalized harmonic gauge condition~\eqref{eqn:3+1_ghg} 
with~$\eta_L=0.4,p=1$ and~$\eta_S=6$. Using then the gauge boundary 
condition~\eqref{eqn:G_BCs} we find that the shift starts to 
grow at the boundary, and the numerics fail 
at~$t\sim 42$. This behavior is perhaps not surprising given 
the large damping coefficents and the understanding obtained for 
the constraint preserving subsystem with damping in 
section~\ref{subsection:CP}. The gauge source functions have 
the same effect on the gauge as the damping terms on the constraints,
namely they cause reflections from the boundary. We expect that it 
will be suppressed as the outer boundary is placed further out so 
that the gauge sources are smaller where the boundary condition is 
applied. Using instead the gauge boundary 
conditions~\eqref{eqn:G_BCs_Freeze} this growth is completely absent,
which is why we do not implement conditions derived explicitly 
to reduce gauge reflections in the present work. This behavior is 
demonstrated in the right panel of Fig.~\ref{fig:pgcp_1_2_16} where one 
sees the magnitude of the shift vector in the outer boundary in each 
case. With the gauge boundaries~\eqref{eqn:G_BCs_Freeze}, at the end 
of the run the harmonic constraint violation~$C_x$ is around~$10^{-14}$ 
and appears not to be growing. Looking at the shift however, it does 
seem that some further improvement may be possible in the future, as 
its peak lies at the outer boundary, with a value around~$10^{-11}$.

%%%%%%%%%%%%%%%%%%%%%%%%%%%%%%%%%%%%%%%%%%%%%%%%%%%%%%%%%%%%%
\subsection{Constraint experiments}\label{section:C_ex}
%%%%%%%%%%%%%%%%%%%%%%%%%%%%%%%%%%%%%%%%%%%%%%%%%%%%%%%%%%%%%

\begin{figure}[t]
  \centering
  \includegraphics[]{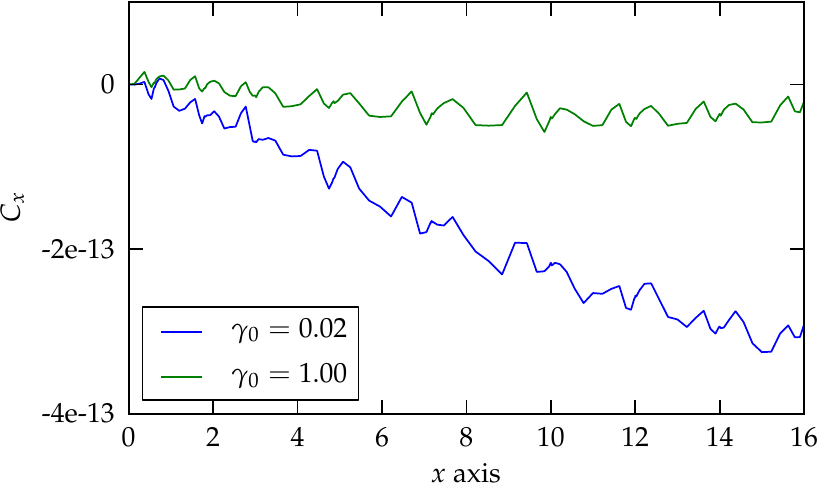}
  \caption{We show the~$C_x$ component of the harmonic constraint
  along the~$x$ at time~$t=100$ for two different sets of constraint
  damping parameters with formulation parameters $\gamma_4=\gamma_5=1/2$.
  in the evolution of a gauge pulse on flat-space as in 
  Fig.~\ref{fig:pgcp_1_2_16}, with the generalized harmonic gauge. 
  On this basis we take these formulation parameters with~$\gamma_0=1$ 
  as our standard choice.
  \label{Fig:Bamps_damping}}
\end{figure}

%%%%%%%%%%%%%%%%%%%%%%%%%%%%%%%%%%%%%%%%%%%%%%%%%%%%%%%%%%%%%
\paragraph*{Simplified subsystem:} We now repeat 
some of the experiments of the previous section with the 
choice~$\gamma_4=\gamma_5=1/2$, and with different choices 
of~$\gamma_0$, using always the gauge boundary 
condition~\eqref{eqn:G_BCs_Freeze}. With the pure harmonic 
gauge~$H_a=0$, we find that the constraint violation at~$t=100$
is again around~$10^{-14}$ if we take~$\gamma_0=1$, and 
slightly larger, but still less than~$10^{-13}$ if we 
choose~$\gamma_0=0.02$, the value suggested by the experiments 
in~\cite{WeyBerHil11} for a related formulation. Moving to the 
generalized harmonic choice~\eqref{eqn:3+1_ghg} once more, we 
find that again that the violation at the end of the experiment
is of the same order as when using the pure harmonic gauge.
The result is plotted in Fig.~\ref{Fig:Bamps_damping}. These results 
may not be representative when evolving different initial data, 
but we cautiously take~$\gamma_4=\gamma_5=1/2$ and~$\gamma_0=1$ as 
our default setting, periodically testing different choices, 
most often playing with~$\gamma_0$ in such experiments.

%%%%%%%%%%%%%%%%%%%%%%%%%%%%%%%%%%%%%%%%%%%%%%%%%%%%%%%%%%%%%
\paragraph*{Constraint preserving conditions:} We performed the
same experiments, with the generalized harmonic gauge and the 
new default formulation parameters, changing to the alternative 
constraint boundaries~\eqref{eqn:geom_cpbc_1} 
or~\eqref{eqn:cpbc_ref_red} and found first that the violation 
throughout is very similar to the initial choice~\eqref{eqn:GHG_11_cpbc}.
Although initially the violation with the reflection reducing 
condition is slightly smaller than with the `geometric' condition,
later on there is practically nothing to choose between them. 
Considering that the violations are in the round-off regime~$10^{-14}$
it is hard to judge from this experiment which of the conditions 
behaves most favorably.

%%%%%%%%%%%%%%%%%%%%%%%%%%%%%%%%%%%%%%%%%%%%%%%%%%%%%%%%%%%%%
%%%%%%%%%%%%%%%%%%%%%%%%%%%%%%%%%%%%%%%%%%%%%%%%%%%%%%%%%%%%%

%%%%%%%%%%%%%%%%%%%%%%%%%%%%%%%%%%%%%%%%%%%%%%%%%%%%%%%%%%%%%
\subsection{Lapse power in constraint damping}
\label{section:Damping}
%%%%%%%%%%%%%%%%%%%%%%%%%%%%%%%%%%%%%%%%%%%%%%%%%%%%%%%%%%%%%

%%%%%%%%%%%%%%%%%%%%%%%%%%%%%%%%%%%%%%%%%%%%%%%%%%%%%%%%%%%%%
\paragraph*{Initial data:} We now evolve centered~$A=2.5$ 
Brill wave initial data, which is subcritical, with an ADM mass
of~$M_{\textrm{ADM}}=0.19$. We evolve on the same grids used in 
the previous section, but with a slightly higher resolution 
($19^3$ rather than~$15^3$ points per cube). We evolve 
using~$\gamma_0=0.2\alpha^l$ with~$l=0$, the standard choice 
elsewhere, or~$l=-1$, a modification which we hope will reduce 
constraint growth in the strongest field region. As above we 
use the generalized harmonic gauge~\eqref{eqn:3+1_ghg}. We use 
only the gauge boundary condition~\eqref{eqn:G_BCs_Freeze}.

%%%%%%%%%%%%%%%%%%%%%%%%%%%%%%%%%%%%%%%%%%%%%%%%%%%%%%%%%%%%%
\paragraph*{Basic dynamics:} The Kretschmann scalar initially 
has a peak at the origin, evaluated around~$2300$ on the \texttt{bamps} 
grid, slightly less than in the previous study~\cite{HilBauWey13}. 
This peak oscillates at the origin, peaking after an initial 
bounce with value around~$500$. The feature then rapidly 
propagates away and by a coordinate time~$t=10$, the peak value 
on the grid is less around~$10^{-2}$. The lapse initially decreases 
at the origin, this feature then propagating out to the outer 
boundary, behind which the lapse drifts back towards its initial 
value, unity. 

%%%%%%%%%%%%%%%%%%%%%%%%%%%%%%%%%%%%%%%%%%%%%%%%%%%%%%%%%%%%%
\paragraph*{Constraint violation:} Examining the~$C_x$ constraint
for the~$A=2.5$ data along the $x$-axis, we see that only very 
small differences in the constraint violation between the~$l=0$
and~$l=-1$ evolutions. The small differences are not surprising 
because the lowest value the lapse function takes is around~$0.78$
having started from~$1$. The peaks of the~$C_x$ constraint 
in the~$l=-1$ evolution are about~$2$-$5\%$ smaller than in 
the~$l=0$ run. Increasing the amplitude of the initial data 
to~$A=4$, one might expect the improvement to be more 
significant as the lowest value of lapse decreases to~$0.37$, 
but the difference still amounts to between~$2$-$5\%$ at the 
peaks of the violation.

%%%%%%%%%%%%%%%%%%%%%%%%%%%%%%%%%%%%%%%%%%%%%%%%%%%%%%%%%%%%%
\subsection{BAM vs. \texttt{bamps} comparison}\label{section:comp}
%%%%%%%%%%%%%%%%%%%%%%%%%%%%%%%%%%%%%%%%%%%%%%%%%%%%%%%%%%%%%

\begin{figure}[t]
  \centering
  \includegraphics[width=\columnwidth]{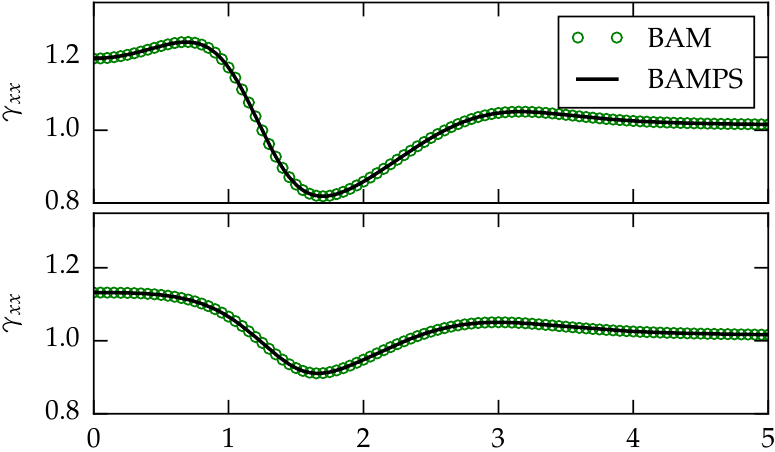}  
  \caption{Comparison of the results of a Brill wave~$A=1$ evolution
  with BAM and \texttt{bamps}. We show snapshots of the metric 
  component~$\gamma_{xx}$ along the~$x$ axis at $t=1.625$.
  In the upper panel we show the pure harmonic gauge, and 
  underneath the damped wave gauge with~$\eta_L=0$ and~$\eta_S=1.0$. 
  The results of the codes are in good agreement in either case.
  \label{Fig:bvb_harmonic}
  }
\end{figure}

Another validation strategy for \texttt{bamps} is to compare the 
numerical results with those of an independent code. For this 
we used BAM~\cite{BruGonHan06}, evolving identical initial data 
with the same gauge conditions. This comparison we performed by 
evolving a centered~$z_0=0$ Brill wave with~$A=1$. We chose this 
weak amplitude because evolving the Brill data accurately with 
BAM rapidly becomes expensive as~$A$ increases in magnitude. We 
used pure harmonic slicing~$\eta_L=0$ with either harmonic 
shift~$\eta_S=0$ or the damped harmonic shift~$\eta_S=1$. 
In the BAM code we evolve with the BSSNOK formulation, for 
completeness, this gauge condition is given by
\begin{align}
\p_t\beta^i&=\alpha^2\chi\left[\tilde{\Gamma}^i
+\tfrac{1}{2}\tilde{\gamma}^{ij}\p_j\ln\chi-\tilde{\gamma}^{ij}\p_j\ln\alpha
\right]\nonumber\\
&\quad-\eta_S\beta^i+\beta^j\p_j\beta^i\,.
\end{align}
in terms of the conformally decomposed BSSNOK variables. For this 
test we did not employ the spherical shells or constraint preserving 
boundary conditions of~\cite{HilBerThi12}. Since the outer boundaries 
were placed at~$x=y=z=12$, the solutions to the continuum PDEs 
being solved are not identical. Therefore we should not hope for 
perfect agreement for long. In Fig.~\ref{Fig:bvb_harmonic} we plot 
the spatial metric component~$\gamma_{xx}$ at $t=1.625$, when the 
agreement is still very good for either choice of the shift, being 
practically indistinguishable by eye. In practice the main source 
of disagreement at the resolution of these runs comes from 
mesh-refinement boundaries in the BAM grid setup. 

%%%%%%%%%%%%%%%%%%%%%%%%%%%%%%%%%%%%%%%%%%%%%%%%%%%%%%%%%%%%%
%%%%%%%%%%%%%%%%%%%%%%%%%%%%%%%%%%%%%%%%%%%%%%%%%%%%%%%%%%%%%

%%%%%%%%%%%%%%%%%%%%%%%%%%%%%%%%%%%%%%%%%%%%%%%%%%%%%%%%%%%%%
\subsection{Octant and Cartoon}\label{section:Oct_Cartoon}
%%%%%%%%%%%%%%%%%%%%%%%%%%%%%%%%%%%%%%%%%%%%%%%%%%%%%%%%%%%%%

%%%%%%%%%%%%%%%%%%%%%%%%%%%%%%%%%%%%%%%%%%%%%%%%%%%%%%%%%%%%%
\paragraph*{Initial data and grids:} To test our implementation 
of symmetry reduced expressions, either octant, Cartoon, or their 
combination we evolve weak~$A=1$ centered pure plus polarization 
initial data as described in section~\ref{subsection:H_plus_data}, 
using once again the generalized harmonic gauge~\eqref{eqn:3+1_ghg}
and the gauge boundary condition~\ref{eqn:G_BCs_Freeze}. We started 
with a base cubed sphere 3d grid with $N=15$ points per direction, 
and the number of subpatches derived 
from~$\mathcal{N}_{\textrm{cu}}=5,\,\mathcal{N}_{\textrm{cs}}=4$
and~$\mathcal{N}_{\textrm{cs}}=3$. The outer boundary was placed 
at~$r=12$ in the units of the code. For ease of comparison, the 
breakdown of the grids was:
%%%%%%%%%%%%%%%%%%%%%%%%%%%%%%%%%%%%%%%%%%%%%%%%%%%%%%%%%%%%%
% table in text:
%%%%%%%%%%%%%%%%%%%%%%%%%%%%%%%%%%%%%%%%%%%%%%%%%%%%%%%%%%%%%
\begin{center}
\vspace{0.25em}
\begin{tabular}{|c|c|c|c|c|c|}
\hline
           & $\mathcal{N}_{\textrm{cu}}^{\textrm{total}}$ 
           & $\mathcal{N}_{\textrm{cs}}^{\textrm{total}}$ 
           & $\mathcal{N}_{\textrm{ss}}^{\textrm{total}}$
           & $\mathcal{N}^{\textrm{total}}$
           & $N^{\textrm{total}}$ \\
\hline
3d         & $125$ & $600$ & $450$ & $1175$ &$4\times10^6$\\
octant     & $27\,\,(12,6,1)$ & $48\,\,(48,12)$ & $81\,(36,9)$ & $216$ & $5\times10^5$\\
\hline
Cartoon    & $25$       & $80$       & $60$       & $165$  & $4\times10^4$ \\
cart. oct. & $9\,(4,1)$ & $24\,(8)$ & $18\,(6)$ & $51$ & $10^4$\\
\hline
\end{tabular}
\vspace{0.25em}
\end{center}
%%%%%%%%%%%%%%%%%%%%%%%%%%%%%%%%%%%%%%%%%%%%%%%%%%%%%%%%%%%%%
where the numbers in parentheses denote the number of those 
grids that were cut in half (at the axis) once, twice, or three 
times respectively, for the 3d grids, and once or twice for the 
Cartoon grids. Note that our current non-octant Cartoon 
implementation is not optimal because we evolve the whole~$x$-$z$ 
plane, wasting effectively a factor of two. Currently we use the 
code most often in Cartoon octant mode, so fixing this does not 
have a high priority. Looking at the table the main observation
is that the expected reduction factor of eight (four) in the total 
number of grid points is present between the 3d (Cartoon) and octant
grids, but that this number is not so closely reflected in the 
grid breakdown, where we get only a factor six (three) in the 
total number of grids. This is obviously because there are many 
grids with fewer points. Since our parallelization does not take
this fact into account, it is possible that one MPI process is 
given {\it all} non-cut grids, and so we can expect that the speedup 
rate is determined to a large extent by ratio in the number of 
grids. As we make the domain larger the relative number of 
cut grids decreases, so we might expect that asymptotically the 
full speedup factors of eight or four can are attained.

%%%%%%%%%%%%%%%%%%%%%%%%%%%%%%%%%%%%%%%%%%%%%%%%%%%%%%%%%%%%%
\paragraph*{Basic dynamics:} Although irrelevant for the octant
Cartoon comparison, since these data have not been used before, 
we give a brief description of their evolution. Initially the 
peak of the Kretschmann scalar occurs at~$\rho=\pm 0.65$ 
with a value~$1.18$. This profile then oscillates about three
times at the origin, attaining a peak value of~$7.25$ before 
rapidly dispersing. Looking at the lapse we see the familiar 
behavior that at the origin it oscillates slightly before 
presenting a longer decrease, although at the minimum is 
only~$0.995$, having started from~$\alpha(t=0)=1$ everywhere. 
Afterwards this pulse propagates out, roughly following the 
disturbance in the Kretschmann. Looking at the shift 
component~$\beta^x$ along the $x$-axis we find that early on 
there is a growth which peaks at~$x=1.06$, with value~$0.0027$.
The development of the shift looks more like a slowly oscillating 
standing wave than a localized propagating feature.

%%%%%%%%%%%%%%%%%%%%%%%%%%%%%%%%%%%%%%%%%%%%%%%%%%%%%%%%%%%%%
\paragraph*{3d, octant, Cartoon and octant-Cartoon comparison:}
Taking first the 3d and octant evolutions, we see near perfect
agreement throughout the evolution. There are small differences 
however, starting from the beginning of the simulation at the 
level of round-off; differences of~$10^{-15}$ in metric components, 
which slowly drift as the evolution goes on. This behavior is 
expected because the derivative approximation differ at this 
level. Similar differences were found between the other setups.
These differences are never larger than the constraint violation, 
in for example~$C_x$, and we have looked at convergence (see 
section~\ref{section:Convergence} for more discussion) with each 
setup, although not for this data, and find no indication of a 
problem. For the speed comparison we ran the code with each setup 
on 24 cores (with hyperthreading) of our local cluster~{\it Core12} 
with Intel Xeon X5650 processors. The octant run was a little 
more than~$6$ times faster than the 3d run, as expected given 
the foregoing discussion. The octant Cartoon run was about~$2.4$ 
times faster than the pure Cartoon test, which is a little 
disappointing. Going from~$\mathcal{N}_{\textrm{ss}}=3$ 
to~$\mathcal{N}_{\textrm{ss}}=6$ radial subdivisions in the outer 
shells, this value increases to~$2.9$, demonstrating the expected
dependence. Comparing the full 3d and octant Cartoon runs, 
there was a gratifying speed up of nearly a factor~$400$. 

%%%%%%%%%%%%%%%%%%%%%%%%%%%%%%%%%%%%%%%%%%%%%%%%%%%%%%%%%%%%%
\subsection{Convergence}\label{section:Convergence}
%%%%%%%%%%%%%%%%%%%%%%%%%%%%%%%%%%%%%%%%%%%%%%%%%%%%%%%%%%%%%

\begin{figure*}[t]
  \centering
  \includegraphics[width=\columnwidth]{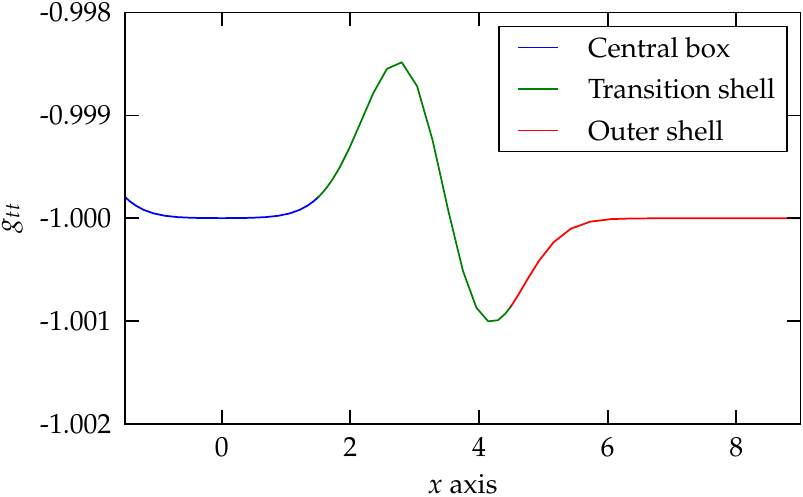}
  \includegraphics[width=\columnwidth]{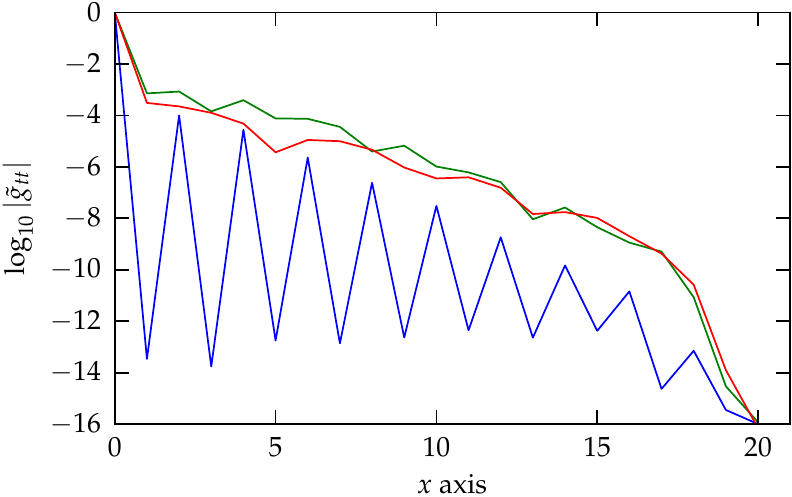}\\
  \includegraphics[width=\columnwidth]{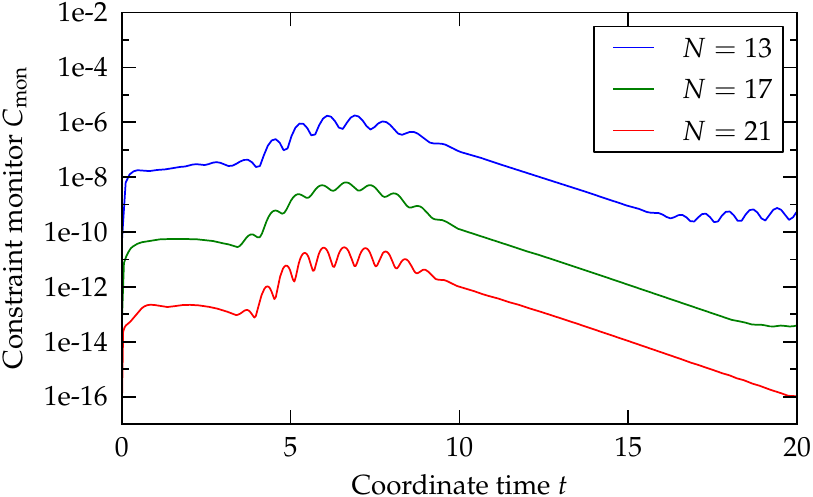}
  \includegraphics[width=\columnwidth]{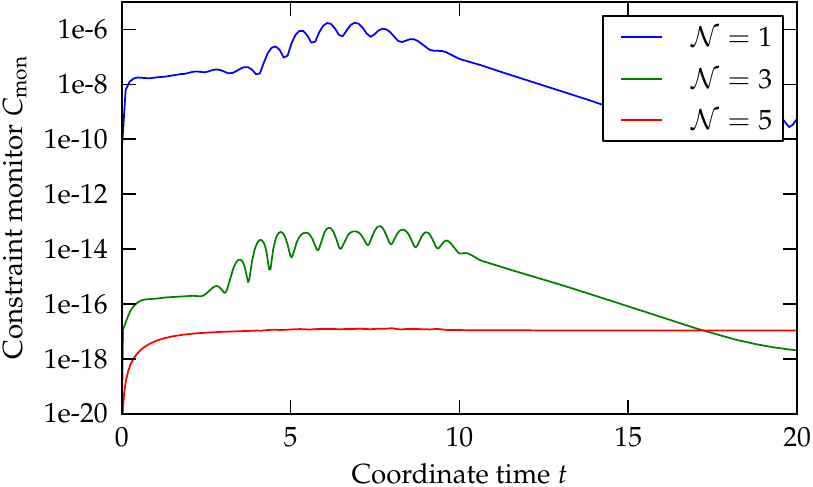}
  \caption{Evolution of a gauge wave with $A=0.01$ and $\sigma=1.0$.
  In the upper panels we used a spatial resolution of~$N=21$ on 
  a grid with~$\mathcal{N}=1$ subpatches. 
  The upper left panel  gives a snapshot of~$g_{tt}$ along
  the~$x$ axes at~t=$3.55$. The upper right shows the Chebyshev
  expansion coefficients at the same time with the same color 
  coding. The lower panels show convergence of the constraints 
  for the same initial data;
  on the left we increase the number of points~$N$ in each grid,
  on the right we increase the number of subpatches~$\mathcal{N}$.
  \label{Fig:Bamps_Conv}
  }
\end{figure*}

The \texttt{bamps} numerical method gives us two options for increasing 
resolution. The first is to add grid-points in every domain, 
the second is to subdivide grids further, keeping the number
of points inside each subpatch fixed. Given fixed finite 
computational resources it is not obvious what is the optimum 
strategy to achieve the smallest possible error, because 
although we would expect that adding points brings spectral
convergence, it also comes with a~$N^{-2}$ dependence in the 
allowed time-step, whereas on the other hand, as we will see, 
adding more subpatches allows the code to scale up to a large
number of processors. Probably the optimal strategy relies on 
a balance between each. To examine the effect of each strategy 
in the simplest possible way, we evolved gauge wave 
initial data on the Minkowski spacetime, which was setup by 
choosing~$\alpha=1+A\exp[-(r/\sigma)^2]$, $\beta^i=0$, 
with~$r=\sqrt{x^2+y^2+z^2}$ as usual, and otherwise the flat 
spatial Cartesian metric and vanishing extrinsic curvature.
The results are plotted in the four panels of 
Fig.~\ref{Fig:Bamps_Conv} and confirm our expectations. 

%%%%%%%%%%%%%%%%%%%%%%%%%%%%%%%%%%%%%%%%%%%%%%%%%%%%%%%%%%%%%
%%%%%%%%%%%%%%%%%%%%%%%%%%%%%%%%%%%%%%%%%%%%%%%%%%%%%%%%%%%%%

%%%%%%%%%%%%%%%%%%%%%%%%%%%%%%%%%%%%%%%%%%%%%%%%%%%%%%%%%%%%%
\subsection{Filtering}\label{section:Filtering}
%%%%%%%%%%%%%%%%%%%%%%%%%%%%%%%%%%%%%%%%%%%%%%%%%%%%%%%%%%%%%

\begin{figure*}[t]
  \centering
  \includegraphics[width=\columnwidth]{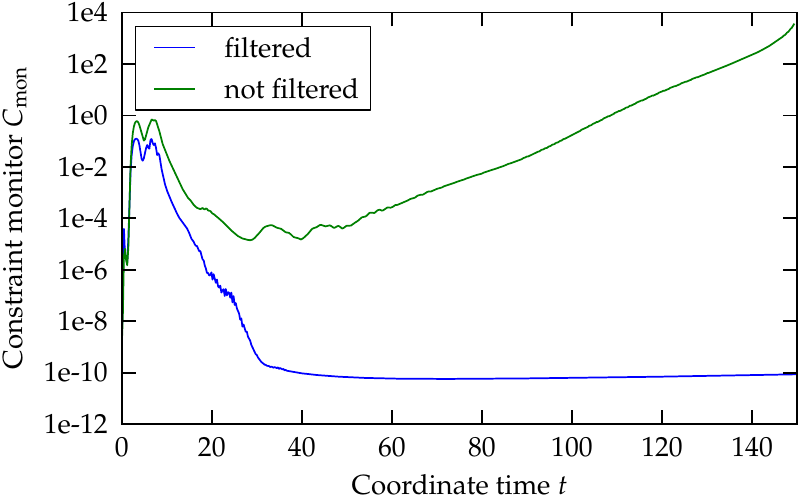}
  \includegraphics[width=\columnwidth]{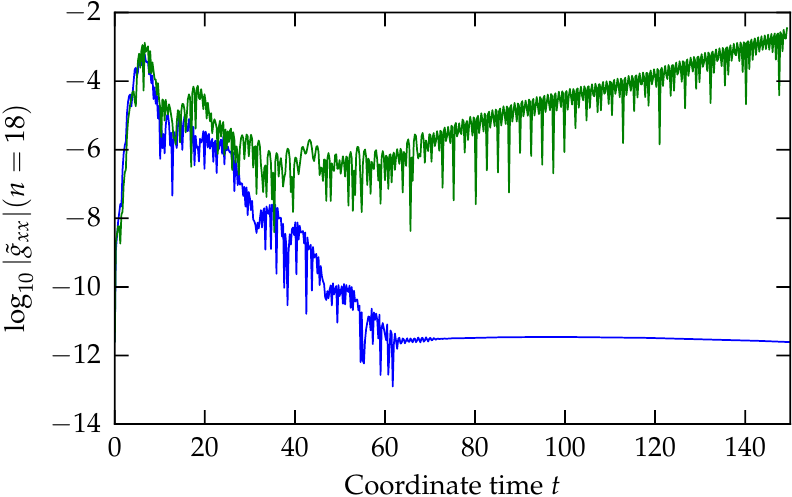}
  \caption{Influence of the filter at example of a~$A=1$ Brill wave evolution.
  On the left we show the time evolution of the constraint monitor~$C_\text{mon}$.
  In the simulation using a filter the constraint violation settle down
  to~$10^{-10}$. Without using a filter the constraint violation grows and lead
  to a failure of the simulation at~$t\approx150$.
  On the right we show the evolution of the fourth highest Chebyshev expansion
  coefficient. It is the highest mode which is not affected by the filter.
  Without the filter the high frequencies grow over time and cause the simulation
  to fail. The filter sets the highest frequency to zero which avoids the growth
  of the high frequency modes.
  \label{Fig:Bamps_filter}
  }
\end{figure*}

To demonstrate the necessity of the filter~\eqref{eqn:filter}
we evolved a centered~$A=1$ Brill wave. The results are plotted 
in Fig.~\ref{Fig:Bamps_filter}. In the left panel we see that 
without filtering the constraint violation starts to grow 
exponentially in time, whereas with filter the growth is completely 
absent and the norm of constraints remains steady at a very 
low value. In the right panel we plot the magnitude of the 
fourth highest spectral coefficient of~$g_{xx}$ in the transition 
shell as a function of time. This coefficient is the first that 
is directly unaffected by the filter. We see that the growth
in the constraints seems to be associated with an explosion
in the higher spectral coefficients. Interestingly we tried 
the same experiment with gauge wave initial data, and did not 
see the effect, at least in the same time-frame. We expect that
the same behavior would manifest if we were to evolve long 
enough. The obvious conclusion we draw from this is that it is 
important to test these methods with several data types to get 
a reliable picture of their properties.

%%%%%%%%%%%%%%%%%%%%%%%%%%%%%%%%%%%%%%%%%%%%%%%%%%%%%%%%%%%%%
%%%%%%%%%%%%%%%%%%%%%%%%%%%%%%%%%%%%%%%%%%%%%%%%%%%%%%%%%%%%%

%%%%%%%%%%%%%%%%%%%%%%%%%%%%%%%%%%%%%%%%%%%%%%%%%%%%%%%%%%%%%
\subsection{Performance}\label{section:Performance}
%%%%%%%%%%%%%%%%%%%%%%%%%%%%%%%%%%%%%%%%%%%%%%%%%%%%%%%%%%%%%

\begin{figure}[t]
  \centering
  \includegraphics[width=\columnwidth]{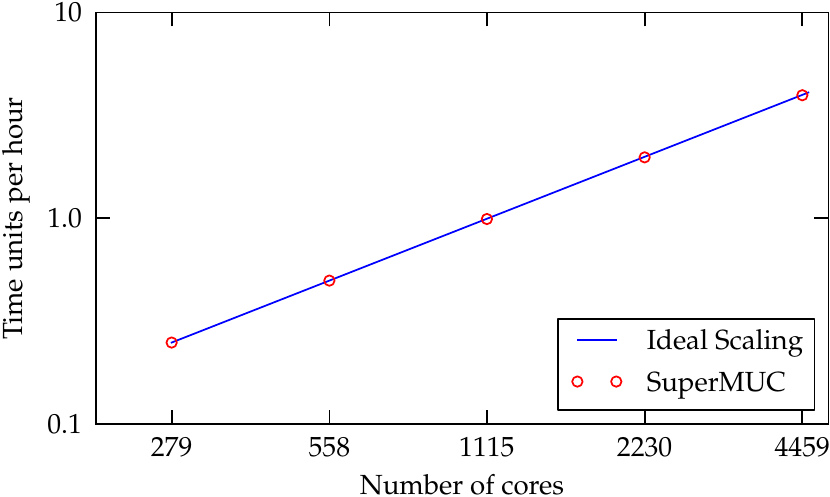}
  \caption{Strong scaling of \texttt{bamps} with $\mathcal{N}=5$
  on the SuperMUC cluster. Here a grid with $\mathcal{N}=5$ sub patches
  was used. In total this grid consists of $4459$ patches. 
  \label{Fig:Bamps_Scaling}
  }
\end{figure}

%%%%%%%%%%%%%%%%%%%%%%%%%%%%%%%%%%%%%%%%%%%%%%%%%%%%%%%%%%%%%
\paragraph*{Strong-scaling:} The current \texttt{bamps} parallelization 
strategy is to obtain perfect scaling using many subpatches, and 
splitting these subpatches across many processors. The key is 
that, contrast to buffer zones required in the decomposition
of a finite differencing grid, only~$2$d surfaces of points need
be passed by network communication, making the relative time 
spent there negligible. In a finite differencing approach the
relative size of the buffer zones decreases with 
resolution, but in practice can still be significant in production
runs. In Fig.~\ref{Fig:Bamps_Scaling} we present strong scaling 
plots performed on the {\it SuperMUC} cluster 
located in LRZ Garching, with Intel Xeon~E$5$-$2680\,8$C 
processors. We ran the code in~$3$d. We took a grid with~$4459$ 
total subpatches, and increased the number of cores used until we 
were computing one patch per core. We find perfect scaling. On the 
other hand \texttt{bamps} is currently not parallelized whatsoever 
at the subpatch level, which means that 
the maximum number of points per subpatch is in principle determined 
by the amount of memory available to one core. At least when 
running the code in Cartoon mode however we do not find, in practical 
terms, that this is problematic. Instead the~$N^{-2}$ restriction in 
the time step makes increasing the number of points infeasible long 
before we are close to filling the available memory. In $3$d this may 
no longer be the case. We leave such considerations for future 
work. 

%%%%%%%%%%%%%%%%%%%%%%%%%%%%%%%%%%%%%%%%%%%%%%%%%%%%%%%%%%%%%
%%%%%%%%%%%%%%%%%%%%%%%%%%%%%%%%%%%%%%%%%%%%%%%%%%%%%%%%%%%%%

%%%%%%%%%%%%%%%%%%%%%%%%%%%%%%%%%%%%%%%%%%%%%%%%%%%%%%%%%%%%%
\section{Single blackholes}\label{section:Blackholes}
%%%%%%%%%%%%%%%%%%%%%%%%%%%%%%%%%%%%%%%%%%%%%%%%%%%%%%%%%%%%%

The main thrust of our development has been towards treating 
collapsing axisymmetric gravitational waves accurately. For 
super-critical data the cubed-ball grid is unsuitable after
the formation of an apparent horizon. Therefore the strategy 
for long-term evolution is to take the data and interpolate 
onto a cubed-shell grid, with the excision surface 
suitably positioned, changing the lapse and shift to be sure 
that the excision surface is a true outflow boundary. A necessary 
requirement is to treat a single blackhole, which is 
what we discuss here.

%%%%%%%%%%%%%%%%%%%%%%%%%%%%%%%%%%%%%%%%%%%%%%%%%%%%%%%%%%%%%
\subsection{Initial data}
\label{subsection:BH_ID}
%%%%%%%%%%%%%%%%%%%%%%%%%%%%%%%%%%%%%%%%%%%%%%%%%%%%%%%%%%%%%

%%%%%%%%%%%%%%%%%%%%%%%%%%%%%%%%%%%%%%%%%%%%%%%%%%%%%%%%%%%%%
\paragraph*{Kerr-Schild coordinates:} 
We evolve the Schwarzschild solution in Kerr-Schild coordinates
as was done with an earlier version~\cite{Bru11} of the present 
code. Although the current numerical method is not particularly 
close to that used previously, some components of the 
older code were inherited. Importantly evolving this data allows 
a simple comparison with the previous method and results. In 
spherical polar coordinates the metric and extrinsic curvature take 
the form,
\begin{align}
g_{ab}\,\textrm{d}x^a\textrm{d}x^b&=
-\left(1-\frac{2M}{r}\right)\textrm{d}t^2
+\frac{4M}{r}\,\textrm{d}t\,\textrm{d}r\nonumber\\
&\quad+\left(1+\frac{2M}{r}\right)\textrm{d}r^2+r^2\textrm{d}\Omega^2\,,
\end{align}
with~$\textrm{d}\Omega^2$ the flat metric on the two-sphere, and
\begin{align}
K_{ij}\textrm{d}x^i\textrm{d}x^j&=\,
-\frac{2M}{\sqrt{1+\frac{2M}{r}}}\left[\frac{1}{r^2}
\left(1+\frac{M}{r}\right)\textrm{d}r^2
-\textrm{d}\Omega^2\right]\,,
\end{align}
respectively. Inside the code the line-element is written in 
Cartesian coordinates in the standard way. More discussion of 
Kerr-Schild coordinates can be found in~\cite{Coo00b,MatHuqSho98}.

%%%%%%%%%%%%%%%%%%%%%%%%%%%%%%%%%%%%%%%%%%%%%%%%%%%%%%%%%%%%%
\paragraph*{Harmonic Killing coordinates:} 
We additionally evolve starting from the harmonic Killing 
slicing described in~\cite{CooSch97}, which serves as a convenient 
starting point when transitioning from one generalized harmonic 
gauge to another. For this initial data, in spherical polar 
coordinates, the metric and extrinsic curvature are 
\begin{align}
g_{ab}\textrm{d}x^a\textrm{d}x^b&=-\left(1-\frac{2M}{r}\right)\textrm{d}t^2
+\frac{8M^2}{r^2}\,\textrm{d}t\,\textrm{d}r\nonumber\\
&\quad+\left(1+\frac{4M^2}{r^2}\right)\left(1+\frac{2M}{r}\right)\textrm{d}r^2
+r^2\textrm{d}\Omega^2\,,
\end{align}
and
\begin{align}
K_{rr}&=-\frac{4M^2}{r^6}\,\frac{4M^3+4M^2r+3Mr^2+2r^3}
{\sqrt{1+\frac{2M}{r}}\,\sqrt{1+\frac{4M^2}{r^2}}}\,,\nonumber\\
K_{\theta\theta}&=\frac{4M^2r^2}{\sqrt{1+\frac{2M}{r}}\,
\sqrt{1+\frac{4M^2}{r^2}}}\,,
\end{align}
with the remaining components vanishing. For this data spatially 
harmonic coordinates are obtained by building Cartesians according 
to,
\begin{align}
x&=(r-M)\sin\theta\cos\phi\,,&\quad 
y=(r-M)\sin\theta\sin\phi\,,\nonumber\\
z&=(r-M)\cos\theta\,,\label{eqn:Harm_Cart_Coords}
\end{align}
The resulting metric has a coordinate singularity at~$r=M$, 
with~$r$ implicitly defined in the obvious way from the new 
coordinates. The coordinate singularity is not a principle 
problem as we could just put the excision surface outside this 
radius. But \texttt{bamps} relies on standard Cartesian coordinates in 
several places. So in the code we could transform in the standard 
way but then choose the gauge source function,
\begin{align}
H^a&=2(\tilde{J}\p \tilde{J})^{(ab)}{}_b\,.\label{eqn:Harm_Killing_sources}
\end{align}
with~$J^a{}_{a'}$ the Jacobian between the standard~$a$-index 
Cartesians and harmonic Cartesian~$a'$ index 
coordinates~\eqref{eqn:Harm_Cart_Coords}, the compound 
object~$(\tilde{J}\p\tilde{J})$ is defined by,
\begin{align}
(\tilde{J}\p\tilde{J})^a{}_{bc}&=(\tilde{J}^{-1})^{a'}_{b}\p_c\tilde{J}^a_{a'}. 
\end{align}
with~$\tilde{J}^a_{a'}=\sqrt{|J|}J^a_{a'}$ and where indices are 
manipulated in the obvious way with~$g_{ab}$ to 
obtain~\eqref{eqn:Harm_Killing_sources}. Instead we just choose the 
gauge source function to be fixed at its initial value, as will 
momentarily be discussed. In this section we use the code exclusively 
in Cartoon mode, on a cubed sphere grid. We start with the excision 
surface at~$r=1.8\,M$, and the outer boundary at~$r=31.8\,M$. In our 
base setup we take~$\mathcal{N}=3$ radial subpatches each with~$N=25$ 
points per direction. The runs were performed on a desktop machine 
with an eight-core intel i7 CPU, which was able to compute at 
about~$250M/\textrm{hour}$, the base run requiring about~$14$~MB 
of RAM.

%%%%%%%%%%%%%%%%%%%%%%%%%%%%%%%%%%%%%%%%%%%%%%%%%%%%%%%%%%%%%
\subsection{Freezing gauge source functions}
\label{subsection:Frozen_Sources}
%%%%%%%%%%%%%%%%%%%%%%%%%%%%%%%%%%%%%%%%%%%%%%%%%%%%%%%%%%%%%

%%%%%%%%%%%%%%%%%%%%%%%%%%%%%%%%%%%%%%%%%%%%%%%%%%%%%%%%%%%%%
\paragraph*{Killing gauge sources:} Given initial data which 
admit a time-like Killing vector, we can ensure that the 
evolution of the system is trivial,  at the continuum level, 
neglecting the effect of outer boundary conditions, 
by choosing the Killing lapse and shift, and taking the gauge 
source functions~$H_a$ so that~$\p_t\alpha=\p_t\beta^i=0$ initially. 
In particular we must choose,
\begin{align}
H_a&=-\Gamma_a(t=0)\,,&\quad \p_tH_a&=0\,.
\end{align}

%%%%%%%%%%%%%%%%%%%%%%%%%%%%%%%%%%%%%%%%%%%%%%%%%%%%%%%%%%%%%
\paragraph*{Kerr-Schild evolutions with SpEC GHG:} We began by 
evolving the Kerr-Schild initial data with the standard 
formulation parameters of~\cite{LinSchKid05}, 
namely~$\gamma_4=\gamma_5=0$ and~$\gamma_0=1$ on our base grid
as just described, using the gauge boundary conditions~\eqref{eqn:G_BCs}.
Immediately we see that the innermost subpatch has the largest 
constraint violation, peaked at around~$10^{-6}$ in the~$C_x$ 
component of the harmonic constraint. This is not surprising 
because the innermost subpatch contains the part of the solution 
with the largest derivatives. The evolution successfully continues 
until the final time~$t=1000\,M$. But after the initial expansion 
to~$10^{-6}$, a slow expansion in~$C_x$ is visible, and this growth 
becomes more rapid as the simulation continues. By the end, the 
maximum value of~$C_x$ is around~$10^{-3}$, with peaks appearing 
at the inner and outer boundary of roughly the same size. We then 
increased resolution from the base grid to~$N=27,29$ and~$N=31$. The~$N=27$
point grid runs at about~$178\,M/\textrm{hour}$, and the initial
peak in the~$C_x$ constraint violation is reduced by a factor of 
about two, with this ratio of improvement slowly declining until 
the end of the evolution. The~$N=29$ grid runs 
at~$129\,M/\textrm{hour}$, with both the initial magnitude of the 
violation and the `slow expansion' of the~$C_x$ constraint 
quashed, the peak being a factor~$2.8$ smaller than in the base 
run at the end of the simulation. The highest resolution~$N=31$
point grid ran at~$96\,M/\textrm{hour}$, with the final 
improvement in~$C_x$ against the base run being a factor 
of~$5.3$. Since the largest constraint violation occurs in the 
excision subpatch an obvious question is whether or not the 
excision and outer boundaries would interact badly if they 
were on the same grid. Although the issue is of little 
practical concern for production runs, for development it 
deserves a little attention, and therefore we evolved our
base grid from before, but cutting the outer two subpatches 
so that the outer boundary lies at~$11.8\,M$. This test is not 
completely fair because the outer boundary conditions are 
expected to perform better as they are applied further out. 
We find that the initial peak in the violation of the~$C_x$
constraint is about five times greater than in the base run 
at~$t=200\,M$. At the end of the evolution again 
at~$t=1000\,M$ by coincidence the constraint violation in the 
restricted domain is smaller, but this is just because the slow 
oscillations in each simulation are out of phase.

\begin{figure}[t]
\centering
\includegraphics[width=\columnwidth]{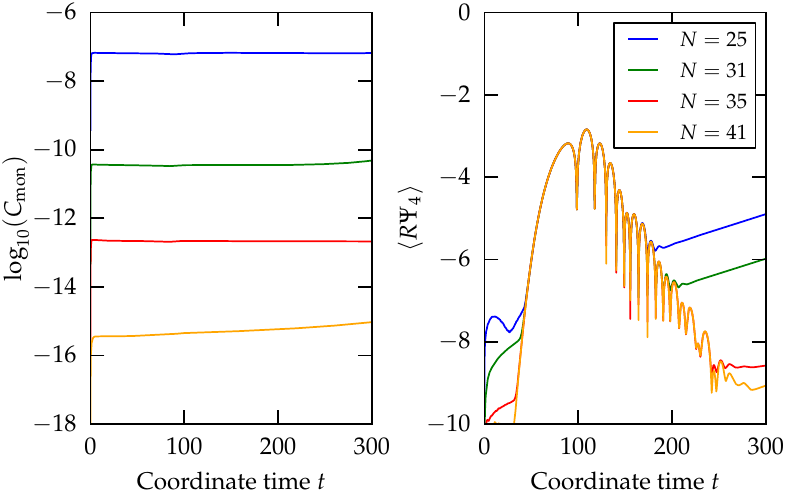}
\caption{The right panel shows the average over the Weyl 
scalar~$\Psi_4$ in the outer boundary in the evolution of 
Schwarzschild perturbed by a small gravitational wave injected 
through the boundary. In the left panel we see convergence of 
the constraints as resolution is increased. At lower resolutions 
a drift is present in the ring-down. There is good agreement with 
Fig.~$3$ of~\cite{LinSchKid05}, and the ring-down frequency agrees 
well with the analytical computation~\cite{ChaDet75}. At the end 
of the test there is some disagreement  with~\cite{LinSchKid05}, 
but since square-roots of very small quantities are being 
taken we expect this is caused by round-off error. It seems 
that on the cubed-sphere grid more resolution is needed to obtain 
clean results than with the spherical harmonic discretization 
used in~\cite{LinSchKid05}. This is perhaps not surprising, since 
the latter discretization is well-suited to the given data. 
\label{fig:Incoming_wave}}
\end{figure}

%%%%%%%%%%%%%%%%%%%%%%%%%%%%%%%%%%%%%%%%%%%%%%%%%%%%%%%%%%%%%
\paragraph*{Kerr-Schild incoming wave evolutions with SpEC GHG:}
Next we evolved the same initial data and gauge, but this time 
with the same domain as in Fig.~$3$ of~\cite{LinSchKid05}. To do
this we took~$\mathcal{N}=2$ radial subpatches, with the same base 
resolution as before, so that the outer boundary is placed 
at~$r=21.8\,M$. We similarly specify exactly the same given data 
for an incoming gravitational wave as in that study, taking in 
particular,
\begin{align}
\p_th_{ab}&=\dot{f}(t)(\hat{x}^a\hat{x}^b+\hat{y}^a\hat{y}^b
-2\hat{z}^a\hat{z}^b)\,,\label{eqn:given_data}
\end{align}
with the vectors here the coordinate vectors defined in the obvious 
way. We take,
\begin{align}
f(t)&=A\exp[-(t-t_p)^2/\omega^2]\,,
\end{align}
with~$A=10^{-3}$, $t_p=60\,M$ and~$\omega=10\,M$. In 
Fig.~\ref{fig:Incoming_wave} we show the results from these experiments, 
obtained with a sequence of different resolutions. We plot the Weyl 
scalar~$\Psi_4$~\eqref{eqn:Weyl_scalars}, averaged over the outer 
boundary,
\begin{align}
4\pi\langle R\Psi_4\rangle^2&=\int|\Psi_4|^2\,\textrm{d}A\,.
\end{align}
The surface area of the outer boundary is~$4\pi R^2$. Fitting the 
highest resolution data between~$t=100$ and~$t=200$ we find a 
ring-down frequency of~$\Re[\,\omega\,M]\sim0.372$ as expected~\cite{ChaDet75}. 
In this evolution we found that the apparent horizon oscillates 
slightly as the gravitational wave is absorbed, increasing the 
horizon mass~\eqref{eqn:M_H_defn} by about~$6\times 10^{-7}\,M$, 
with~$M$ the ADM mass of the analytic initial data. Note that 
the gauge boundary condition being employed here is not identical 
to that used in~\cite{LinSchKid05}, so the agreement is 
remarkable. The effect of much larger pulses of gravitational radiation
falling onto a blackhole using similar gauge conditions was studied 
in~\cite{ChuPfeCoh10}.

%%%%%%%%%%%%%%%%%%%%%%%%%%%%%%%%%%%%%%%%%%%%%%%%%%%%%%%%%%%%%
%%%%%%%%%%%%%%%%%%%%%%%%%%%%%%%%%%%%%%%%%%%%%%%%%%%%%%%%%%%%%

%%%%%%%%%%%%%%%%%%%%%%%%%%%%%%%%%%%%%%%%%%%%%%%%%%%%%%%%%%%%%
\paragraph*{Discussion of and comparison with~\cite{Bru11}:} The 
prior \texttt{bamps} study focussed on obtaining numerical stability in 
the evolution of a single Schwarzschild blackhole with the 
Kerr-Schild slicing. The numerical method used a 
Chebyschev-Fourier-Fourier spatial discretization on a single 
shell with a spin weighted spherical harmonic filter to 
prevent high frequency growth of the error. In that study 
the outer boundary condition employed simply fixed 
the incoming characteristic variables~\eqref{eqn:GHG11_char_vars} 
to some given data, namely their initial values. This approach is 
possible only when the analytic solution is known, otherwise 
incoming constraint violations are generated. Placing the inner 
boundary at~$r=1.8\,M$ and the outer boundary at~$r=11.8\,M$, 
very long evolutions, say until at least~$t=200\,000\,M$, could be 
performed with little resolution, in accordance 
with~\cite{LinSchKid05}. On the other hand, using this method, 
the naive boundary conditions rapidly deteriorated as the outer 
boundary was pushed out, and, crucially resolution did not help 
but rather made the problem worse. A possible explanation for the
latter effect is that no filter was being applied in the radial
(Chebyschev discretized) direction, which have already seen 
is a crucial ingredient for stability with the current method.
The likely cause of the boundary problem is that, as explained 
in~\cite{RinLinSch07}, boundary conditions that just freeze the 
incoming GHG characteristic variables are orders of magnitude
more reflecting than the Sommerfeld like choice contained 
in~\eqref{eqn:G_BCs}. Evidence for this is obtained in the current 
code by changing from the gauge boundary condition~\eqref{eqn:G_BCs} 
to use instead,
\begin{align}
\perp^{(G)cd}_{ab}\big[\p_tu^{\hat{-}}_{cd}\big]&\,\hat{=}\,0\,,
\label{eqn:G_BCs_ruined}
\end{align}
evolving once more the Kerr-Schild initial data on the base grid.
Placing the outer boundary further out then results in greater 
reflections. However rather than trying to improve a condition 
only suitable for evolving known data, we immediately moved to 
the constraint preserving, radiation controlling conditions, with 
which this issue is completely absent. The first attempted 
implementation of a regular center in the \texttt{bamps} code was to use the 
Chebyschev-Fourier-Fourier discretization with a double covering 
in the radial direction, similar to that employed in~\cite{Tre00}.
The approach was not successful, as we always eventually found 
irregularities in the numerical solution at the origin. An 
exponential filter was applied to the Chebyschev coefficients in 
the radial direction, but to little effect. Eventually we settled 
on the cubed sphere approach, in part because of the expectation 
that they will later be more convenient for mesh-refinement. 
Other possible solutions to the problems we faced would be to use 
one-sided Jacobi polynomials as in SpEC~\cite{MuhNouDue14} or to 
employ a filter that projects the solution in another basis onto 
these polynomials.

%%%%%%%%%%%%%%%%%%%%%%%%%%%%%%%%%%%%%%%%%%%%%%%%%%%%%%%%%%%%%
\paragraph*{Kerr-Schild evolutions with simplified constraint
subsystem:} Using our standard choice for the formulation 
parameters~$\gamma_4=\gamma_5=1/2$, and taking~$\gamma_0=0.2$,
returning to our base resolution from the tests with the SpEC
version of GHG, we find that by~$t=200$ the~$C_x$ constraint is 
about~$5$ times larger than that we obtained before, and by the 
end of the simulation the new run has accrued a~$C_x$ constraint 
violation with a sharp peak at the outer boundary of order~$10^{-1}$.
This result seems to be in contradiction to those of 
section~\ref{section:C_ex}, until we remember that there the 
gauge boundary condition~\eqref{eqn:G_BCs_Freeze} was employed. 
Increasing the constraint damping to~$\gamma_0=1$, the initial 
violation is comparable to the SpEC GHG evolution previously
described throughout the evolution, and the spike at the outer 
boundary is suppressed by roughly an order of magnitude. At the end 
of this run the maximum of the~$C_x$ constraint occurs at the 
excision boundary with a value around~$10^{-3}$. This experiment 
thus highlights that the choice of the damping parameters and 
boundary conditions can be rather subtle.

%%%%%%%%%%%%%%%%%%%%%%%%%%%%%%%%%%%%%%%%%%%%%%%%%%%%%%%%%%%%%
\begin{figure}[t]
\centering
\includegraphics[width=\columnwidth]{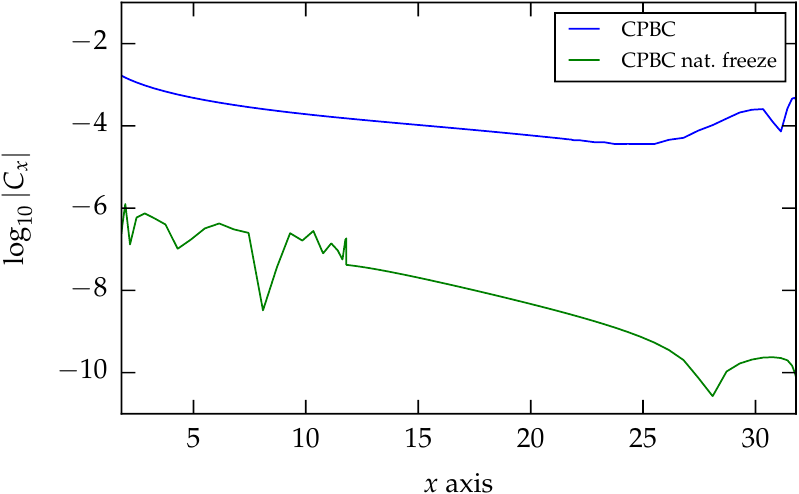}
\includegraphics[width=\columnwidth]{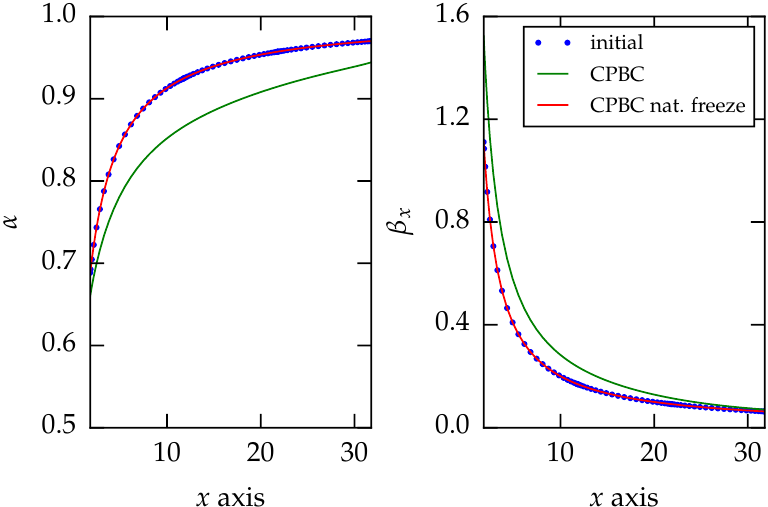}
\caption{Comparison of the evolution of Schwarzschild with 
Killing-Kerr-Schild gauge sources with either the gauge boundary 
condition~\eqref{eqn:G_BCs} or the alternative~\eqref{eqn:G_BCs_Freeze} 
at the end of the simulation~$t=1000\,M$. In the upper panel we plot 
the logarithm of the constraint violation~$C_x$. In the latter 
case the violation is greatly reduced. In the lower two panels 
we show the lapse and shift; the drift present when 
using~\eqref{eqn:G_BCs} is practically absent 
with~\eqref{eqn:G_BCs_Freeze}.\label{fig:bh_bndy_cmp}}
\end{figure}
%%%%%%%%%%%%%%%%%%%%%%%%%%%%%%%%%%%%%%%%%%%%%%%%%%%%%%%%%%%%%

%%%%%%%%%%%%%%%%%%%%%%%%%%%%%%%%%%%%%%%%%%%%%%%%%%%%%%%%%%%%%
\paragraph*{Kerr-Schild evolutions with alternative boundary
conditions:} Next we returned to the base grid, and switched to 
the alternative gauge boundary conditions~\eqref{eqn:G_BCs_Freeze}, 
with~$\gamma_4=\gamma_5=1/2$ and~$\gamma_0=1$. We find that the 
aforementioned growth in the constraints is completely eradicated, 
and the drift in the lapse and shift is also suppressed. Evolving 
the same data with the same formulation and gauge boundary condition, 
but using the modified constraint preserving boundary 
condition~\eqref{eqn:geom_cpbc_1} gives almost identical results.
Using instead the reflection reducing conditions~\eqref{eqn:cpbc_ref_red} 
we see a small improvement in the violation throughout the simulation. 
Repeating the experiment with the incoming gravitational wave injected 
through the boundary with the standard constraint preserving 
condition~\eqref{eqn:GHG_11_cpbc} and the gauge boundary 
conditions~\eqref{eqn:G_BCs_Freeze}, the growth visible in 
Fig.~\ref{fig:Incoming_wave} is also completely absent, even on 
the base resolution~$N=25$ grid. 

%%%%%%%%%%%%%%%%%%%%%%%%%%%%%%%%%%%%%%%%%%%%%%%%%%%%%%%%%%%%%
\paragraph*{Harmonic Killing slice evolutions:} We now returned 
to our base grid and resolution, taking the formulation 
parameters~$\gamma_4=\gamma_5=1/2$, and~$\gamma_0=1$, evolving 
the Harmonic Killing slice with the gauge boundary 
condition~\eqref{eqn:G_BCs}. The test successfully runs 
to~$t=1000\,M$. Comparing with the equivalent evolution of 
Kerr-Schild data, we see that initially near the excision boundary 
the~$C_x$ constraint violation is significantly greater in the 
Harmonic Killing test. By~$t=200\,M$ this difference 
has accrued to around two orders of magnitude. Later however, 
as the violation in the Kerr-Schild Killing evolution starts to grow, it 
overtakes that of the Harmonic Killing evolution. At~$t=1000\,M$ 
the peak of the constraint violation in the Harmonic Killing run 
is about an order of magnitude smaller than in the earlier test. 
As remarked before, in the Kerr-Schild test the inner and outer 
boundaries have roughly the same magnitude in the~$C_x$ constraint 
violation. Interestingly the twin peaks are not present in the 
Harmonic Killing data because the outer boundary is hugely 
improved. This finding is consistent with the gauge wave tests 
presented in section~\ref{section:Exp_G_BCs}, although this test 
is somewhat easier for the gauge boundary conditions because of 
the complete lack of dynamics present in the gauge wave test. In 
the Harmonic Killing evolution we are evolving with pure harmonic 
slicing, and some non-zero spatial gauge source functions, which 
suggests perhaps that the growth at the outer boundary is predominantly 
caused by the use of a non-trivial gauge source function for the 
lapse function, as it interacts with the boundary. Indeed looking 
once more at the lapse function towards the end of the Kerr-Schild 
evolution we see that it is drifting from its initial value, but 
that this effect converges away with resolution. In any case, the 
peak in the constraint violation at the outer boundary in the Killing 
Kerr-Schild data is suppressed as the outer boundary is placed 
further out.

%%%%%%%%%%%%%%%%%%%%%%%%%%%%%%%%%%%%%%%%%%%%%%%%%%%%%%%%%%%%%
\paragraph*{Harmonic Killing slice with gauge perturbation:} 
A desirable property for a set of dynamical coordinates is that 
in the presence of a, perhaps approximate, timelike Killing vector 
they quickly asymptote to a time-independent state. For an 
arbitrary physical or gauge perturbation there is no hope that 
this will occur, and nor can any finite set of numerical experiments 
prove that that there is a basin of attraction to a stationary state. 
We can however look for some indication of this behavior. To do
so we start by taking the initial data for the Killing harmonic 
coordinates, and then perturb the initial lapse function by 
Gaussian as in the previous gauge wave evolutions. In terms of 
the first order GHG variables this is a slightly fiddly procedure, as 
compared with the use of lapse, shift and spatial metric, so we 
give a quick summary:
\begin{itemize}
\item Set spatial metric and extrinsic curvature from the exact
solution.  
\item Take the Killing lapse and shift. Use the 
conditions~$\p_t\alpha=0$ and~$\p_t\beta^i=0$ to set the gauge 
source functions~$H_a$.
\item Add the desired perturbation to the lapse (or shift) and 
then transform to the first order GHG variables. 
\end{itemize}
We perturbed the lapse by a Gaussian,
\begin{align}
\Delta\alpha=A \exp\big[-2\,(r-r_0)^2\big]\,,
\end{align}
with~$A=0.3\,M$ and~$r_0=4\,M$. A similar experiment was made 
in~\cite{LinSzi09}, but starting from a Maximal slice of 
Schwarzschild to test the gauge driver system. We find that the 
perturbation in the lapse propagates away, rapidly leaving 
behind the solution with the harmonic Killing data with unperturbed 
spatial coordinates, or at least negligibly perturbed. The greatest 
danger to the evolution is probably that the excision boundary 
fails to be outflow, but at least with this perturbation that 
does not occur.

%%%%%%%%%%%%%%%%%%%%%%%%%%%%%%%%%%%%%%%%%%%%%%%%%%%%%%%%%%%%%
\paragraph*{Harmonic evolutions with incoming gravitational 
wave:} Giving the same gravitational wave 
data~\eqref{eqn:given_data} as previously, evolving with the 
standard boundary conditions~\eqref{eqn:GHG_11_cpbc} 
and~\eqref{eqn:G_BCs} but using the harmonic Killing gauge source 
functions. It is not obvious how, if at all the spacetime 
computed is related to that considered before, but in any case 
we find a very similar decay in~$\Psi_4$. Remarkably the growth 
present in Fig.~\ref{fig:Incoming_wave} is absent even in this 
low resolution~$N=25$ test. 

%%%%%%%%%%%%%%%%%%%%%%%%%%%%%%%%%%%%%%%%%%%%%%%%%%%%%%%%%%%%%
\subsection{Phasing-in the damped wave gauge}
\label{subsection:Phasing}
%%%%%%%%%%%%%%%%%%%%%%%%%%%%%%%%%%%%%%%%%%%%%%%%%%%%%%%%%%%%%

%%%%%%%%%%%%%%%%%%%%%%%%%%%%%%%%%%%%%%%%%%%%%%%%%%%%%%%%%%%%%
\paragraph*{The transition function:} As elsewhere, we 
follow~\cite{SziLinSch09} to transform from one generalized 
harmonic gauge~$H^1_a$ to another~$H^2_a$. The composite source 
function is simply,
\begin{align}
H_a(t)&=T(t)\,H^1_a+[1-T(t)]\,H^2_a\,.
\end{align}
The transition function is,
\begin{align}
T(t)&=\left\{
\begin{array}{cc}
0\,, &\quad t<t_d\,,\\
\exp\big(-(t-t_d)^2/\sigma_d^2\,\big)\,, &\quad t\geq t_d\,.
\end{array}\right.
\end{align}
In the following experiments we choose~$t_d=0$ and~$\sigma_d=10\,M$. 
Note that care must be taken to construct the time and space 
derivatives of~$H_a$ with the transition function. This choice 
results in gauge source functions that are only~$C^1$ at~$t=t_d$, 
which could be avoided with a different transition function. It 
is not clear if this finite differentiability will have a large 
effect on extracted physical quantities from a simulation.

%%%%%%%%%%%%%%%%%%%%%%%%%%%%%%%%%%%%%%%%%%%%%%%%%%%%%%%%%%%%%
\paragraph*{Kerr-Schild initial slice:} For our first phase-in
test, we started with the Kerr-Schild slicing of Schwarzschild 
and evolved with~$\gamma_4=\gamma_5=1/2$ and~$\gamma_0=1$, on our 
base resolution grid. We took the gauge boundary 
condition~\eqref{eqn:G_BCs} and the constraint preserving 
condition~\eqref{eqn:geom_cpbc_1} (including a~$1/r$ term). 
We used the wave gauge parameters~$p=r=1$ and~$\eta_L=\eta_S=0.1\,M$.
The value of~$\eta_S$ here is much smaller than in our wave 
collapse evolutions. The reason for this is that when evolving 
a blackhole it is crucial that the excision boundary is pure 
outflow in the PDEs sense. In other words the characteristic 
speeds must all have the same outward pointing sign. Since the 
speeds in the~$s^i$ direction are like~$-\beta^s\pm\alpha$ this 
means that the shift can not become too small or else the excision 
boundary will fail, which in turn means that~$\eta_S$ can not be
chosen too large. We therefore place the excision boundary deeper
into the blackhole so that~$r_{\textrm{min}}=M$ and carefully monitor 
the coordinate lightspeeds at the inner boundary. Note that this 
requirement is likely to cause difficulties when computing extreme 
gravitational waves, because on the one hand large shifts can result 
in poor resolution of important features, but on the other they 
may be required in some other region so that we may successfully 
excise the blackhole region. In the evolution we immediately see 
significant dynamics and that for example the peak of the~$C_x$ 
constraint violation along the~$x$-axis is two orders of magnitude
greater than in our initial Kerr-Schild base run with Killing 
gauge sources. The reason for this is presumably the presence 
non-trivial dynamics, plus the fact that we are excising nearer
the physical singularity similar to the effect we saw with the 
harmonic Killing slice. Regardless, by~$t=100\,M$ the data seem 
very close to stationary. The simulation then evolves to the target
time $t=1000\,M$, and remarkably at the end of simulation the 
constraint violation in~$C_x$ along the $x$-axis has a maximum value 
which is an order of magnitude smaller than in the base run. At no 
point does the excision boundary fail to be outflow. As a check of 
the axisymmetric apparent horizon finder we compare the results 
obtained with the simpler algebraic condition,
\begin{align}
H&=\frac{1}{\sqrt{g_{rr}}}\p_r\log(\gamma_{\theta\theta})
-2 K^\theta{}_\theta=0\,.
\end{align}
which characterizes the position of the apparent horizon in 
spherical symmetry. We find near perfect agreement throughout.
The apparent horizon moves from its initial radius $r_H=2.00$
inwards until it reaches $r_H=1.44$ around $t=25$. From there
the horizon starts to grow again and seems to settle down to
$r_H=1.48$. However in our lowest resolution run,
a small drift of the horizon outwards is visible. At late time
of the simulation, around $t=800$, this drift accelerates and
we observe that the horizon becomes aspherical. Higher resolution runs 
show that this effect converges away.

%%%%%%%%%%%%%%%%%%%%%%%%%%%%%%%%%%%%%%%%%%%%%%%%%%%%%%%%%%%%%
\begin{figure}[t]
\centering
\includegraphics[width=\columnwidth]{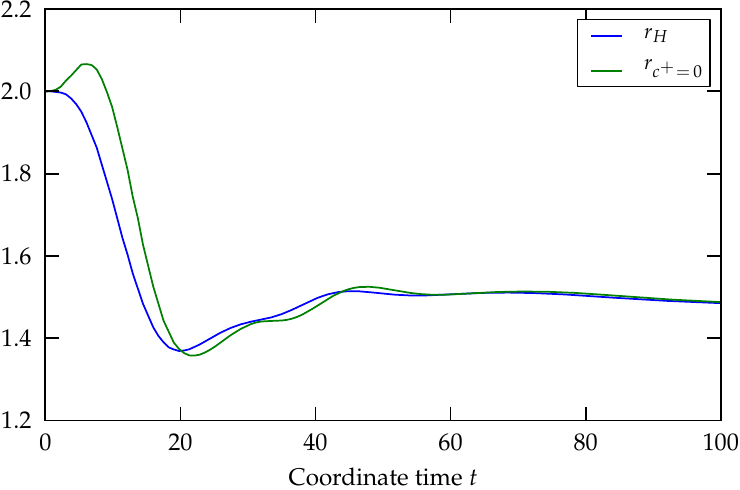}
\caption{The radius of the apparent horizon~$r_H$, and the radius 
at which the outward lightspeed vanishes~$r_{c^+=0}$, computed on 
our base grid with inner boundary at~$r=1.2\,M$. To successfully 
excise, the speed must be negative at the inner boundary. Observe
that excision {\it exactly} on the apparent horizon is not possible 
throughout all of the run.\label{fig:Harm_Phase-in}}
\end{figure}
%%%%%%%%%%%%%%%%%%%%%%%%%%%%%%%%%%%%%%%%%%%%%%%%%%%%%%%%%%%%%

%%%%%%%%%%%%%%%%%%%%%%%%%%%%%%%%%%%%%%%%%%%%%%%%%%%%%%%%%%%%%
\paragraph*{Harmonic initial slice:} Since the stationary 
fully harmonic coordinates are singular at~$r=M$, one might guess 
that the stationary spatial generalized harmonic coordinates with 
gauge source functions~\eqref{eqn:3+1_ghg} are also singular at 
some radius on the Killing slice, at least for some range of the 
parameters~$\eta_L,\eta_S$. Given the broad experience in using 
these coordinates in binary blackhole simulations, the naive 
expectation would be that, if present, this coordinate singularity 
is pushed further towards the physical singularity rather than out 
towards the event horizon for standard choices of the gauge source 
functions. But this behavior is not clear. To truly resolve the 
issue one could simply solve for such coordinates along the lines 
of~\cite{HanHusPol06}, but this we defer for the future. Instead 
we performed simulations varying the initial excision surface from 
the base grid excision radius~$r_{\textrm{min}}=1.8\,M$ down 
to~$r_{\textrm{min}}=1.0\,M$ in steps of~$0.2\,M$. Unsurprisingly we 
find that initially the constraint violation, is greater in the 
excision subpatch as the inner boundary is placed closer to the 
singularity, amounting to about an order of magnitude in 
the~$C_x$ constraint between the~$r_{\textrm{min}}=M$ 
and~$r_{\textrm{min}}=1.2\,M$ boundary runs by~$t=50$. Besides this 
there is little to distinguish between the five runs, and at least
down to this excision radius no sign of a coordinate singularity
forming. By eye the lapse function in the shared part of the 
domain agrees very well throughout the evolution. Although a slight 
drift between them is present towards the end of the test, this is 
acceptable since the outer boundary conditions are being imposed at 
different radii, the solutions need not agree everywhere. There is 
however a time around~$t=20$ above which the runs with inner 
boundary~$r\geq1.4\,M$ fail to be outflow at the excision surface. 
Assuming that this is not caused by numerical error this means 
that boundary conditions are required at the surface. It furthermore 
means that convergence of the numerical scheme as resolution is increased 
is impossible. The fact that this does not correspond to a catastrophic 
failure of the code is inconvenient, because it indicates that great 
care must be taken in monitoring the excision surface. On the other hand, 
since placing the excision boundary very far in has a large cost in accuracy,  
a careful balance must be struck. In the SpEC code this is 
taken care dynamically of by a control mechanism~\cite{HemSchKid13,SchGieHem14} 
which \texttt{bamps} does not yet have. In Fig.~\ref{fig:Harm_Phase-in} the 
relationship between the character of the excision boundary and the 
apparent horizon is examined. Comparing the initially harmonic and 
Kerr-Schild slice evolutions with excision radius~$r_{\textrm{min}}=M$ we find 
that although the lapse functions initially disagree, by 
about~$t=125\,M$ they have exactly the same profile and lie almost 
on top of one another. After this time the agreement is maintained.

%%%%%%%%%%%%%%%%%%%%%%%%%%%%%%%%%%%%%%%%%%%%%%%%%%%%%%%%%%%%%
\section{Evolution of Supercritical Waves}
\label{section:Supercritical}
%%%%%%%%%%%%%%%%%%%%%%%%%%%%%%%%%%%%%%%%%%%%%%%%%%%%%%%%%%%%%

In this section we present the numerical evolution of 
a centered Brill wave, see section~\ref{subsection:Brill}, 
with~$A=8$. This highly supercritical initial data is used as 
a test case for our excision algorithm for a dynamically forming 
blackhole. 

%%%%%%%%%%%%%%%%%%%%%%%%%%%%%%%%%%%%%%%%%%%%%%%%%%%%%%%%%%%%%
\subsection{Dynamical excision strategies}
\label{subsection:dynamical_excision}
%%%%%%%%%%%%%%%%%%%%%%%%%%%%%%%%%%%%%%%%%%%%%%%%%%%%%%%%%%%%%

Our dynamical excision method currently consists of the following
steps:

%%%%%%%%%%%%%%%%%%%%%%%%%%%%%%%%%%%%%%%%%%%%%%%%%%%%%%%%%%%%%
{\bf 1. Evolve to collapse:} Evolve on cubed ball grid, running 
the apparent horizon finder in `daemon' mode. The finder then 
triggers a \texttt{bamps} checkpoint once a horizon is found.

%%%%%%%%%%%%%%%%%%%%%%%%%%%%%%%%%%%%%%%%%%%%%%%%%%%%%%%%%%%%%
{\bf 2. Go-to excision grid:} Next interpolate the checkpoint 
data onto a cubed-sphere grid. In this step we want to place 
excision boundary just inside the apparent horizon, but as we 
have already seen in the single blackhole evolutions this may 
not always be possible, as some wiggle room is needed to allow 
for dynamical behavior of the horizon. This can require some 
experimentation, although fine-tuning does not seem necessary. 

%%%%%%%%%%%%%%%%%%%%%%%%%%%%%%%%%%%%%%%%%%%%%%%%%%%%%%%%%%%%%
{\bf 3. Regauge:} Adjust the lapse and shift to ensure that the
excision boundary is pure outflow. As a particular choice, we 
take lapse and shift from Kerr-Schild slicing of Schwarzschild,
\begin{align}
\quad\alpha=\Big(1+\frac{2m}{r}\Big)^{-1/2},\quad\quad 
\beta^r=\frac{2m}{r}\Big(1+\frac{2m}{r}\Big)^{-1} \,,
\end{align}
and translate to Cartesian components in the obvious way. It is 
desirable that the radial coordinate light-speeds are close to 
zero, preferably positive, at the apparent horizon, since this 
determines the direction of motion of the horizon. Therefore we 
choose the~$m$ parameter to satisfy this condition reasonably well,
although again without particular fine tuning.

%%%%%%%%%%%%%%%%%%%%%%%%%%%%%%%%%%%%%%%%%%%%%%%%%%%%%%%%%%%%%
{\bf 4. Safety-net evolution:} We then use single blackhole gauge source 
parameters like~$\eta_L=0.1$ and~$\eta_S=0.2$. During the evolution 
we use a safety net. If any coordinate light-speed on the excision 
boundary reaches a given threshold, typically~$c_{*}=-0.05$ we again 
regauge to guarantee the outflow character is maintained. We monitor 
the apparent horizon, and if it falls off of the numerical domain we 
return to an earlier checkpoint, regauging with a smaller~$m$ to avoid 
this behaviour. As the horizon expands we monitor the position and periodically 
return to the {\bf Go-to} step above, excising further out and regauging 
with a greater~$m$. 

%%%%%%%%%%%%%%%%%%%%%%%%%%%%%%%%%%%%%%%%%%%%%%%%%%%%%%%%%%%%%
\paragraph*{Discussion:} As currently implemented this procedure requires 
that some steps be performed by hand. The numerical results in the 
following subsection serve to demonstrate `proof of principle' of this 
algorithm. On the other hand it seems at least clear how those 
steps should be automated. At the regauge step the use of the first 
order GHG variables is again a little fiddly. Much more convenient 
would be if the lapse and shift were readily available as variables. 
But the procedure is similar to that described in the gauge 
perturbation tests in section~\ref{subsection:Frozen_Sources}, so we 
do not give full details. Also at the regauge step, it might be good 
to choose lapse and shift by abandoning the spherical ansatz and 
imposing that the coordinate light-speeds at the apparent horizon 
vanish. The SpEC approach to controlling the excision surface is much 
more sophisticated, employing a control mechanism~\cite{HemSchKid13}, we 
hope to avoid that investment in the near future. Because we are 
interested in the collapse of waves to form, presumably, a single 
blackhole it seems reasonable to use a simple approach if at all 
possible. One aspect of the method that is not very aesthetically 
appealing, is that by changing the lapse and shift in discrete steps 
we are computing a spacetime, or patch of spacetime in coordinates 
that are not globally smooth. Another issue associated with this is 
that of geometric uniqueness, which for the IBVP is an open question. 
Nevertheless one expects that the differences to the computed spacetime 
with one choice of regauging parameters or another will be rather small 
in practice, so this does not represent an immediate practical concern. 

%%%%%%%%%%%%%%%%%%%%%%%%%%%%%%%%%%%%%%%%%%%%%%%%%%%%%%%%%%%%%
\subsection{Supercritical Brill wave evolution}
\label{subsection:A=8_Brill}
%%%%%%%%%%%%%%%%%%%%%%%%%%%%%%%%%%%%%%%%%%%%%%%%%%%%%%%%%%%%%

%%%%%%%%%%%%%%%%%%%%%%%%%%%%%%%%%%%%%%%%%%%%%%%%%%%%%%%%%%%%%
\paragraph*{Initial data and grid setup:} We evolved a centered 
Brill wave as described in section~\ref{subsection:Brill}, with 
seed function~\eqref{eqn:brill_seed}. We chose a centered~$\rho_0=0$ 
wave with~$A=8$. The ADM mass of this initial data 
is~$M_{\textrm{ADM}}=1.77$. The maximum of the Kretschmann scalar 
in the initial data occurs at the origin, taking the value~$1.7\times10^4$. 
Following the algorithm just outlined, we began on a cubed-ball grid 
with~$\mathcal{N}_{\textrm{cu}}=11,\,\mathcal{N}_{\textrm{cs}}=13,\,
\mathcal{N}_{\textrm{ss}}=20$, and~$55^3$ points per cube, with internal
boundaries~$r_{\textrm{cu}}=1.5$, ~$r_{\textrm{cs}}=6.5$ and the 
outer boundary placed at~$r=30\simeq17\,M$. We ran the code in 
Cartoon mode on our local cluster~{\it Quadler} with~$240$ cores. 
We evolved with the generalized harmonic gauge, as in 
section~\ref{section:Experiments} in the evolution of a {\it much} 
weaker~$A=2.5$ Brill wave, now with the gauge parameters~$\eta_L=0$ 
and~$\eta_S=6$. At coordinate time~$t=1.95$ we first found an 
apparent horizon with mass~$M_H=1.59\simeq0.9\,M$. 

%%%%%%%%%%%%%%%%%%%%%%%%%%%%%%%%%%%%%%%%%%%%%%%%%%%%%%%%%%%%%
\paragraph*{Continuation to code crash:} If we continue this 
evolution without going to an excision grid after the apparent 
horizon forms, we find that the constraints inside the apparent 
horizon rapidly grow along with the Kretschmann scalar. The run 
then crashes at roughly~$t=3.9$. This gives the clear signal 
that if we are to examine the final masses of blackholes formed 
during collapse, using the GHG formulation, a robust excision 
algorithm will be essential. In fact at~$t=3.85$ the 
horizon has a mass of~$M_H=1.64$ on the cubed-ball grid, but at 
the end of our excision simulation, to be described momentarily,
we find that~$40\,M$ after apparent horizon formation it 
has mass~$M_H=1.70$. In the first critical gravitational wave 
collapse paper~\cite{AbrEva92}, the blackhole masses were evaluated 
roughly~$t=17\,M$ after apparent horizon formation, according to a 
prescription based on the quasinormal modes of Schwarzschild. 
Comparing those values with ours is difficult because we use different 
time coordinates, but the basic expectation is that the Maximal 
slicing condition is more ``singularity avoiding'' than one of our 
generalized harmonic gauges, and therefore we might expect to obtain 
comparable results if we can evolve for a similar coordinate time 
after the appearance of a horizon. This is, however, not clear and 
deserves further investigation. In any case without excising
the blackhole region, the meager~$\sim2\,M$ after collapse is 
clearly insufficient. We have seen in~\cite{HilBauWey13} that with the 
moving-puncture method this type of data also did not result in 
successful evolutions beyond apparent horizon formation. But here at 
least a concrete improvement has been made, in that we find an apparent 
horizon {\it before} the method fails!

%%%%%%%%%%%%%%%%%%%%%%%%%%%%%%%%%%%%%%%%%%%%%%%%%%%%%%%%%%%%%
\paragraph*{Evolution on excision grid:} Checkpointing the 
solution at~$t=3.6$ we then interpolating, again with barycentric 
Lagrange interpolation as used in the apparent horizon finder, onto 
a cubed-sphere grid with excision radius at~$r=0.73\,M$ with the outer 
boundary position fixed, and with~$\mathcal{N}_{\textrm{ss}}=27$ with~$9$ 
angular patches, now with~$35^3$ points per cube, naturally again 
evolving in Cartoon mode. In the regauge step we choose here~$m=0.4$. 
This step immediately removes most of the constraint violation 
from the computational domain, and the largest spatial derivatives, 
so that the constraint monitor is~$\sim10^{-8}$ as compared 
to~$\sim10^{3}$ on the original cubed-ball. This difference seems very 
troublesome until we take into account that, for example the peak of 
the Kretschmann scalar on the cubed ball grid is~$\sim10^3$ whereas on 
the cubed sphere it is~$\sim1$. 
So the reduction in the constraints obviously occurs because we 
are removing the most extreme part of the domain. Note also that our 
definition of the constraint monitor does not include a normalization 
by the size of the solution, as in for example~\cite{LinSchKid05} and 
subsequent papers. In view of this our reduction in resolution is 
justified. The evolution then proceeded, now on~$120$ cores 
using~$\eta_L=\eta_S=0.1$. The regauge safety-net was triggered~$3$ times 
up to~$t=5.9\,M$, having fixed~$c_*=-0.05$, but the apparent horizon 
remains on the computational domain throughout the calculation. 
At~$t=5.9\,M$ we perform the ``{\bf Go-to}'' step of our algorithm 
again, this time excising at~$r=1.0\,M$ choosing~$m=0.8$. After this 
the regauge safety-net was not called before~$t=17\,M$, 
when we changed cubed-sphere grid once more, keeping the same grid 
parameters but excising at~$r=1.12\,M$, and regauging with~$m=1$. 
The evolution continued~$t=24.7\,M$, at which time we changed grid 
for the final time, before which the safety-net was again not called.
In the last grid we took the excision radius to be~$r=1.24\,M$ and 
regauged with~$m=1.2$. After this the regauge safety-net 
was not called, and the evolution was terminated at~$t=50\,M$ 
after apparent horizon formation. Note that in this evolution the 
``{\bf Go-to}'' step also employed the phase-in for the generalized 
harmonic gauge, as described in our single blackhole evolutions in
section~\ref{subsection:Phasing}, taking the same parameters employed 
in those earlier tests, but now with the initial source functions
chosen so that the lapse and shift were frozen as the evolution starts 
on the new grid. Other experiments show that this procedure is not strictly
necessary. It may be that some refinement is required to this method to 
allow the evolution of supercritical data {\it indefinitely} after the 
collapse, but examining the mass of the apparent horizon, we interpret 
the solution as having mostly settled down, which should be good 
enough to diagnose a final mass of the blackhole.

%%%%%%%%%%%%%%%%%%%%%%%%%%%%%%%%%%%%%%%%%%%%%%%%%%%%%%%%%%%%%
\begin{figure}[t]
\centering
\includegraphics[width=\columnwidth]{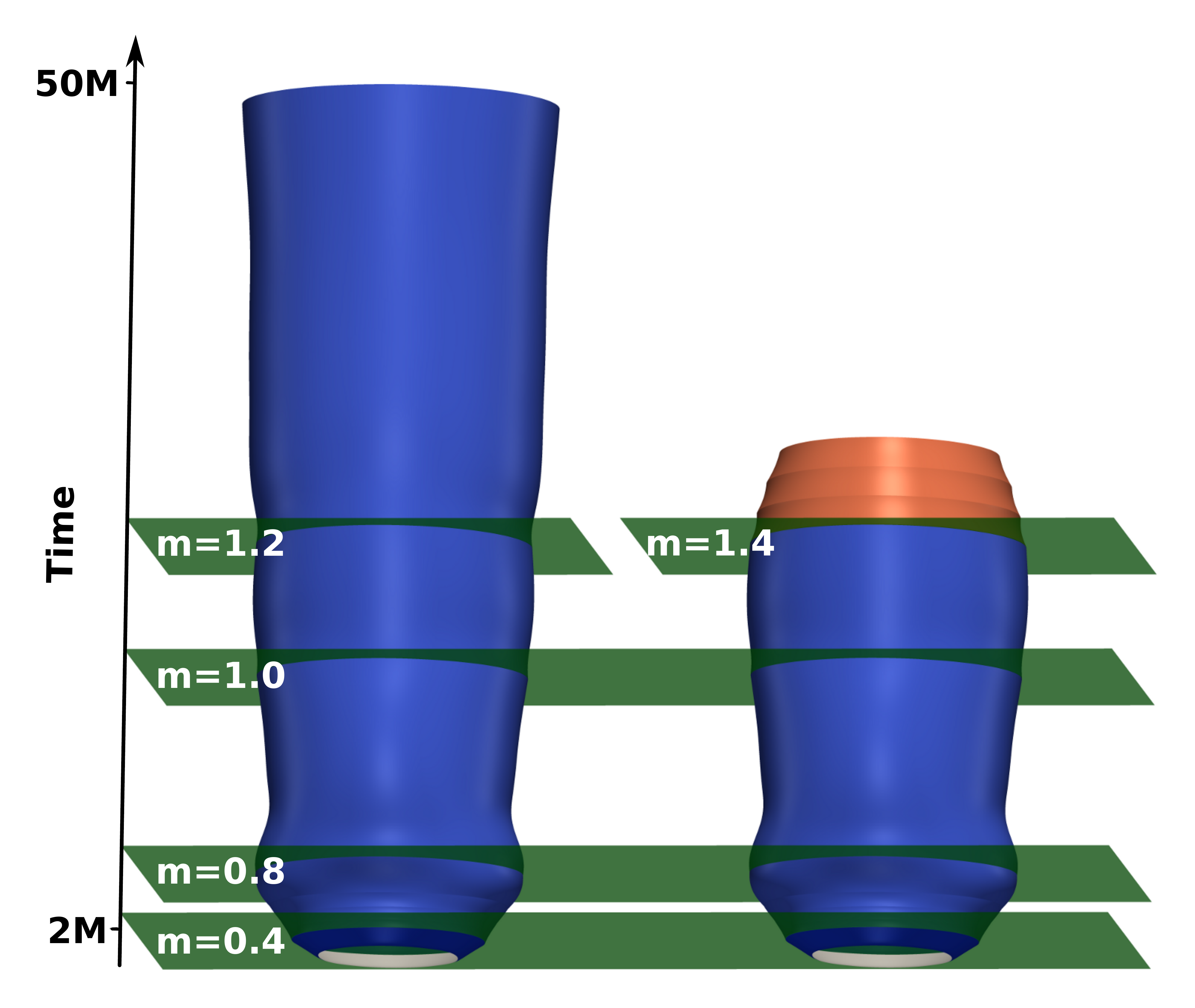}
\caption{The dynamics of the apparent horizon with our dynamical
excision strategy for an~$A=8$ centered Brill wave. The green 
planes indicate the times at which the ``{\bf Go-to}'' step 
was applied, and what parameter~$m$ was chosen in that procedure. 
The left plot shows a successful choice, and on the right 
what happens if this parameter is chosen less carefully. In the 
upper part of the right hand plot one sees that the horizon 
contracts, and also sees that the `regauge' step is frequently 
applied, resulting in kinks in the horizon. 
\label{fig:AH_dynamics}}
\end{figure}
%%%%%%%%%%%%%%%%%%%%%%%%%%%%%%%%%%%%%%%%%%%%%%%%%%%%%%%%%%%%%

%%%%%%%%%%%%%%%%%%%%%%%%%%%%%%%%%%%%%%%%%%%%%%%%%%%%%%%%%%%%%
\paragraph*{Dynamics of the apparent horizon:} In the computation 
described above, as can be seen in in the left panel of 
Fig.~\ref{fig:AH_dynamics}, the apparent horizon is always present 
on the computational domain. The horizon mass initially rapidly grows 
to a value around~$M_H=1.7$ where it remains roughly constant. 
Throughout we see that when the regauge safety-net is 
triggered a slight oscillation in the horizon mass follows. On 
the other hand when we change grid we see that the horizon mass 
exhibits a more prominent kink. In the right-hand panel of 
Fig.~\ref{fig:AH_dynamics} we plot the apparent horizons obtained 
when, less-wisely, the parameter~$m=1.4$ is chosen in the last 
``{\bf Go-to}'' at~$t=24.7\,M$. With this choice the apparent horizon 
rapidly contracts, although the code fails before it leaves the domain. 
The safety-net is called ever-more frequently as the method insists on 
forcing the inner boundary to remain pure outflow, until eventually the 
code crashes at~$t=31.6\,M$. The physical interpretation of this is that 
the excision boundary is falling off of the domain, which starts to 
drift outside the blackhole region, and that the safety-net then forces 
the worldline of the excision boundary to be spacelike. This interpretation 
would be clearer if we had an event horizon finder, but is given credence 
by performing evolutions of a Schwarzschild blackhole with the~$m$ gauge 
parameter similarly poorly chosen. In such tests we see that the areal 
radius of the excision boundary can indeed fall outside of the event 
horizon at~$r=2\,M$. 

%%%%%%%%%%%%%%%%%%%%%%%%%%%%%%%%%%%%%%%%%%%%%%%%%%%%%%%%%%%%%
\section{Conclusions}\label{section:Conclusions}
%%%%%%%%%%%%%%%%%%%%%%%%%%%%%%%%%%%%%%%%%%%%%%%%%%%%%%%%%%%%%

We have developed a pseudospectral numerical relativity code, 
\texttt{bamps}, and in so doing have made a series of improvements and 
investigations into the approach employed in the SpEC code. 
We presented a set of experiments carefully performed so that 
direct comparison with either published work, or independent 
computations of the BAM finite differencing code could be made.
These included evolutions of gauge waves, convergence tests, 
the use of different constraint damping and GHG formulation 
parameters, scaling tests, perturbed blackhole evolutions and 
the treatment of supercritical gravitational waves. Ultimately 
we conclude that the \texttt{bamps} code is working efficiently, scales 
as desired up to large numbers of processors, and works on
sufficiently general grid setups to evolve initial data of 
interest. Particularly surprising to us was the sensitivity of 
the method to our modifications of the GHG boundary conditions, 
even within the class of constraint preserving conditions. This was 
the case even in our simple evolutions of the Schwarzschild spacetime, 
so it would be very interesting to see the extent to which such results 
carry over to compact binary evolutions, be it in SpEC, or in the 
more distant future in \texttt{bamps}. From the physics point of view, however,
our focus is presently on the collapse of axisymmetric gravitational 
waves. Much of the development reflects this fact. Most notably
the implementation of octant symmetry with the Cartoon method gives 
orders of magnitude speedups over evolving the same data in full 3d. 
For a recent complimentary approach see~\cite{SchRin14}. We have additionally 
written a bespoke axisymmetric apparent horizon finder, which already 
proved a valuable diagnostic tool, crucial in the evolution of supercritical 
data, where the existence of an apparent horizon was used as the criterion 
for moving to an excision grid. 

Naturally further developments to the code may be desirable. For 
physical interpretation, an event horizon finder would complement our 
apparent horizon finder. A control system like that of SpEC~\cite{HemSchKid13} 
would be useful in controlling the positions of the apparent horizons.
But the highest priority will likely be in generalizing available grid 
setups to enable dynamical mesh-refinement. 

We have also considered various different types of axisymmetric 
moment of time-symmetry gravitational wave initial data. In forthcoming 
work we use \texttt{bamps} to evolve this initial data, close to the 
critical amplitude separating dispersion and collapse to a blackhole.  

%%%%%%%%%%%%%%%%%%%%%%%%%%%%%%%%%%%%%%%%%%%%%%%%%%%%%%%%%%%%%
\acknowledgments
%%%%%%%%%%%%%%%%%%%%%%%%%%%%%%%%%%%%%%%%%%%%%%%%%%%%%%%%%%%%%

We are grateful to Sebastiano Bernuzzi, David Garfinkle, Enno Harms,
Sascha Husa, Nathan Kieran Johnson-McDaniel, Harald Pfeiffer and
Hannes R\"uter for interesting discussions. DH would like to express 
special gratitude to Helmut Friedrich for interesting discussions and 
for his warm encouragement. This work was supported in part by the Deutsche 
Forschungsgemeinschaft (DFG) through its Transregional Center SFB/TR7 
``Gravitational Wave Astronomy'', by the DFG Research Training Group 
1523/1 ``Quantum and Gravitational Fields'', and the Graduierten-Akademie 
Jena. Computations were performed primarily at the LRZ (Munich). 

%%%%%%%%%%%%%%%%%%%%%%%%%%%%%%%%%%%%%%%%%%%%%%%%%%%%%%%%%%%%%
\begin{appendix}
%%%%%%%%%%%%%%%%%%%%%%%%%%%%%%%%%%%%%%%%%%%%%%%%%%%%%%%%%%%%%

\end{appendix}

%%%%%%%%%%%%%%%%%%%%%%%%%%%%%%%%%%%%%%%%%%%%%%%%%%%%%%%%%%%%%
\bibliographystyle{apsrev}

%%%%%%%%%%%%%%%%%%%%%%%%%%%%%%%%%%%%%%%%%%%%%%%%%%%%%%%%%%%%%

\end{document}